\documentclass[showpacs,amsmath,amssymb,onecolumn,pra,aps,superscriptaddress,notitlepage,twocolumn,longbibliography]{revtex4-2}

\usepackage{amsmath,amssymb,amsthm,mathrsfs,amsfonts,dsfont}
\usepackage{qcircuit}
\usepackage[dvips]{graphicx}
\usepackage{subfigure, epsfig}
\usepackage{braket}
\usepackage{bm}
\usepackage{enumerate}
\usepackage{color}
\usepackage{comment}
\usepackage[normalem]{ulem}

\begin{document}
\title{Quantum error mitigation as a universal error-minimization technique: applications from NISQ to FTQC eras}

\author{Yasunari Suzuki}
\email{yasunari.suzuki.gz@hco.ntt.co.jp}
\affiliation{NTT Computer and Data Science Laboratories, NTT Corporation, Musashino 180-8585, Japan}
\affiliation{JST, PRESTO, 4-1-8 Honcho, Kawaguchi, Saitama, 332-0012, Japan}
\author{Suguru Endo}
\email{suguru.endou.uc@hco.ntt.co.jp}
\thanks{Y.S and S.E contributed to this work equally}
\affiliation{NTT Computer and Data Science Laboratories, NTT Corporation, Musashino 180-8585, Japan}

\author{Keisuke Fujii}
\affiliation{Graduate School of Engineering Science, Osaka University, 1-3 Machikaneyama, Toyonaka, Osaka 560-8531, Japan}
\affiliation{Center for Quantum Information and Quantum Biology, Institute for Open and Transdisciplinary Research Initiatives, Osaka University, Japan}
\affiliation{Center for Emergent Matter Science, RIKEN, Wako Saitama 351-0198, Japan}

\author{Yuuki Tokunaga}
\email{yuuki.tokunaga.bf@hco.ntt.co.jp}
\affiliation{NTT Computer and Data Science Laboratories, NTT Corporation, Musashino 180-8585, Japan}

\begin{abstract}
In the early years of fault-tolerant quantum computing (FTQC),  it is expected that the available code distance and the number of magic states will be restricted due to the limited scalability of quantum devices and the insufficient computational power of classical decoding units. Here, we integrate quantum error correction and quantum error mitigation into an efficient FTQC architecture that effectively increases the code distance and $T$-gate count at the cost of constant sampling overheads in a wide range of quantum computing regimes. For example, while we need $10^4$ to $10^{10}$ logical operations for demonstrating quantum advantages from optimistic and pessimistic points of view, we show that we can reduce the required number of physical qubits by $80\%$ and $45\%$ in each regime. From another perspective, when the achievable code distance is up to about 11, our scheme allows executing $10^3$ times more logical operations. This scheme will dramatically alleviate the required computational overheads and hasten the arrival of the FTQC era.
\end{abstract}
\maketitle

\section{Introduction}
Quantum computers are believed to be capable of implementing several tasks such as factoring and Hamiltonian simulations, in exponentially smaller computational times than those of classical computers~\cite{shor1999polynomial, harrow2009quantum}. However, quantum systems generally interact with their environments, which leads to physical errors in the system that may destroy their quantum advantages. Since the physical error rates of quantum computers are still much higher than those of classical computers, it is vital to suppress these errors. As a solution, fault-tolerant quantum computing~(FTQC) using quantum error-correcting codes has been studied~\cite{nielsen2002quantum,lidar2013quantum,fowler2012surface,horsman2012surface,fowler2018low}. The long-term FTQC allows executing conventional quantum algorithms such as Hamiltonian simulation algorithms~\cite{lloyd1996universal}. According to the current state-of-the-art resource estimations~\cite{kivlichan2020improved,babbush2018encoding}, the logical quantum operation count will be in the order of $10^{10}$ to observe clear quantum advantages based on the computational complexity theory.

\begin{figure*}[t!]
    \centering
    \includegraphics[clip,width=2\columnwidth]{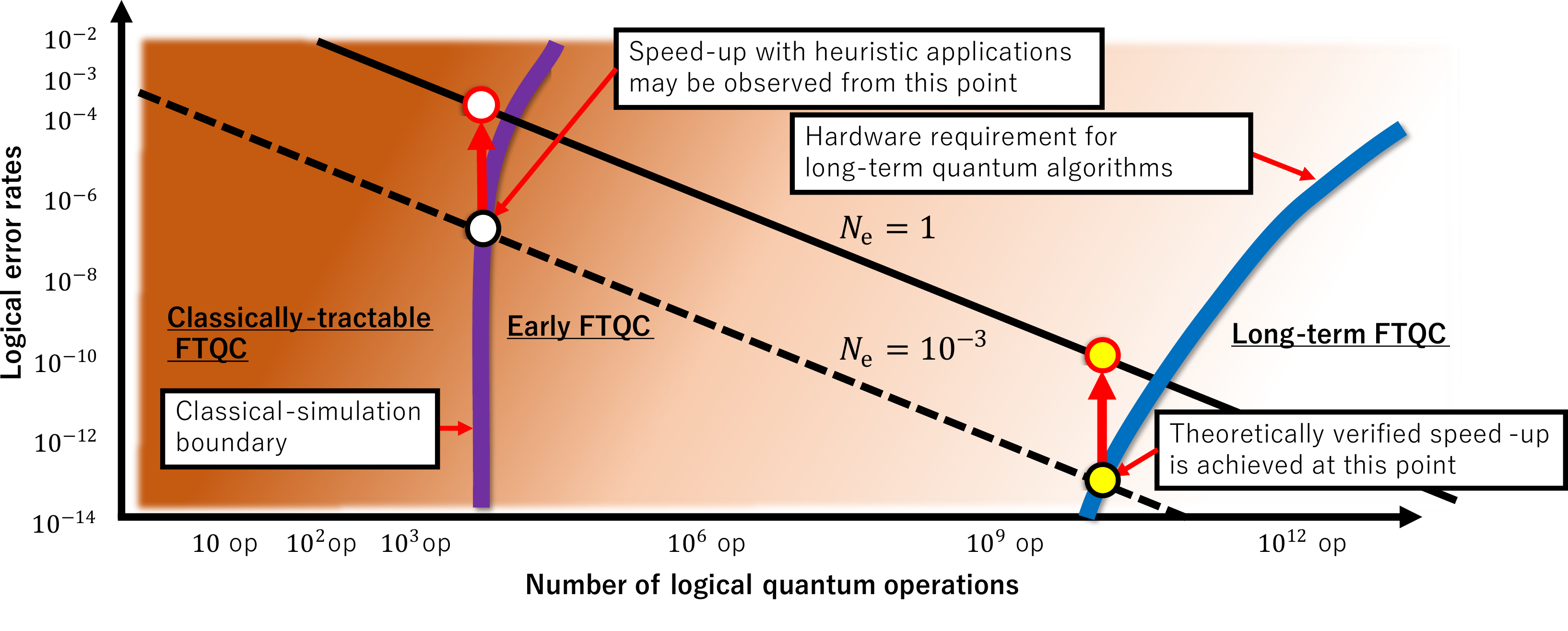}
    \caption{A schematic picture representing the transitional period towards the realization of the long-term FTQC. In the figure, the purple line indicates the hardware requirement for performing classically intractable tasks with a realistic time whereas the blue line corresponds to the requirement for demonstrating quantum advantages with conventional long-term quantum algorithms. To estimate these lines, we refer to the quantum supremacy experiments~\cite{arute2019quantum} and the existing state-of-the-art resource estimation~\cite{kivlichan2020improved,babbush2018encoding,note_blue_line}. The early-FTQC regime is defined as a region between these lines. In the main text, we assume that the number of error events during FTQC $N_{\rm e}$ is required to be smaller than $10^{-3}$, which is shown as the dotted black line. Our technique allows for FTQCs with the number of error events in the order of unity $N_{\rm e} \sim 1$, which is shown as a solid black line, to execute applications that originally require a much smaller error event count. For example, at the beginning and the end of the early-FTQC regime, our technique allows simulating applications (white and yellow circles with black rims) with the relaxed hardware requirement (white and yellow circles with red rims).}
    \label{fig:regime_overview}
\end{figure*}

Towards the realization of the long-term FTQC, we will experience several intermediate regimes as shown in Fig.\,\ref{fig:regime_overview} because high-level encoding is not allowed due to restrictions of quantum resources such as qubit and magic-state count\textcolor{black}{~\cite{fowler2012surface,fowler2018low}}. Since quantum error correction~(QEC) requires massive classical computation for repetitive error estimations, the available code distance would also be strongly limited in the near future~\cite{holmes2020nisq+,ueno2021qecool,das2021lilliput}.
As quantum technologies become mature, computational quantum supremacy\textcolor{black}{~\cite{arute2019quantum}} will be achieved in the logical space. We will refer to the intermediate regime from the realization of logical quantum supremacy to the demonstration of long-term applications as an early-FTQC regime. The number of physical qubits will go beyond one thousand in this region, and we anticipate that more than about $10^4$ reliable logical operations on $10^2$ logical qubits are available. Even at the beginning of the early-FTQC regime, we may observe a quantum speed-up with heuristic quantum algorithms, for example, with the variational quantum eigensolver~\cite{peruzzo2014variational,kandala2017hardware,mcclean2016theory}.

% introduction
In this paper, to realize efficient and high-accuracy quantum computation in the early-FTQC era, we propose a novel framework of FTQC, where QEC and quantum error mitigation~(QEM) are combined on an equal footing. 
While QEM has been considered to be an alternative error minimization technique for noisy intermediate-scale quantum~(NISQ) devices due to its low hardware overhead at the expense of the sampling cost, we show that, by integrating probabilistic error cancellation~\cite{temme2017error,endo2018practical} into the FTQC framework, we can mitigate all the dominant types of errors in the logical space. We also note that our scheme can efficiently mitigate Pauli errors by virtually updating the quantum states with a classical memory called the Pauli frame\textcolor{black}{~\cite{fowler2012surface}}.
In the conventional QEM formalism, the sampling cost of QEM increases exponentially with the number of physical error events~\cite{takagi2021fundamental,wang2021can}. Therefore, the sampling overheads of QEM become unrealistic in NISQ computing when the number of physical operations increases for a fixed error rate per quantum gate; and the number of error events that QEM can efficiently suppress is limited to the order of unity.
In our framework, the sampling cost of QEM increases exponentially with the number of {\it logical error events} in the encoded space. Note that we can tune the number of logical error events by adjusting several parameters such as the code distance, distillation levels, and precision of approximations for Solovay-Kitaev decomposition. Thus, it is highly likely that we can find regions where the QEM techniques are the most effective, i.e., the number of logical error events is the order of unity. Accordingly, we can relax the hardware requirement with constant sampling overheads. Even after the scalable FTQC is realized, taking QEM into account, we can optimize quantum computation by allocating computation resources at will to perform even more efficient quantum computing.

We need to overcome several fundamental difficulties for applying QEM in the logical space because the costs and restrictions of logical operations and dominant sources of errors are different from the NISQ formalism. We resolve them in the affirmative by giving a solution one by one. For example, solutions to major problems are as follows. 
In FTQC, logical Clifford operations and Pauli measurements can be efficiently applied while non-Clifford operations are costly because it involves a number of $T$-gate injection, distillation, and teleportation procedures\textcolor{black}{~\cite{fowler2012surface,fowler2018low}}. These logical operations are affected by three types of logical errors: logical errors in each elementary gate operation due to restricted code distances, noise in non-Clifford logical gates deriving from shortage of magic-state distillation processes, and errors induced in the Solovay-Kitaev decomposition~\cite{kitaev1997quantum,dawson2005solovay}. We call the first two logical errors decoding errors and the last one approximation errors. We will discuss what types of errors are present when implementing logical operations, and provide a hierarchical way to mitigate noisy and costly operations with clean and less costly ones.
To detect and correct physical errors during computation, we store the estimated errors in the Pauli frame instead of physically applying recovery operations\textcolor{black}{~\cite{fowler2012surface}}. This means that actual physical states are almost never in the code space. We will provide concrete procedures for a universal set of logical operations incorporating QEM which are compatible with the Pauli frame. To apply probabilistic error cancellation, we need a good characterization of the noise model to construct QEM operations. We show that decoding errors can be efficiently characterized with gate set tomography~\cite{blume2013robust,greenbaum2015introduction} on the code space. Note that the approximation errors of the Solovay-Kitaev algorithm can be characterized efficiently on classical computers. Finally, while probabilistic error cancellation is a QEM technique to mitigate errors in the algorithms for calculating the expectation values, many FTQC algorithms are sampling algorithms using the phase estimation\textcolor{black}{~\cite{kitaev1995quantum,kivlichan2020improved,babbush2018encoding}}. We show that probabilistic error cancellation is compatible with the phase estimation algorithm. See Appendix.\,\ref{sec:apply_qem_to_bqp} for details.

We perform resource estimation of FTQC under realistic scenarios with and without QEM, and we show that our scheme can dramatically alleviate the required computational overheads in FTQC. We assume that the mean number of logical error events $N_{\rm e}$ is required to reach $N_{\rm e}=10^{-3}$, and the sampling overhead by QEM is restricted to a reasonable level, i.e., within $10^2$ times greater samples for achieving a certain accuracy. We expect at least $10^4$ logical operations are required to demonstrate classically intractable applications. In this case, the required number of qubits is reduced to approximately one-fifth with QEM compared to the original qubit count. We also expect that $10^{10}$ logical operations are at least necessary to perform conventional long-term applications. The required number of qubits is reduced to $55\%$ in this regime. 
From another perspective, our scheme can be used for increasing the number of available logical operations when the available code distance is strongly restricted. The lifetime of current superconducting qubits is about up to $1~{\rm ms}$, and a cycle of error estimations during FTQC must be sufficiently faster than the lifetime, i.e., about $1~{\rm \mu s}$\textcolor{black}{~\cite{holmes2020nisq+,gidney2021factor}}. To cope with this strong restriction, an efficient implementation of classical error-decoding architectures has been studied. According to the recent state-of-the-art proposals~\cite{holmes2020nisq+,ueno2021qecool,das2021lilliput}, the available code distance would be limited up to about $11$ in the near future even with simplified decoding algorithms. When the available code distance is limited up to $11$, our scheme enables $10^3$ times more logical operations with the same hardware requirement. 
Thus, our technique can clearly accelerate the realization of applications in early- and long-term FTQC regimes. This improvement is illustrated by red arrows in Fig.\,\ref{fig:regime_overview}. It is also worth noting that, to the best of our knowledge, these are the first examples where the performance of useful quantum algorithms with clear quantum advantages is enhanced via QEM under realistic conditions since QEM has been investigated for near-term heuristic quantum algorithms dependent on numerical optimization.

% organization of the manuscript
This paper is organized as follows. In Sec.\,\ref{sec:preliminary}, we review probabilistic error cancellation and the architecture of fault-tolerant quantum computing. In Sec.\,\ref{sec:method}, we describe how to evaluate decoding errors and approximation errors. Then we show our novel FTQC architecture with an analytical argument of the cost of QEM and explain the effect of model estimation errors. 
In Sec.\,\ref{sec:result}, we numerically analyze the sampling cost of QEM for decoding errors and approximation errors and demonstrate that we can effectively increase the code distance and the number of $T$-gates via QEM even when there are finite estimation errors. 
Finally, we conclude our paper with a discussion in Sec.\,\ref{sec:conclusion}.

\section{Preliminaries}
\label{sec:preliminary}
\subsection{Quantum error mitigation and probabilistic error cancellation}
\label{sec:pec}
Quantum processors are affected by a number of physical noise sources, which should be mitigated to obtain correct results. Here, for simplicity, we will assume that the gate errors are Markovian, i.e., the noise process $\mathcal{N}$ for a gate is totally independent of other gate errors. In this case, we have
\begin{align}
\rho_{\mathrm{out}}= \mathcal{N}_{N_G} \circ \mathcal{U}_{N_G} \circ \mathcal{N}_{N_G-1} \circ \mathcal{U}_{N_G-1} \cdots  \mathcal{N}_{1}\circ \mathcal{U}_{1} (\rho_{\mathrm{in}}),
\label{noisy}
\end{align}
where $\rho_{\mathrm{out}}$ and $\rho_{\mathrm{in}}$ are the output and input quantum states, $\mathcal{U}_k$ and $\mathcal{N}_k$ denote the ideal and noisy part of the process of the $k$-th gate, and ${N_G}$ is the number of gates. To ensure correct computations, it is necessary to mitigate the effect of $\mathcal{N}_k,~(k=1,2,..., N_G)$ and obtain 
\begin{align}
\rho_{\mathrm{out}}^{\mathrm{ideal}}= \mathcal{U}_{N_G} \circ \mathcal{U}_{N_G-1} \cdots \circ \mathcal{U}_{1} (\rho_{\mathrm{in}}).
\end{align}

Quantum error mitigation (QEM) has been proposed as a method for suppressing errors without encoding, and it is useful especially for NISQ devices with a restricted number of qubits~\cite{temme2017error,li2017efficient,endo2018practical}. Generally speaking, QEM methods recover not the ideal density matrix  $\rho^{\mathrm{ideal}}_{\mathrm{out}}$ itself, but rather the ideal expectation value of an observable $\braket{\hat{M}}_{\mathrm{ideal}}=\mathrm{Tr}(\rho^{\mathrm{ideal}}_{\mathrm{out}} \hat{M})$
via classical post-processing. Note that QEM is not a scalable technique because it needs exponentially increasing circuit runs with the number of error events in the quantum circuit~\cite{endo2018practical,temme2017error}.

Now let us explain the concept of probabilistic error cancellation with which we can eliminate a bias from the expectation value of the observables completely given the complete information on the noise model~\cite{temme2017error,endo2018practical}. (Later, we will use this method to suppress errors in FTQC.) 
First, we identify the noise map $\mathcal{N}$ via either process or gate set tomography~\cite{blume2013robust,greenbaum2015introduction}, and calculate the inverse $\mathcal{N}^{-1}$. Then, by finding a set of processes $\{\mathcal{B}_i \}$ such that $\mathcal{N}^{-1}=\sum_i \eta_i \mathcal{B}_i$ where $\eta_i \in \mathbb{R}$ and $\sum_i \eta_i=1$, we have 
\begin{equation}
\begin{aligned}
\mathcal{U}&= \mathcal{N}^{-1} \mathcal{N} \mathcal{U} \\
&=\sum_i \eta_i  \mathcal{B}_i \mathcal{N} \mathcal{U}.
\label{Eq: inverse}
\end{aligned}
\end{equation}
Note that arbitrary operations can be represented as linear combinations of tensor products of single-qubit Clifford operations and Pauli measurements~\cite{endo2018practical}. Here, we can rewrite Eq.\,(\ref{Eq: inverse}) as
\begin{align}
\mathcal{U}=\gamma_Q \sum_i q_i \mathrm{sgn}(\eta_i)  \mathcal{B}_i \mathcal{N} \mathcal{U},
\label{Eq:quasimonte}
\end{align}
where $\gamma_Q=\sum_i |\eta_i|$, $q_i=\frac{|\eta_i|}{\gamma_Q}$, $\gamma_Q \geq 1$ and $\mathrm{sgn}(\eta_i)$ is a parity which takes $\pm 1$, corresponding to the operation $\mathcal{B}_i$. We refer to $\gamma_Q$ as the QEM cost because it is related to the sampling overhead.

Now let us suppose that we have measured an observable $\hat{M}$ and obtain
\begin{equation}
\begin{aligned}
\braket{\hat{M}}_{\mathcal{U}}= \gamma_Q \sum_i q_i \braket{\hat{\mu}_i^{\rm eff}}.
\end{aligned}
\end{equation}
Here, $\hat{\mu}_i^{\rm eff} =\mathrm{sgn} (\eta_i) \hat{m}_i$, and $\hat{m}_i$ is a measurement outcome for a process $\mathcal{B}_i \mathcal{N} \mathcal{U}$. We generate the process $\mathcal{B}_i$ with a probability $q_i$ and multiply the corresponding parity with the measurement result, which is denoted as $\hat{\mu}^{\rm eff}$. Then, the expectation value of the random variable $\hat{\mu}^{\rm mit}= \gamma_Q \hat{\mu}^{\rm eff}$ approximates the error-free expectation value $\braket{\hat{M}}_{\mathcal{U}}$. Note that since $\mathrm{Var}[\hat{\mu}^{\rm mit}]=\gamma_Q^2 \mathrm{Var}[\hat{\mu}^{\rm eff}]$ and a measurement outcome without QEM, which we denote $\hat{\mu}^{\rm nmit}$ has a similar variance, the variance of the error-mitigated value is approximately amplified as $\Gamma_Q=\gamma_Q^2$. Therefore we need to have $\Gamma_Q$ times more samples to achieve a similar accuracy before applying QEM.

In practice, we use probabilistic error cancellation for each gate in  quantum circuits. The ideal process for the entire quantum circuit is described as $\prod_{k=1}^{N_G} \mathcal{U}_k$. Denoting $\mathcal{U}_k=\gamma_Q^{(k)} \sum_{i_k} q_{i_k} \mathrm{sgn} (\eta_{i_k}) \mathcal{B}_{i_k} \mathcal{N}_k \mathcal{U}_k$, we have
\begin{equation}
\begin{aligned}
&\prod_{k=1}^{N_G} \mathcal{U}_k = \prod_{k=1}^{N_G}\gamma_Q^{(k)} \sum_{i_1 i_2...i_{N_G}} \prod_{k=1}^{N_G}q_{i_k} \prod_{k=1}^{N_G}\mathrm{sgn} (\eta_{i_k}) \prod_{k=1}^{N_G} \mathcal{B}_{i_k} \mathcal{N}_k \mathcal{U}_k.
\label{Eq:quasiincircuit}
\end{aligned}
\end{equation}
From Eq.\,(\ref{Eq:quasiincircuit}), we can see that, in each gate, a process $\mathcal{B}_{i_k}$ is generated with probability $q_{i_k}$, and the product of parities $\prod_{k=1}^{N_g}\mathrm{sgn} (\eta_{i_k})$ is multiplied with the measurement results to obtain the outcome $\hat{\mu}^{\rm eff}$. This procedure is repeated, and the product of the mean of the outcomes $\braket{\hat{\mu} ^{\rm eff}}$ and $\gamma^{\rm tot}_Q=\prod_{k=1}^{N_g}\gamma_Q^{(k)}$ approximates the correct expectation value. Note that here $\gamma^{\rm tot}_Q$ is the QEM cost for the entire quantum circuit. Let us assume the cost for each gate is uniform and can be approximated as $\gamma_Q^{(k)} = \gamma_Q = 1+ a \varepsilon$ with $a$ and $\varepsilon$ being a positive constant value and the effective error rate, respectively.
Now the QEM cost and sampling overhead can be approximated as $\gamma^{\rm tot}_Q \simeq e^{a \varepsilon N_G}= e^{(\gamma_Q-1) N_G}$ and $\Gamma^{\rm tot}_Q =  (\gamma^{\rm tot}_Q)^2$, which increase exponentially with the mean number of error events in the quantum circuit $\varepsilon N_G$. Note that for $\varepsilon N_G =O(1)$ and $\varepsilon \rightarrow 0$, since $\varepsilon^k N_G=0 ~(k \geq 2)$, the QEM cost can be exactly described as $\gamma_Q^{\rm tot}= e^{(\gamma_Q-1) N_G}$.

\subsection{Fault-tolerant quantum computing}
\label{sec:ftqc}
\subsubsection{Stabilizer formalism}
In the framework of fault-tolerant quantum computing (FTQC), one prepares a redundant number of physical qubits and performs quantum computing in a code space defined as a subspace of the whole Hilbert space. By repetitively performing quantum error detection and correction, we can protect the logical qubits defined in the code space against physical errors. The state of the logical qubits is manipulated in a fault-tolerant manner with a set of logical operations. 

The stabilizer formalism~\cite{gottesman1997stabilizer,nielsen2002quantum} is the most standard way to construct quantum error-correcting codes. Here, supposing that we construct $k$ logical qubits with $n$ physical qubits, a $2^k$-dimensional code space $\mathcal{C}$ is specified with a subgroup of $n$-qubit Pauli operators called the stabilizer group.
Let the $n$-qubit Pauli group be 
\begin{equation}
\begin{aligned}
\mathcal{G}_n = \{\pm 1, \pm i\} \times \{I,X,Y,Z\}^{\otimes n},
\end{aligned}
\end{equation}
where $I$ is the identity operator and $X = \left( \begin{matrix} 0 & 1 \\ 1 & 0 \end{matrix}\right) , Y = \left( \begin{matrix} 0 & -i \\ i & 0 \end{matrix}\right), Z = \left( \begin{matrix} 1 & 0 \\ 0 & -1 \end{matrix}\right)$ are Pauli operators.
The set of Pauli operators $\mathcal{S} \subset \mathcal{G}_n$ is called a stabilizer group if $\mathcal{S}$ is a commutative subgroup, the number of elements in $\mathcal{S}$ is $2^{n-k}$, and $-I \not\in \mathcal{S}$. We denote the $(n-k)$ generator set of a stabilizer group as $\mathcal{G} = (g_1, \cdots, g_{n-k})$.  The code space $\mathcal{C}$ is defined as an eigenspace with $+1$ eigenvalues for all the operators in the stabilizer group, i.e., $\mathcal{C} = \{\ket{\psi} | ~\forall s_i \in \mathcal{S}, s_i\ket{\psi} = \ket{\psi}\}$. In the code space, we can introduce a logical basis as $\{\ket{0}_L, \ket{1}_L\}^{\otimes k}$ and logical Pauli operators as $\{I_L, X_L, Y_L, Z_L\}^{\otimes k}$. The code distance $d$ is defined as the minimum number of physical qubits on which an arbitrary logical operator, except the logical identity $I_L^{\otimes k}$, acts. 

During a quantum computation, physical errors that occur in the encoded state are detected by using $(n-k)$ Pauli measurements $P_{s} = \frac{1}{2}(I + (-1)^s g_i)$ for $s \in \{0,1\}$. These measurements are called stabilizer measurements and their binary outcomes $s$ are called syndrome values. The original state is restored by applying appropriate feedback operations that are estimated from the syndrome values. 
These stabilizer measurements are performed repeatedly during a computation. One repetition of the stabilizer measurements is called a code cycle of fault-tolerant quantum computing. If the effective error probability per physical qubit during a cycle is smaller than a certain threshold, we can estimate the Pauli operator that restores the original state with an exponentially small failure probability with the code distance $d$. Since the required number of physical qubits $n$ increases polynomially with the code distance $d$ in typical quantum error-correcting codes, we can exponentially decrease the error probability of logical qubits with a polynomial qubit overhead.

\subsubsection{Logical operations}
We must not only correct physical errors but also update the logical quantum state for performing quantum computation. To this end, a universal set of logical operations should be performed in a fault-tolerant manner. According to the Solovay-Kitaev theorem~\cite{kitaev1997quantum,dawson2005solovay}, we can approximate an arbitrary one- and two-qubit gates with a finite set of local operations. For example, the Hadamard gate $H = \frac{1}{\sqrt{2}}\left( \begin{matrix}1 & 1 \\ 1 & -1\end{matrix} \right)$, controlled-not (CNOT) gate $\Lambda = \ket{0}\bra{0} \otimes I + \ket{1}\bra{1} \otimes X$, and $T$-gate $T = \exp\left(i \frac{\pi}{8} Z\right)$ form a universal gate set. Several logical operations can be performed by transversally operating the same one- or two-qubit operations on physical qubits. 
Since transversal operations constantly increase the effective physical error rate per qubit during a cycle, we can fault-tolerantly achieve transversal logical operations. 
However, it is known that there is no stabilizer code for which the set of transversal gates is universal~\cite{eastin2009restrictions}. Thus, we need an additional technique to achieve fault-tolerant and universal quantum computing. The most promising solution is to create a quantum state called a magic state and perform non-transversal logical operations with gate teleportation~\cite{fowler2012surface}. For example, $\ket{A}_{\rm L} = T H \ket{0}_{\rm L} = \frac{1}{\sqrt{2}}(e^{i\pi/8} \ket{0}_L + e^{-i\pi/8} \ket{1}_L)$ is a typical magic state and $T$-gate operations can be performed by consuming this state. This magic state encoded in a logical qubit can be constructed with a process called magic-state injection. While the infidelity of a magic state created by magic-state injection is generally larger than the logical error rate, we can create a high-fidelity magic state from several noisy magic states by using another quantum error-correcting code implemented on the logical space, which is called magic-state distillation.
Since the application of $T$-gates requires a longer time than the other operations, the number of $T$-gates is the dominant factor affecting the computation time of FTQC. 

\begin{figure}[t!]
    \centering
    \includegraphics[width=7.5cm]{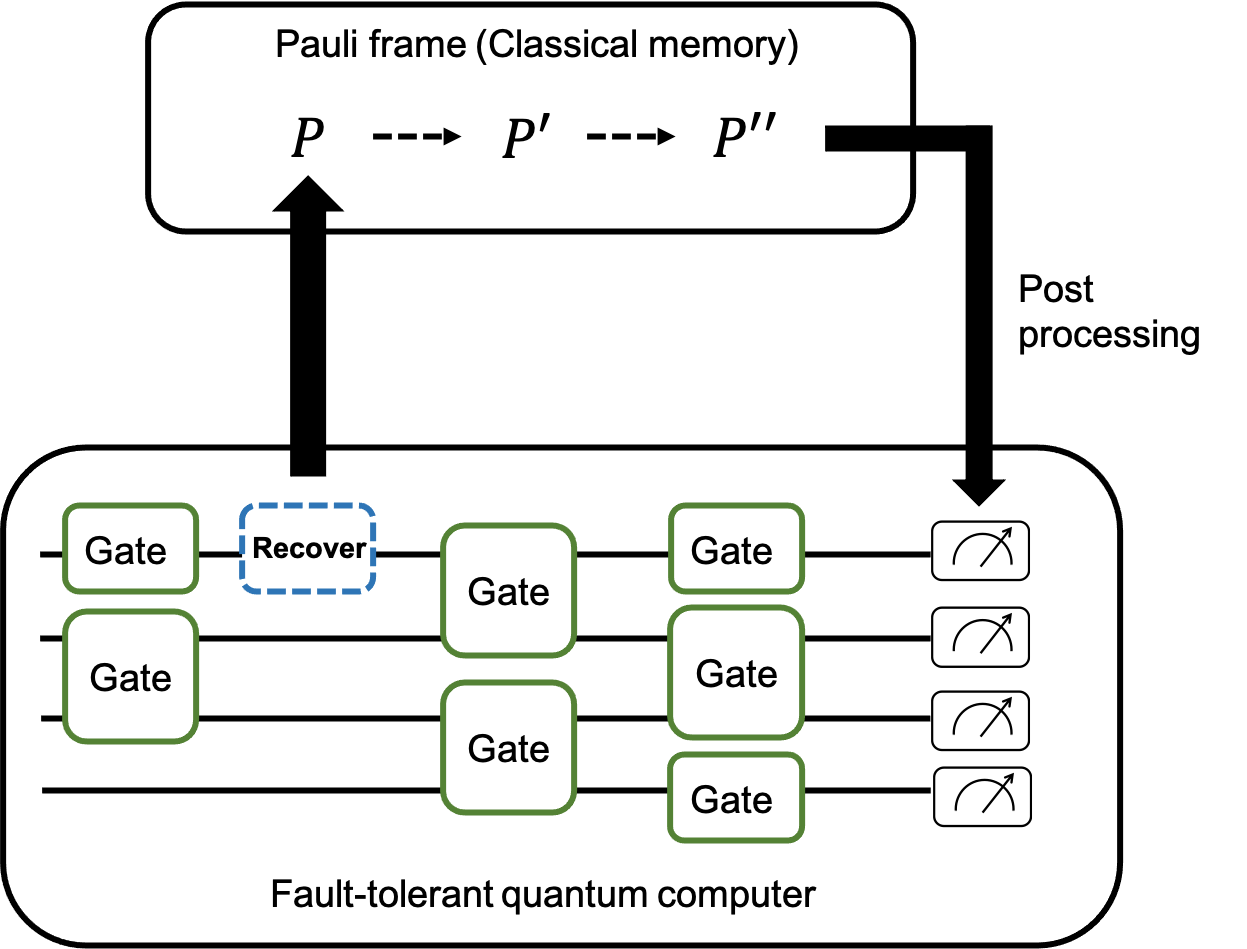}
    \caption{Schematic figure of the Pauli frame. The recovery operations are not physically applied to quantum computers but rather are stored in the Pauli frame and efficiently updated after each Clifford gate operation. The measurement outcomes are flipped depending on the state of the Pauli frame. }
    \label{fig:pauliframe}
\end{figure}

Although we can estimate a Pauli operation for recovery from syndrome values, we do not directly apply it immediately after estimation. Instead, we store the Pauli operations that should be applied to the physical qubits for recovery in a classical memory called the Pauli frame~\cite{fowler2012surface,riesebos2017pauli}. The stored operations will be taken into account when the logical measurements are performed; the outcome of a logical measurement is flipped according to the Pauli frame. A schematic figure is shown in Fig.\,\ref{fig:pauliframe}. 
In the above construction of logical operations, the whole process, except for magic-state injection, consists only of Clifford operations and Pauli channels in the code space. Since a Pauli operator conjugated by a Clifford operator is also a Pauli operator, we can always track a recovery operator as a Pauli operator during a computation. In addition, when we can apply a logical Pauli operator to a quantum state, we can perform it simply by updating the Pauli frame, since a logical Pauli operator is a transversal physical Pauli operation. 
As far as classical computers are reliable, this operation is effectively noiseless.

\section{Quantum error mitigation for fault-tolerant quantum computing}
\label{sec:method}
In this section, we discuss how to integrate QEM into the FTQC architecture. Here, we consider two types of errors in FTQC: decoding errors due to failures in the error estimation and insufficiency of magic-state distillation and approximation errors in the Solovay-Kitaev decomposition. 
In Sec.\,\ref{sec:error_in_ftqc}, we explain how these errors in FTQC can be modeled. In Sec.\,\ref{sec:mitigate_error_in_ftqc}, we discuss how these errors can be canceled and evaluate their QEM costs. Probabilistic error cancellation requires the errors to be estimated in advance. In Sec.\,\ref{sec:effect_estimation_error}, we also discuss the effect of estimation errors on probabilistic error cancellation and the characterization efficiency.

\subsection{Errors in fault-tolerant quantum computing}
\label{sec:error_in_ftqc}
\subsubsection{Decoding error}
Here, we describe noise due to the failures of error estimation in elementary logical operations, i.e., stabilizer measurements and magic-state distillation. The first obstacle to applying probabilistic error cancellation to FTQC is how to characterize an effective map of noise due to the failures of error estimation. If we suppose that the physical errors can be modeled as a stochastic physical Pauli map and assume that there are no errors on the ancillary qubits for syndrome measurements, we can define a logical noise map for decoding errors that is Markovian and a logical stochastic Pauli map. Yet, these assumptions do not hold in practice. Nevertheless, here we will assume that we can define an effectively Markovian logical error map for each logical operation and also assume that this noise map is a stochastic logical Pauli map. 
It is known that even if noise is unitary, a noise map in a logical space of surface codes can be well-approximated as stochastic Pauli noise when the code distance is sufficiently large~\cite{bravyi2018correcting}. Furthermore, the remaining coherent errors can be canceled by using pulse optimization techniques. Thus, it is reasonable to suppose that the decoding errors due to the failure of error estimations in surface codes are almost stochastic Pauli errors. In addition, we numerically verified that we can regard the decoding errors as Markovian errors even in the presence of measurement errors. See Appendix.\,\ref{sec:decoding_error_approx} for the details. 
While we mainly describe and analyze the decoding errors in the surface codes, a similar idea can be applied to the decoding errors due to insufficient magic-state distillation. As for the logical noise map on a prepared magic state due to insufficient magic-state distillation, we can twirl the noise map by logical Clifford operations, and it can be also assumed to be a stochastic Pauli noise.

Under the above assumptions, we can describe a noise map for a $l$-qubit logical operation $\mathcal{N}_{\rm dec}$ as the following stochastic Pauli noise:
\begin{equation}
\begin{aligned}
\mathcal{N}_{\rm dec}(\rho) =& \sum_{g \in \{I_{\rm L},X_{\rm L},Y_{\rm L},Z_{\rm L}\}^{\otimes l}} p_g g \rho g^{\dag},
\label{Eq:logicalerror}
\end{aligned}
\end{equation}
where $p_g \in {\mathbb R} $, $\sum_g p_g = 1$ and $p_g \ge 0$. 
The sum of probabilities of non-identity logical operations is called the logical error probability $p_{\rm dec}$, i.e., $p_{\rm dec} = \sum_{g \neq I^{\otimes m}} p_g$. 
It is known that when the physical error rate $p$ is smaller than a value called the threshold $p_{\rm th}$, the effective logical error probability decreases exponentially with respect to the code distance $d$. For the effective logical error probability per syndrome-measurement cycle of surface codes $p_{\rm cyc}$, it decreases as 
\begin{equation}
    \begin{aligned}
    p_{\rm cyc} \simeq C_1 \left(C_2 \frac{p}{p_{\rm th}} \right)^{(d+1)/2},
    \label{Eq:threshold}
    \end{aligned}
\end{equation}
where $C_1, C_2$ are constants~\cite{jones2012layered}. While the constant values depend on the details of the error correction schemes, $C_1 \simeq 0.13$ and $C_2 \simeq 0.61$ are expected in a typical construction of surface codes and the noise model~\cite{jones2012layered,fowler2012towards}.
Suppose that a logical operation requires $m$ cycles; then, the logical error probability for the logical operation can be approximated as $p_{\rm dec}$ as 
\begin{equation}
\begin{aligned}
p_{\rm dec} = 1 - (1-p_{\rm cyc})^m \simeq mp_{\rm cyc}.
\label{Eq:pdec_def}
\end{aligned}
\end{equation}
Note that the number of cycles per logical gate increases at most linearly with the code distance $d$.

In order to apply probabilistic error cancellation, we need to know the logical error probabilities $\{p_g\}$ in advance. While we can estimate $\{p_g\}$ by using gate set tomography in the logical space, the estimations are not exact. The effect of estimation errors is discussed in Sec.\,\ref{sec:effect_estimation_error}, while the efficiency of our proposal, including noise characterization, is discussed in Appendix.\,\ref{sec:PECwithgateset}.

\subsubsection{Approximation error}
Since we are only allowed to use a limited set of logical operations for achieving fault-tolerance, we need to decompose an arbitrary unitary gate into a sequence of available gates. Any unitary operator can be decomposed into a product of CNOT gates and single-qubit gates. Thus, we need to approximate single-qubit gates with a given gate set to the desired accuracy. By using the improved Solovay-Kitaev algorithm~\cite{ross2014optimal}, given a universal gate set such as $\{T, H, S\}$ and the single-qubit gate $U$ to be approximated, we can construct an approximated gate $\tilde{U}$ which satisfies $\varepsilon=\|\tilde{U} - U\|$ to an arbitrary accuracy $\varepsilon$ as a sequence of given gate set with length $\tilde{O}(\mathrm{log}(\varepsilon^{-1}))$ with $\| \cdot \|$ being an operator norm. The error of approximated map is given by 
\begin{equation}
    \begin{aligned}
    \mathcal{N}_{\rm SK}(\rho) = \tilde{U} U^\dag \rho U \tilde{U}^{\dag}.
    \label{Eq:approerror}
    \end{aligned}
\end{equation}
Since this decomposition involves only single-qubit operations, this error channel can be efficiently and exactly evaluated in advance.

\subsection{Quantum error mitigation for fault-tolerant quantum computing}
\label{sec:mitigate_error_in_ftqc}

\subsubsection{Overview of our framework}
Here, we show that decoding errors and approximation errors can be mitigated with probabilistic error cancellation. 
When we insert recovery operations for probabilistic error cancellation, it is assumed that the noise level of the recovery operations for QEM is much lower than that of the error-mitigated gates. In NISQ computing, for example, it is reasonable to assume that the error probabilities of two-qubit gates are much larger than those of single-qubit gates and measurements; therefore, the errors of two-qubit gates can be mitigated by using single-qubit recovery operations. However, this is not a reasonable assumption in FTQC, since the operations that are noisy and time-consuming are different from those of NISQ architecture. More concretely, even Clifford operations involving only one logical qubit suffer decoding errors.

Here, we show an architecture of FTQC that implements QEM with small overheads. The keys are the two significant properties of FTQC architecture: logical Pauli operations are error-free and instantaneous due to the Pauli frame, and the noise map of the decoding errors can be assumed as stochastic Pauli noise. Thanks to these properties, we can mitigate errors in all the elementary logical operations simply by updating the Pauli frame. This means the error-mitigated Clifford operations and Pauli measurements are available for computation. Because they form a complete basis for mitigating arbitrary errors~\cite{endo2018practical}, we can mitigate approximation errors due to the Solovay-Kitaev decomposition. Since the approximation errors can be exactly known in advance, an unbiased estimator free from approximation error can be obtained, as will be explained in Sec.\,\ref{sec:mitigate_approximation_error}.

To make our QEM procedure to work, the accuracy and efficiency of the decoding error estimation are vital. We show that the decoding errors can be estimated with gate set tomography under an appropriate choice of the gauge, considering state-preparation and measurement errors. We also show that the cost of gate set tomography is acceptable compared with the main computation of FTQC for estimating expectation values in Sec.\,\ref{sec:effect_estimation_error}. In this section, we further show a refined gate set tomography suited to our framework that significantly improves the estimation for logical Clifford gates.

\subsubsection{Quantum error mitigation for decoding errors}
\label{sec:mitigate_decoding_error}
We can express the inverse channel of the non-uniform depolarizing channel Eq.\,(\ref{Eq:logicalerror}) as a linear combination of Pauli operations. Thus, we can express the inverse channel as
\begin{equation}
\begin{aligned}
\mathcal{N}_{\rm dec}^{-1} (\rho) &= \sum_{g \in \{I_{\rm L},X_{\rm L},Y_{\rm L},Z_{\rm L}\}^{\otimes m}} \eta_g g \rho g^{\dag} \\
&= \gamma_{\rm dec} \sum_{g \in \{I_{\rm L},X_{\rm L},Y_{\rm L},Z_{\rm L}\}^{\otimes m}} q_g \mathrm{sgn}(\eta_g) g \rho g^\dag.
\label{Eq:inverse}
\end{aligned}
\end{equation}
Refer to Appendix.\,\ref{sec:coef} for a concrete expression of each coefficient $\eta_g$, $\gamma_{\rm dec}$, and $q_g$. Thus, we can suppress the errors by applying probabilistic error cancellation only with Pauli operators after the decoding processes. 
The QEM cost for decoding errors in the entire circuit can be expressed as $\gamma^{\rm tot}_{\rm dec} = \prod_{k=1}^{N_{\rm dec}}\gamma_{\rm dec}^{(k)}$, where $N_{\rm dec}$ is the number of logical gates, and $\gamma_{\rm dec}^{(k)}$ is a QEM cost of the $k$-th operation.

\begin{figure}[t!]
    \centering
    \includegraphics[width=\columnwidth]{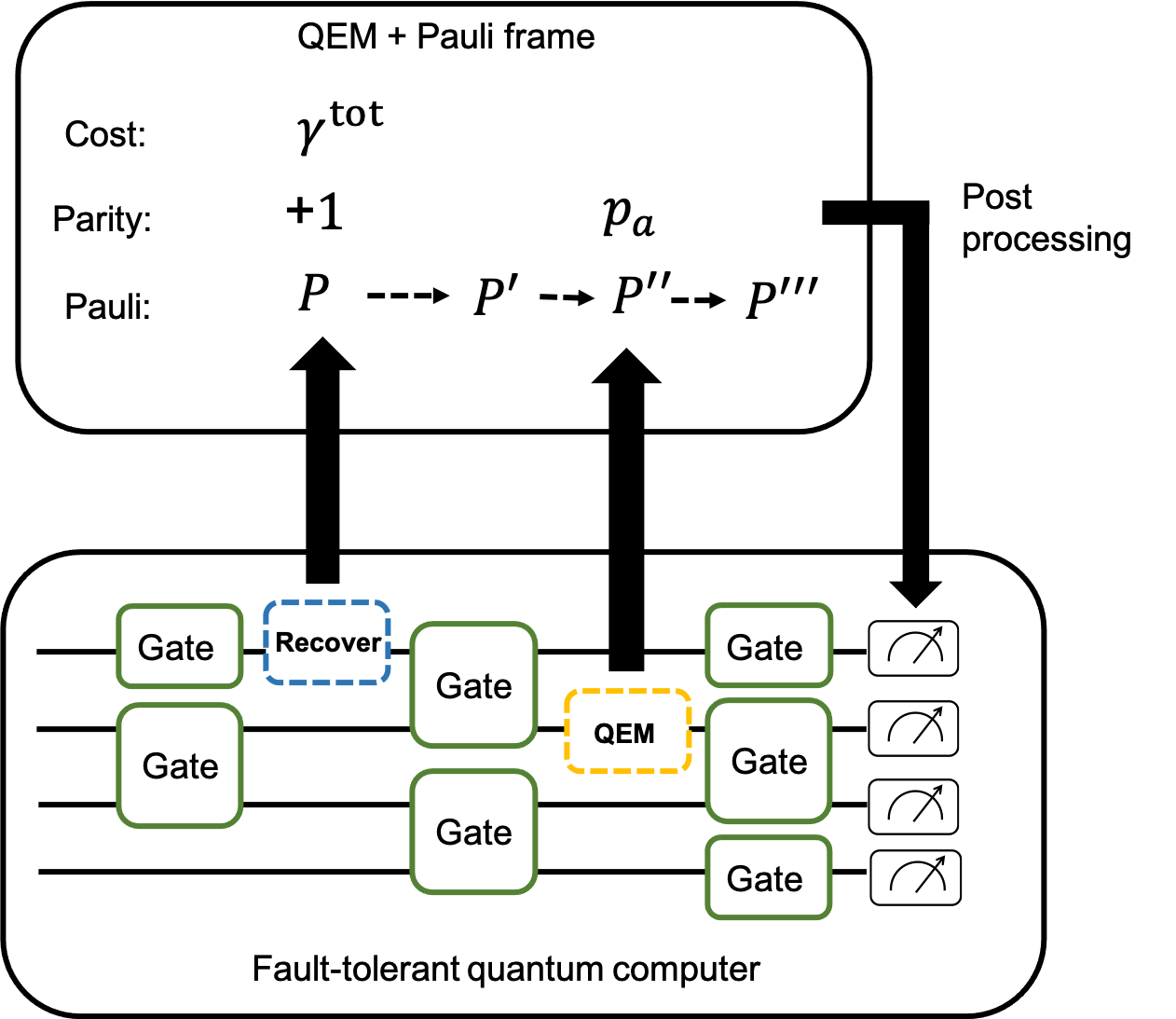}
    \caption{Schematic figure for the Pauli frame incorporating QEM. If a QEM recovery operation is a Pauli operation, it is not directly applied to the quantum computer but rather the Pauli frame is updated instead. The parity is also updated in accordance with the generated recovery operations of QEM. Here, we denote the parity corresponding to the QEM recovery operation as $p_a$ in the figure. If a QEM recovery operation is not a Pauli operator, it is performed physically. Then measurement outcomes are then post-processed depending on the Pauli frame, parity, and QEM cost. }
    \label{fig:pauliframe2}
\end{figure}

Note that probabilistic error cancellation usually applies the recovery operations of QEM immediately after the noisy gates ~\cite{temme2017error,endo2018practical}; however, because we perform only logical Pauli operations as the recovery operations for decoding errors, they can be done simply by updating the Pauli frame instead of directly applying them after noisy gates. Finally, the measurement result is post-processed according to the state of the Pauli-frame, the parity corresponding to the applied recovery operations, and the QEM cost. Thus, unlike in probabilistic error cancellation for NISQ devices, the logical noise due to decoding errors can be mitigated without any additional noise due to the recovery operation.
A schematic figure is shown in Fig.\,\ref{fig:pauliframe2}.
Note that the information on the QEM cost and the parity is used only when the final measurement result is obtained; the outcome of a destructive logical Pauli measurement is flipped depending only on the state of the Pauli frame. Whether we can mitigate decoding errors of complicated logical operations such as magic-state preparation, gate teleportation, and adaptive Clifford gates by simply updating the Pauli frame is not trivial; therefore we provide a concrete procedure for actual devices and Pauli frames in Appendix.\,\ref{sec:concrete_logical_operations}. 

In the case of decoding errors in surface codes, by approximating the QEM cost to the first order of the logical error, we have
\begin{equation}
    \begin{aligned}
    \gamma_{\rm dec} \simeq  1 + 2p_{\rm dec}.
    \label{Eq:costandlogicalerror}
    \end{aligned}
\end{equation}
Refer to Appendix.\,\ref{sec:coef} for details. Under the assumption that the logical error rate is the same for all the logical operations and $p_{\rm dec} N_{\rm dec}=O(1)$ with $p_{\rm dec} \rightarrow +0$, the QEM cost $\gamma_{\rm dec}^{\rm tot}$ for the entire quantum circuit can be shown to be exactly equal to $e^{2 p_{\rm dec} N_{\rm dec}}$ on the basis of the argument in Sec.\,\ref{sec:pec}.
Thus, by using Eqs.\,(\ref{Eq:threshold}) and (\ref{Eq:pdec_def}), we obtain
\begin{equation}
    \begin{aligned}
    \gamma_{\rm dec} - 1 =  2 m C_1 \left(C_2 \frac{p}{p_{\rm th}}\right)^{(d+1)/2},
    \end{aligned}
    \label{Eq:gammadec}
\end{equation}
which results in the total QEM sampling overhead
\begin{align}
\Gamma^{\rm tot}_{L}=e^{2 (\gamma_{\rm dec}-1) N_{\rm dec}}= \exp \left( 4 m C_1 N_{\rm dec} \left( C_2 \frac{p}{p_{\rm th}}\right)^{(d+1)/2} \right).
\label{Eq:tradeoff}
\end{align}
Notice that Eq.~(\ref{Eq:tradeoff}) clearly shows a trade-off relationship between the sampling overhead and the code distance, i.e., the number of physical qubits.

\subsubsection{Quantum error mitigation for approximation errors}
\label{sec:mitigate_approximation_error}
Unlike decoding errors due to the failure of error correction, we cannot describe errors due to the Solovay-Kitaev decomposition as stochastic Pauli errors. Nevertheless, we can still apply probabilistic error cancellation with negligible overheads. 
Denote $\mathcal{N}_{\rm SK}(\rho) = \tilde{U} U \rho (\tilde{U} U)^{\dag}$; we invert this approximation error by
\begin{equation}
\begin{aligned}
\label{Eq:sv_decomp}
\mathcal{N}_{\rm SK}^{-1}&=\sum_i \eta_i \mathcal{B}_i^{(L)} \\
&=\gamma_{\rm SK} \sum_i q_i \mathrm{sgn}(\eta_i) \mathcal{B}_i^{(L)},
\end{aligned}
\end{equation} 
where $\{\mathcal{B}_i^{(L)}\}$ denotes recovery operations in the logical space. Note that we can represent any map as a linear combination of Clifford operations and Pauli channels~\cite{endo2018practical}, and thus, we do not need $T$-gates for mitigating approximation errors. Recovery operations are randomly chosen and applied immediately after each single-qubit logical operation if they are not Pauli operations. In the case of Pauli operations, we can again use the Pauli frame, and physical operations on quantum computers are not required, in a similar vein to QEM for decoding errors. Since a single-qubit logical unitary operation consists of several repetitions of Clifford gates and $T$-gate teleportation, the insertion of the recovery operation for probabilistic error cancellation negligibly increases the length of the quantum circuit.
In the numerical simulations described in the next section, we will verify that the QEM costs can be approximated with the following equation:
\begin{equation}
    \begin{aligned}
        \gamma_{\rm SK}-1 = \beta_1 e^{-\beta_2 N_T },
    \end{aligned}
    \label{Eq:SVcost}
\end{equation}
where $\beta_1$ and $\beta_2$ are constants dependent on the quantum gate and $N_T$ is the number of available $T$-gates. 

The QEM cost due to approximation errors can also be represented as $\gamma^{\rm tot}_{\rm SK}=\prod_{k=1}^{N_{\rm SK}} \gamma_{\rm SK}^{(k)}$, where $N_{\rm SK}$ is the total number of recovery operations for mitigating approximation errors in the quantum circuit with the cost $\gamma_{\rm SK}^{(k)}$ corresponding to the $k$-th recovery operation. 
By assuming that the cost does not depend on gates, we have the following QEM sampling overhead:
\begin{align}
\Gamma_{\rm SK}^{\rm tot} \simeq \exp \left(2 \beta_1 N_{\rm SK} e^{- \beta_2 N _T}\right).
\end{align}
This shows there is a trade-off relationship between the sampling overhead and the number of available $T$-gates.

\subsection{Effect of estimation errors of the noise map}
\label{sec:effect_estimation_error}
\subsubsection{Effect of estimation errors on expectation values}
While approximation errors can be exactly determined in advance, decoding errors have to be characterized. Since the logical error probabilities of the decoding errors are small, it is unavoidable that the characterization will contain finite and non-negligible estimation errors. Thus, we need to care about QEM with estimation errors and the efficiency of the characterization of decoding errors.

Let us discuss how estimation errors affect the performance of QEM. Given a perfect characterization of the noise model $\mathcal{N}_k$ for the $k$-th gate, we can realize the inverse operation $\mathcal{N}_k^{-1}$ with probabilistic error cancellation to achieve $\mathcal{N}_k^{-1} \mathcal{N}_k= \mathcal{I}$. If we obtain an incorrect estimation for the error process $\mathcal{N}_k' \neq \mathcal{N}_k$, it leads to an estimation error $\Delta \mathcal{N}_k \equiv \mathcal{N}_k^{\prime -1} \mathcal{N}_k \neq \mathcal{I}$. 

Now, denoting the ideal process of the $k$-th gate as $\mathcal{U}_k$, the difference of the the error-mitigated process and the error-free process for the entire quantum circuit can be described by the diamond norm:
\begin{equation}
\begin{aligned}
\bigg\|\prod_{k=1}^{N_G} \Delta \mathcal{N}_k \mathcal{U}_k - \prod_{k=1}^{N_G}  \mathcal{U}_k \bigg\|_{\diamond} \leq \Delta \varepsilon N_G  
\end{aligned}
\end{equation}
where we used the fact that the diamond norm is subadditive and we denote $\Delta \varepsilon = \mathrm{max}_k \|\Delta \mathcal{N}_k -\mathcal{I} \|_\diamond$. Similarly, the discrepancy of the noisy and ideal process can be upper-bounded as $\|\prod_{k=1}^{N_G} \mathcal{N}_k \mathcal{U}_k - \prod_{k=1}^{N_G}  \mathcal{U}_k \|_{\diamond} \leq \varepsilon N_G$, where $\varepsilon = \mathrm{max}_k \| \mathcal{N}_k -\mathcal{I} \|_\diamond$. Because the deviation of the expectation values of an observable $M$ for two processes $\mathcal{E}_1$ and $\mathcal{E}_2$ with the input state $\rho$ can be described as $\delta M = \mathrm{Tr}[M (\mathcal{E}_1(\rho)- \mathcal{E}_2(\rho))] \leq  \|M \| \|\mathcal{E}_1-\mathcal{E}_2 \|_\diamond$~\cite{campbell2019random}, where $\|\cdot \|$ is an operator norm, we have $\delta M_{\rm QEM} \leq  \|M \| \Delta \varepsilon N_G $ and $\delta M_{\rm noise} \leq  \|M \| \varepsilon N_G $. Here, $\delta M_{\rm QEM}$ and $\delta M_{\rm noise}$ are the deviation of the observable with and without error mitigation.

Thus, we can see that QEM is beneficial when we can achieve $r<1$ for
\begin{equation}
\begin{aligned}
\label{def:coef_est_error}
\Delta \varepsilon = r \varepsilon.
\end{aligned}
\end{equation}
Note that this discussion does not include sampling errors; i.e., $\delta M$ is the error of the expectation value given infinite samples. 

\subsubsection{Efficiency of characterization of decoding errors}
As a cause of model estimation errors, when we use gate set tomography to characterize the noise model for decoding errors, we need to consider state preparation and measurement (SPAM) errors and the finite statistical error arising from an insufficient number of samples. It has been shown that the effect of SPAM errors can be eliminated in the case of probabilistic error cancellation based on gate set tomography~\cite{endo2018practical}. While the general choice of the gauge is not compatible with the Pauli frame, we can modify the scheme of gate set tomography so that this method is compatible with QEM with the Pauli frame. Refer to Appendix.\,\ref{sec:PECwithgateset} for details.

To achieve an accuracy $r$ given in Eq.\,(\ref{def:coef_est_error}), we need to perform $N_{\rm GST} = O((r\varepsilon)^{-2})$ samplings with gate set tomography~\cite{greenbaum2015introduction,nielsen2020gate}. Here, we show this efficiency is acceptable compared with the main part of FTQC, i.e., the time required for gate set tomography corresponds to $O(r^{-2} n_q N_G)$ runs of the whole quantum logical circuits to obtain expectation values, where $n_q$ is the number of logical qubits.
Let the time for a single run of the logical circuit of FTQC be $\tau$.
The depth of logical quantum circuit is estimated as $O(N_G n_q^{-1})$, and the time per gate can be roughly approximated as $\tau_{\mathrm{gate}}=O(\tau n_q N_G^{-1})$. Then, the time for gate set tomography can be estimated as $\tau_{\rm GST} = O(\tau_{\rm gate} N_{\rm GST}) = O(\tau N_G^{-1} n_q (r \varepsilon)^{-2})$. 
In a situation where QEM is useful, we have $\varepsilon N_G = O(1)$~\cite{endo2020hybrid}. Thus, we can conclude that to use QEM to decrease the logical error rate $p_{\rm dec}$ to $r p_{\rm dec}$ by QEM, we need gate set tomography as a pre-computation that takes $\tau_{\rm GST} = O(\tau N_G n_q r^{-2})$, which is $\tau_{\rm GST}/\tau = O(N_G n_q r^{-2})$ times longer than a single circuit run of FTQC.
The numbers of logical gates $N_G$ and logical qubits $n_q$ are expected to grow polynomially with the problem size, and FTQC circuits will be repeated on the order of $O(r^{-2})$ to make the statistical fluctuation of expectation values smaller than the reduced bias. Accordingly, while the estimation costs of the noise map cause another overhead to FTQC depending on the required accuracy, it is performed with a time that grows polynomially with the problem size and without requiring additional physical qubits.
We remark that when we assume the noise properties of the quantum devices are uniform, we can perform the sampling for gate set tomography in parallel. If we use all the logical qubits for characterization, the time for gate set tomography is reduced to $\tau_{\rm GST} = O(\tau N_G r^{-2})$. Note that, in the scenario that we can fully parallelize the sampling procedure, i.e., when we have $O((r\varepsilon)^{-2})$ distinct uniform quantum gates, we have $\tau_{\rm GST} = O(\tau N_G^{-1})$.

To further make the characterization of noise more efficient, we propose an improved gate set tomography for decoding errors of the Clifford process that is fast and compatible with the Pauli frame. See Appendix.\,\ref{sec:PECwithgateset} for the details of this scheme. The number of measurements $N_{\rm GST}$ is reduced to $N_{\rm GST} = O(r^{-2} \varepsilon^{-1})$, which makes the costs of pre-computation $O(n_q r^{-2})$. Thus, as long as $r$ is not too small, the time for characterization is expected to be relatively short.
While our efficient gate set tomography cannot be applied to the characterization of the $T$-gate preparation, several ways to reduce the costs for estimating errors of $T$-gates can be considered. Since the error of logical $T$-gate depends on physical $T$-gate and the process of injection and distillation is constructed by a few $T$-gate circuit, there may be an efficient way to numerically estimate the noise of logical $T$-gate from the characterization of physical $T$-gate and efficient simulation for quantum circuits dominated by Clifford gates~\cite{bravyi2016improved}. There may be a way to mitigate $T$-gate errors by temporally expanding the code distance or increasing the distillation depth for $T$-gate. The cost of gate set tomography might be also reduced by utilizing long-sequence GST~\cite{nielsen2020gate}, i.e., repeating several $T$-gates to amplify a small error rate to a large value. Ref.\,\cite{piveteau2021error} shows that if decoding errors of logical Clifford gates are negligible, one can reliably twirl the noise of $T$-gate and perform efficient process tomography on that by repeating $T$-gates. Nevertheless, it is still an open problem whether there exists a more efficient gate set tomography on the logical space with imperfect logical Clifford gates.

\subsubsection{Effective increase in code distance by quantum error mitigation under estimation errors}
We can regard that QEM effectively increases the code distance. Suppose that we can effectively achieve an $r$ times smaller logical error rate $p_{\rm eff} = r p_{\rm dec}$ via QEM. Since the logical error rate is roughly approximated with the code distance as $p_{\rm dec}(d)=p(p/p_{\rm th})^{(d-1)/2}$, QEM effectively achieves a larger code distance $d'$ where $p_{\rm eff} = p_{\rm dec}(d')$ without increasing the number of physical qubits. The effective increase in the code distance via QEM can be derived as 
\begin{equation}
d'-d=2 \frac{\mathrm{ln}~r}{\mathrm{ln}\big(\frac{p}{p_{\rm th}}\big)}.
\end{equation}
Therefore, by setting $r=(p/p_{\rm th})^{x}$, we can effectively increase the code distance by $2x$. Note that, as discussed in the previous sections, we need $\mathrm{exp}(O (N_{\rm dec} p_{\rm dec}))=\mathrm{exp}(O(1))$ times more repetitions to achieve the same precision as the error-free case. 
It is worth noting that we can increase the number of available logical qubits via QEM. If we are allowed to use a fixed number of physical qubits, the decrease in the code distance indicates that we can allocate more logical qubits; therefore, we can convert the code distance into the number of logical qubits.

\section{Numerical analysis}
\label{sec:result}
We numerically evaluated how well error mitigation suppresses the qubit overhead in FTQC. (See Appendix.\,\ref{sec:numercal_detail} for the detailed settings and the definitions of the terms used in the numerical analysis.)

\subsection{Quantum error mitigation for decoding errors}
\subsubsection{Cost analysis}
We evaluated the performance of QEM on decoding errors occurring during logical operations, where we assumed FTQC with surface codes and lattice surgery. (See Appendix.\,\ref{sec:surface_code_and_lattice_surgery} for details about surface codes.) For simplicity, we assumed a single-qubit depolarizing noise model for each data and measurement qubit at the beginning of each cycle, which corresponds to a phenomenological noise model~\cite{wang2003confinement,delfosse2017almost}. To determine the failure probability of decoding with faulty syndrome-measurement cycles, we further assumed perfect syndrome measurements in the $0$-th and $d$-th cycles. Then, we checked whether any logical Pauli errors occurred during $d$ cycles. The recovery operations were estimated from the syndrome values by using the minimum-weight perfect matching decoder~\cite{dennis2002topological,edmonds1965paths}. We evaluated the logical error probabilities of Pauli-$X,Y,Z$ and computed the QEM cost for $d$ cycles according to Eq.\,(\ref{QEM_cost_Pauli_1q}). Despite our assumption of perfect syndrome measurements of the $0$-th and $d$-th cycle, we expect that the numerical results are asymptotically equivalent to those without the assumption when $d$ is sufficiently large.

\begin{figure}
    \centering
    \begin{tabular}{c}
    \subfigure[]{
        \includegraphics[width=7.5cm]{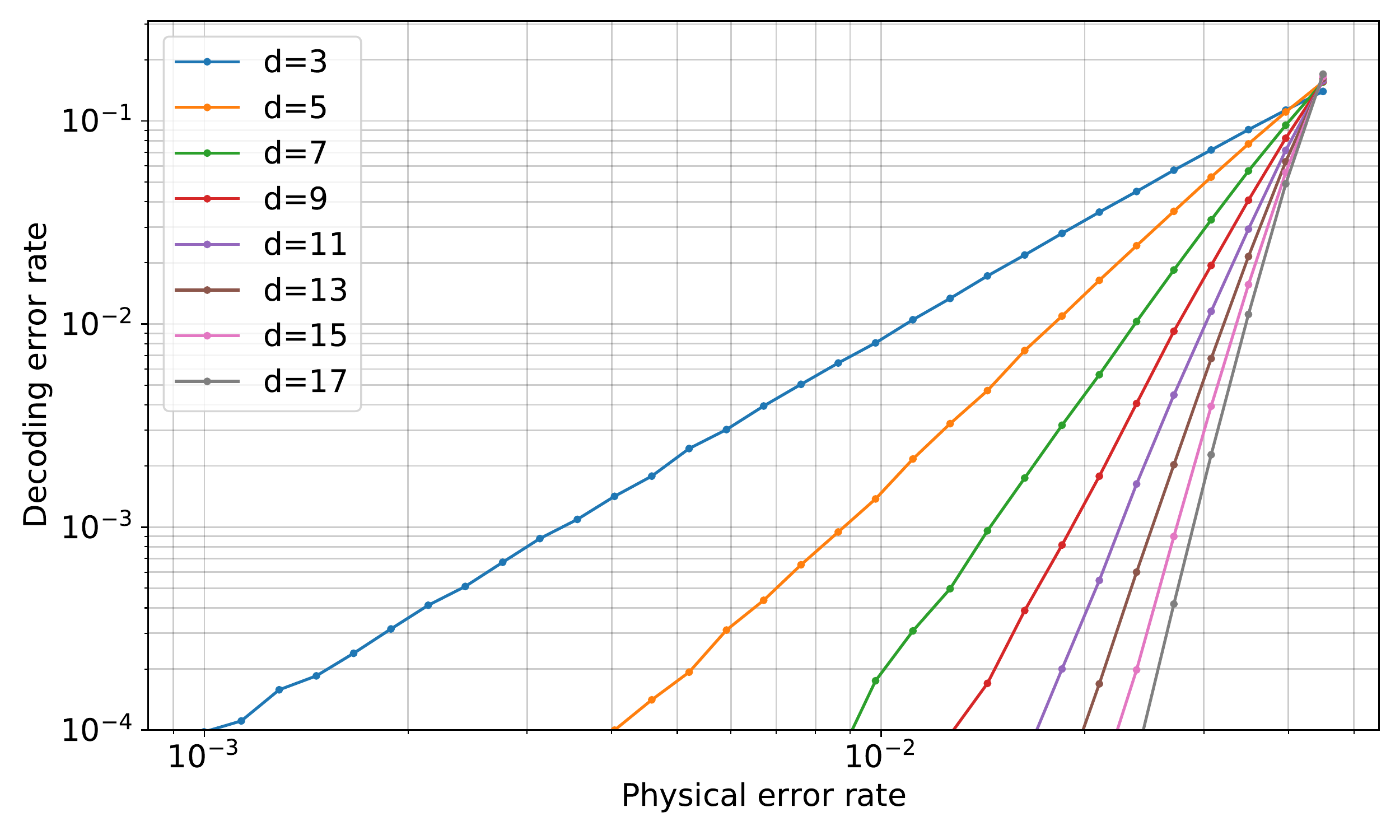}
        \label{fig:logical_error_lowp}
    } \\
    \subfigure[]{
        \includegraphics[width=7.5cm]{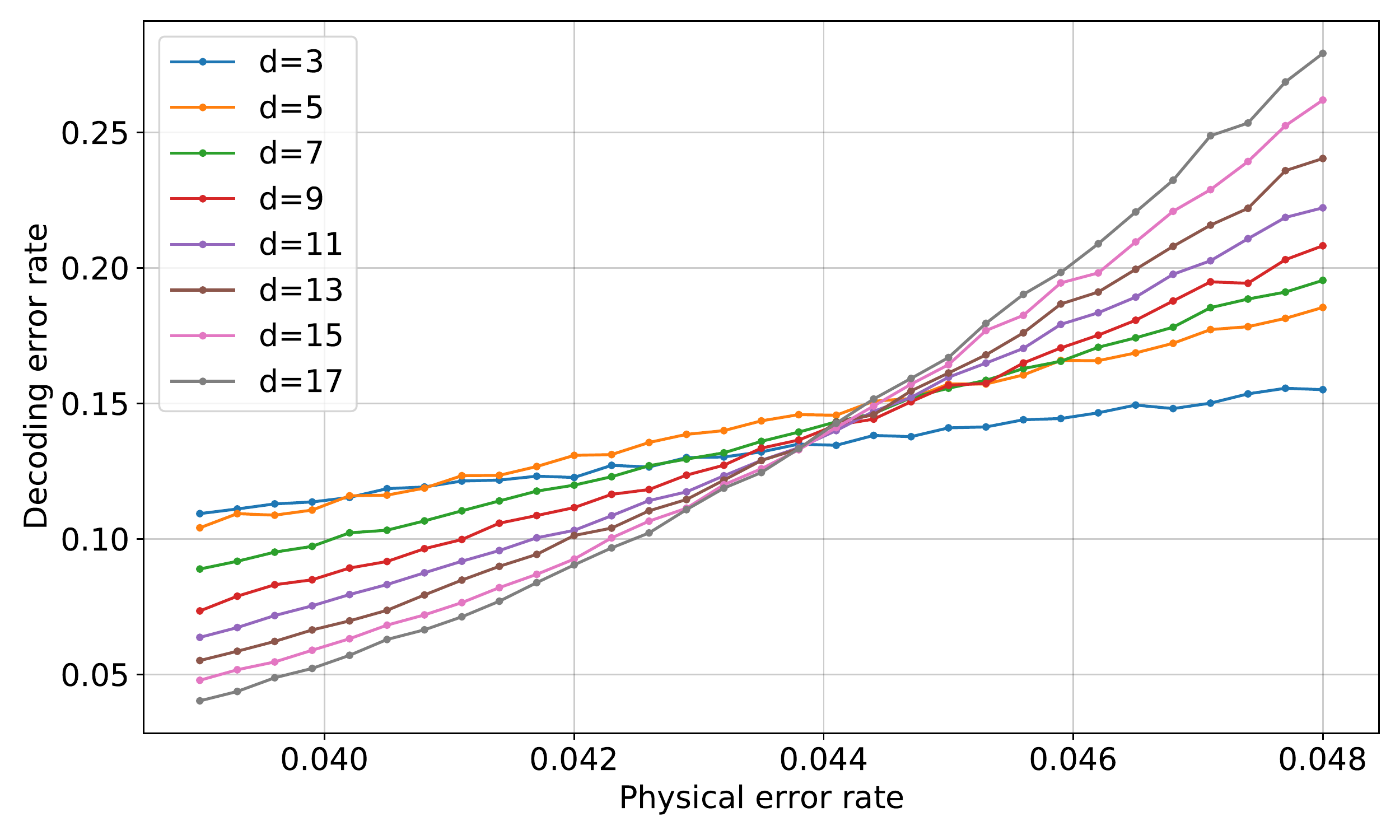}
        \label{fig:logical_error_zoom}
    } \\
    \subfigure[]{
        \includegraphics[width=7.5cm]{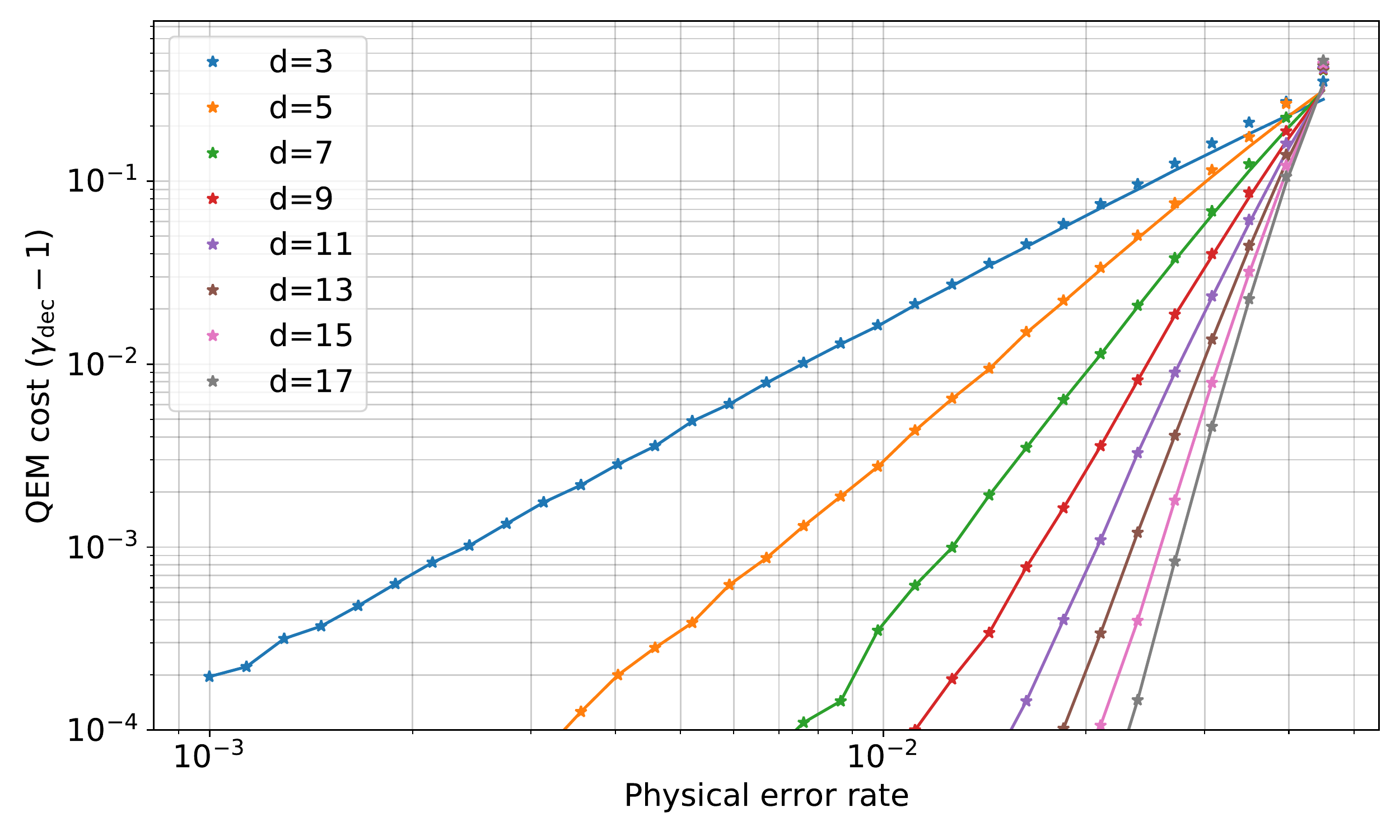}
        \label{fig:logical_error_gamma}
    }
    \end{tabular}
    \caption{Logical error probability and QEM cost for $d$-cycle syndrome measurements plotted against the physical error rates at several code distances. (a) Logical error probability as a function of physical error rate. (b) The same figure zoomed in around the threshold value. (c) QEM costs for $d$-cycle idling operations. The first-order approximations from Eq.\,(\ref{Eq:costandlogicalerror}) are shown as solid lines.}
\end{figure}

The logical error probabilities of Pauli-$X,Y,Z$ of a single logical qubit for several code distances were calculated. The sum of their probabilities are plotted according to physical error rates in Fig.\,\ref{fig:logical_error_lowp}. 
The logical error probability exponentially decreases according to the code distance when the physical error probability $p$ is smaller than a threshold value. Fig.\,\ref{fig:logical_error_zoom} plots the logical error probabilities around the threshold value, which is around $p_{\rm th} = 0.044$. 

We computed the QEM costs for decoding errors $\gamma_{\rm dec}$ corresponding to $d$ cycles and different code distances and compared them with the first-order approximation shown in Eq.\,(\ref{Eq:costandlogicalerror}). The numerical results are plotted in Fig.\,\ref{fig:logical_error_gamma}. In this figure, the solid lines correspond to the approximation of the QEM cost in Eq. (\ref{Eq:costandlogicalerror}).
We can see that the QEM costs decay exponentially depending on the code distances and show a threshold behavior like the logical error probabilities. They coincide well when the physical error rate is sufficiently small.

\subsubsection{Performance analysis}
\label{sec:logical_error_performance_analysis}
Next, we examined the performance of QEM on decoding errors in large-scale quantum circuits with a 100-qubit logical random Clifford circuit with 100 layers. We remark that since a linear combination of Clifford operations can represent arbitrary quantum operations, it is sufficient to demonstrate the performance of QEM for Clifford operations~\cite{strikis2020learning}.
We simultaneously applied randomly generated single-qubit Clifford gates to each layer, and then we applied 50 CNOT gates to two randomly chosen qubits. 
We can simulate these protocols efficiently by using an efficient algorithm for stabilizer circuits~\cite{gottesman1997stabilizer,aaronson2004improved}.
As an observable, we chose a Pauli operator whose measurement outcome is always unity for the final state vector if there are no physical errors; i.e., the final state is a +1 eigenstate of the chosen observable.
The numerical simulations assumed a non-uniform single-qubit depolarizing logical error in the form of Eq.~(\ref{Eq:logicalerror}) for each layer. The logical error probabilities of depolarizing channels were determined according to the numerical results of the last section. We chose $p=0.01$ and obtained the logical error probabilities by extrapolation. The estimated logical error probabilities are summarized in Table.\,\ref{tab:logical_probs}. 
\begin{table}[htb]
  \centering
  \begin{tabular}{|l|l|l|}
    \hline
    code distance $d$ & $p_{X_{\rm L}}, p_{Z_{\rm L}}$ & $p_{Y_{\rm L}}$ \\ 
    \hline \hline
    $5$ & $1.80 \times 10^{-4}$ & $1.96 \times 10^{-6}$ \\
    \hline
    $7$ & $1.39 \times 10^{-5}$ & $4.11 \times 10^{-8}$ \\
    \hline
    $9$ & $1.08 \times 10^{-6}$ & $8.64 \times 10^{-10}$ \\
    \hline
    $11$ & $8.35 \times 10^{-8}$ & $1.81 \times 10^{-11}$ \\
    \hline
  \end{tabular}
  \caption{Estimated logical error probability for code distances with $p = 0.01$ and $p_{\rm th} \sim 0.044$.}
  \label{tab:logical_probs}
\end{table}
Without QEM, the final state converges to a highly mixed state due to physical errors. Thus, it is expected that expectation values decay to zero. By employing QEM, they are taken back to unity, sacrificing statistical accuracy and requiring a greater number of experiments accordingly. 

Note that while the required number of cycles for Clifford operations scales linearly with the code distance, the actual number of cycles and logical error probability per logical gate are dependent on the Clifford operations. In particular, logical CNOT gates with lattice surgery may induce correlated logical Pauli errors on multiple logical qubits. Nevertheless, we used a simplified error model, since we expect this evaluation captures the basic properties of QEM performance.

\begin{figure*}
    \centering
    \subfigure[]{
        \includegraphics[width=\textwidth]{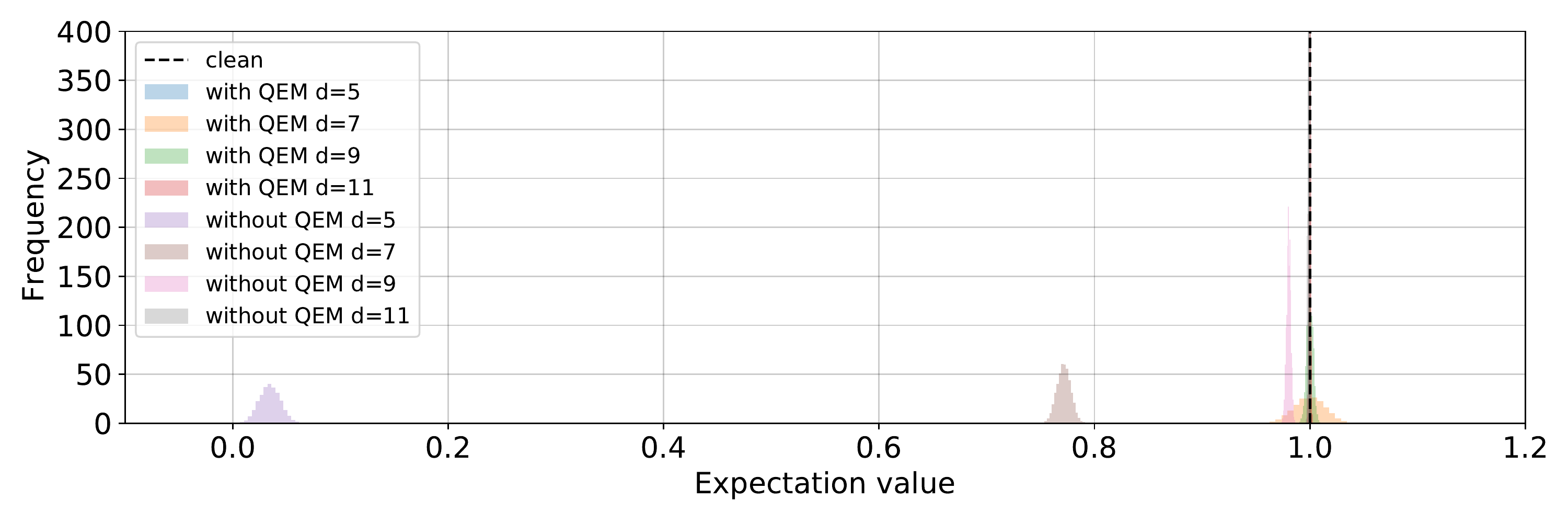}
        \label{fig:demonstrate_clifford}
    } \\
    \begin{tabular}{cc}
    \subfigure[]{
        \includegraphics[width=7.5cm]{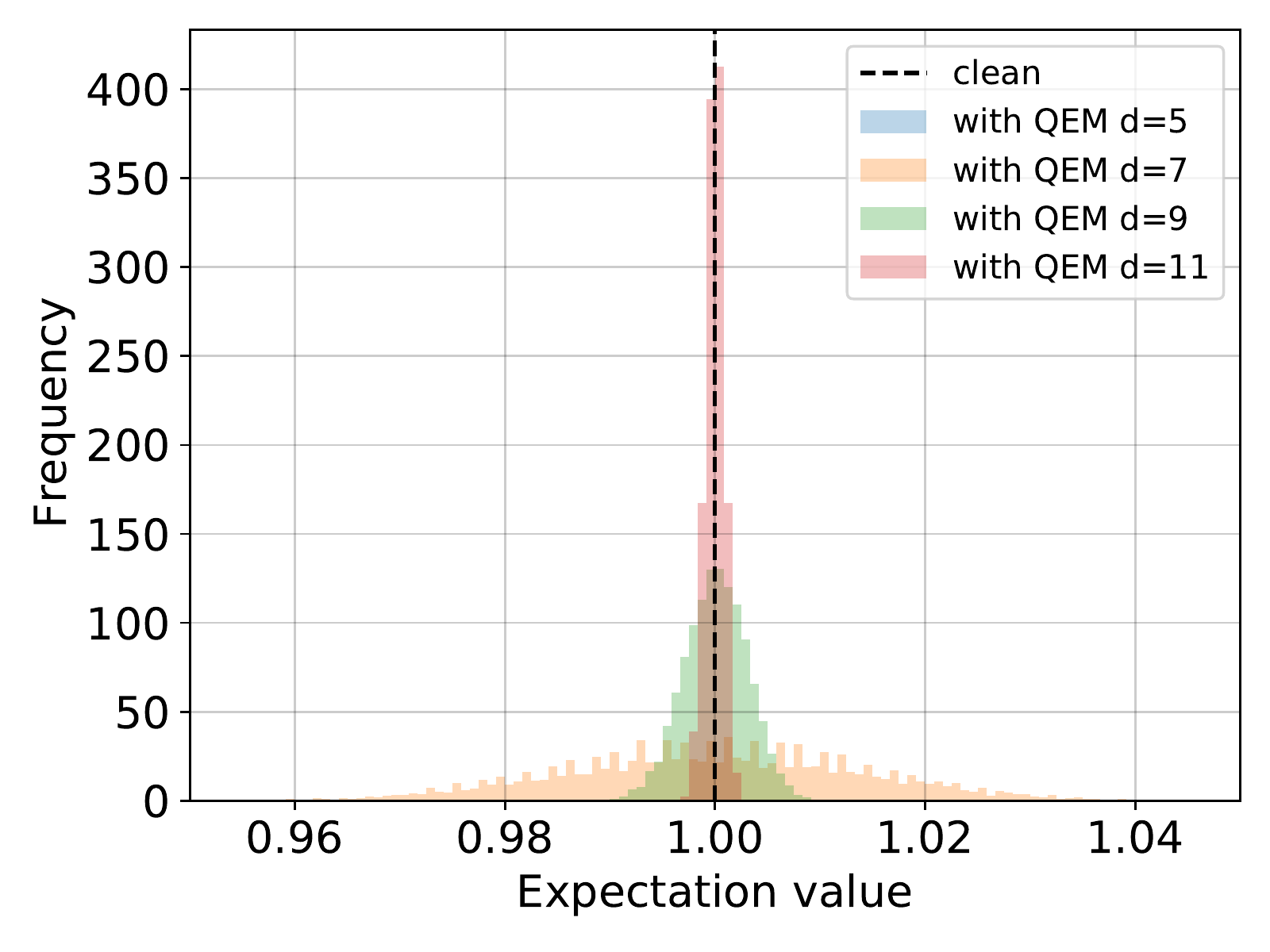}
        \label{fig:demonstrate_clifford_mitigated}
    } & 
    \subfigure[]{
        \includegraphics[width=7.5cm]{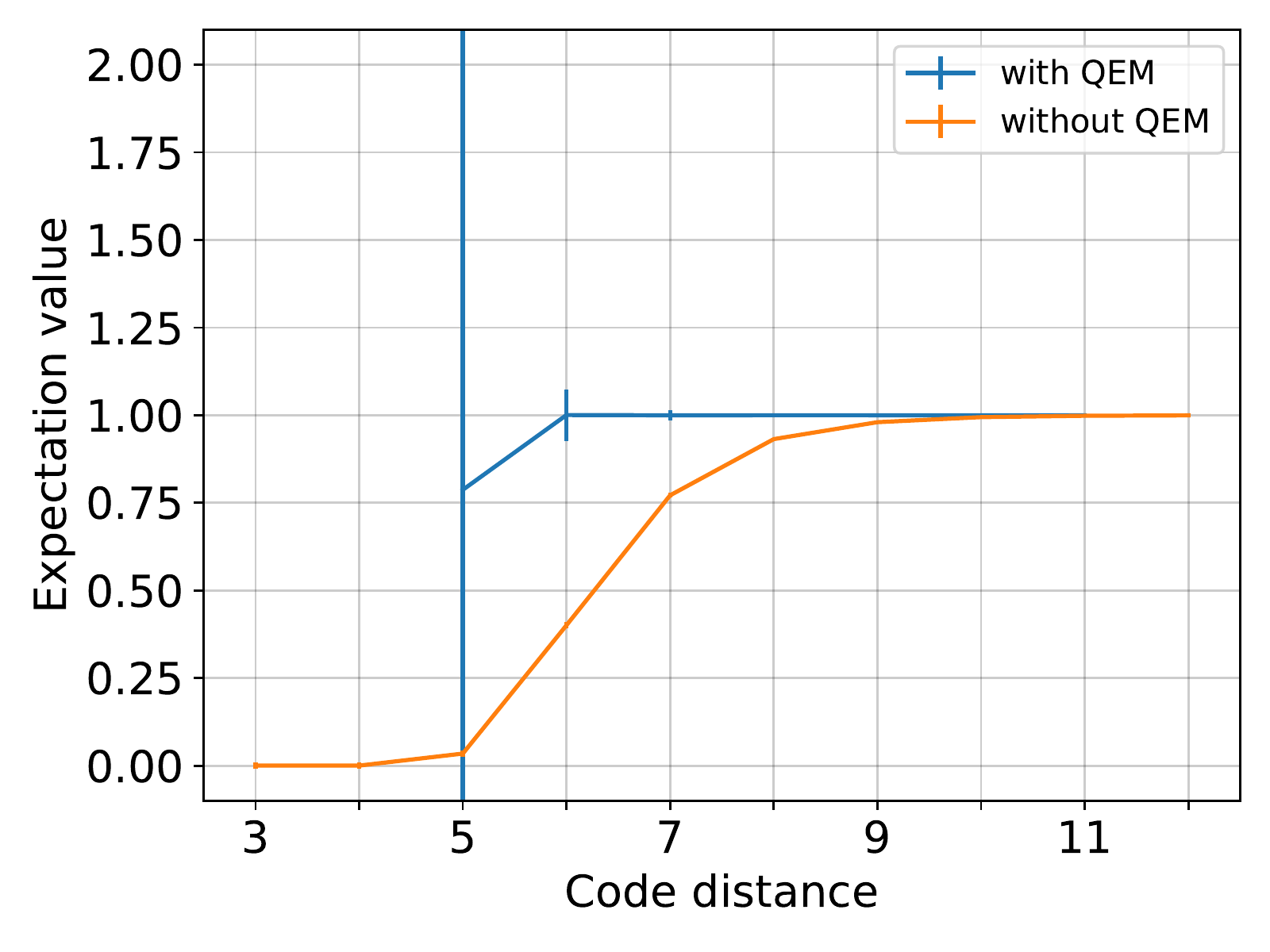}
        \label{fig:demonstrate_clifford_compare}
    }
    \end{tabular}
    \caption{(a) Histogram of expectation values for 100-qubit random Clifford circuits. (b)  Histogram of expectation values with QEM for 100-qubit random Clifford circuits. (c) Sample averages and standard deviation of 100-qubit random Clifford circuits are plotted as a function of code distances.}
\end{figure*}

We numerically performed a series of $10^4$ experiments, each of which computed an expectation value from $10^4$ single-shot measurements.
The results are shown in Fig.\,\ref{fig:demonstrate_clifford}, while data around the ideal expectation value are shown in Fig.\,\ref{fig:demonstrate_clifford_mitigated}. 
Fig.\,\ref{fig:demonstrate_clifford_compare} shows the mean value of $10^4$ samples for each logical error probability together with its standard deviation (the error bar).
We can see that there was a large bias in the expectation value without QEM, but no bias when the QEM technique was employed, while its standard deviation was amplified. The standard deviation of the expectation value for $d=5$ was $14.2$, and thus, it is not visible in the histogram because it is too large. 
The mean number of Pauli errors in the whole quantum circuit was $3.6$ for $d=5$ and $0.28$ for $d=7$. Thus, as explained in Sec.\,\ref{sec:pec}, QEM is useful when the number of Pauli errors in the circuit is less than unity.
These results show that the QEM technique is effective for large-scale quantum computing and it enables us to increase the effective code distance.

\subsection{Quantum error mitigation for approximation errors}
\subsubsection{Cost analysis}
\begin{figure}
    \centering
    \begin{tabular}{c}
    \subfigure[]{
        \includegraphics[width=7.4cm]{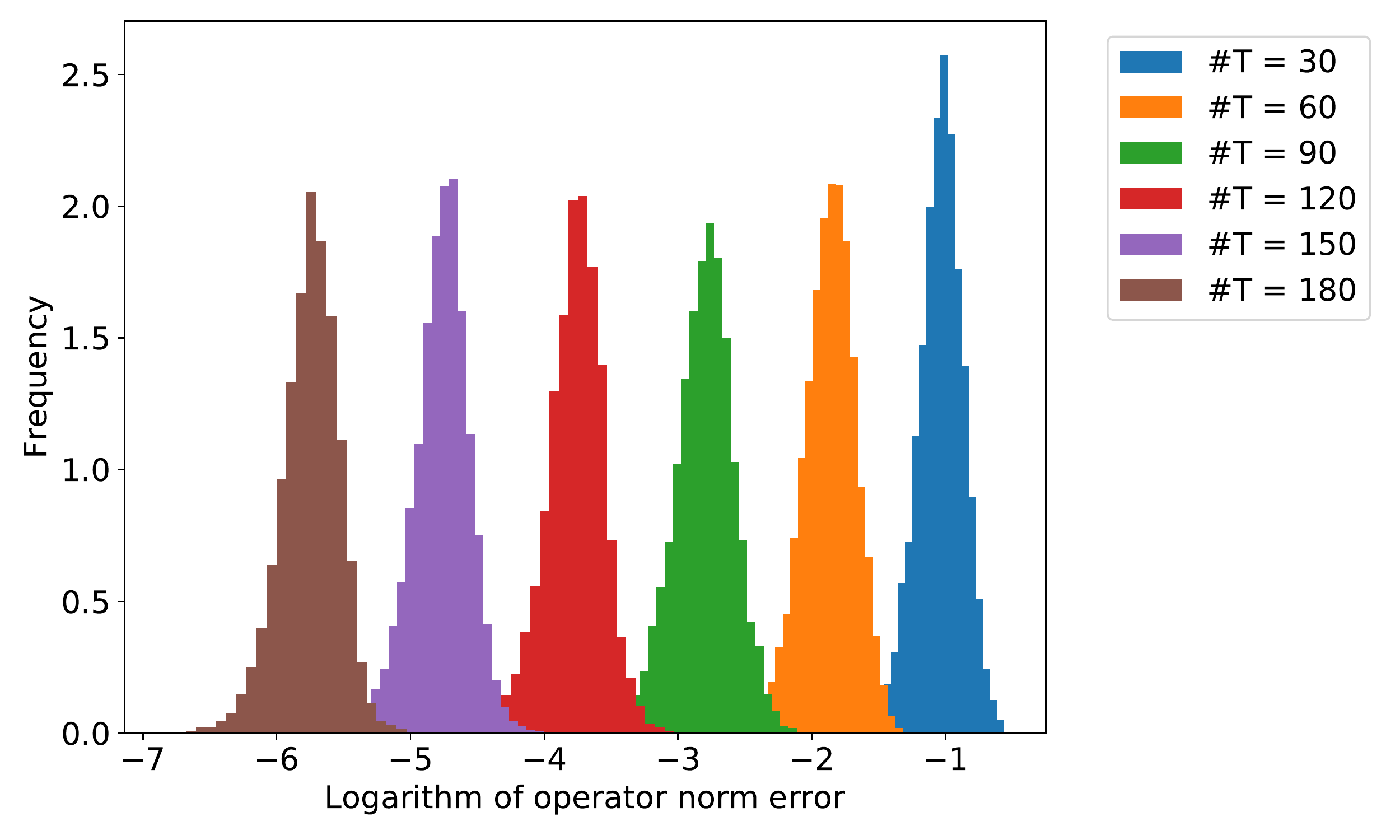}
        \label{fig:solovay_opnorm}
    }\\
    \subfigure[]{
        \includegraphics[width=7.4cm]{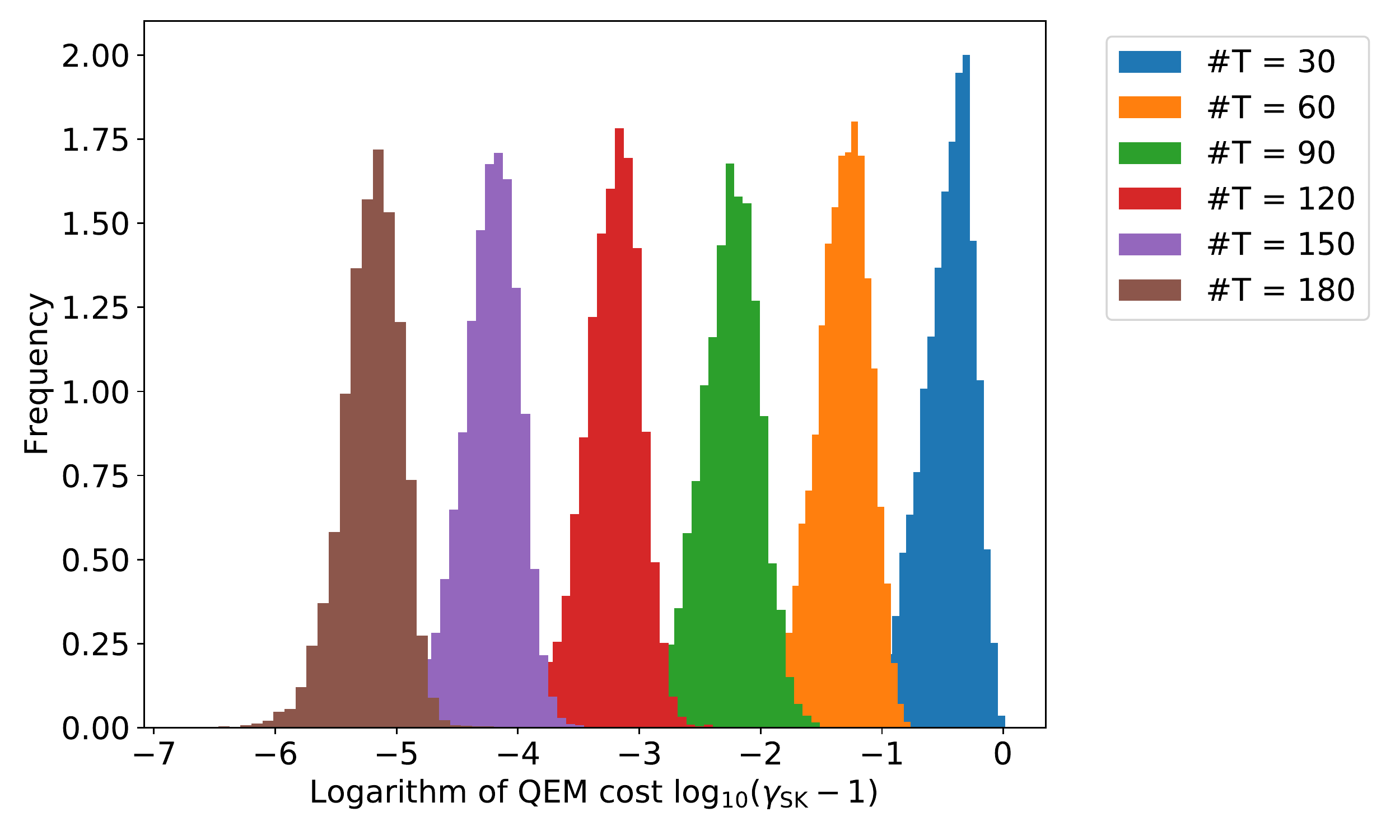}
        \label{fig:solovay_gamma}
    }\\
    \subfigure[]{
        \includegraphics[width=7.4cm]{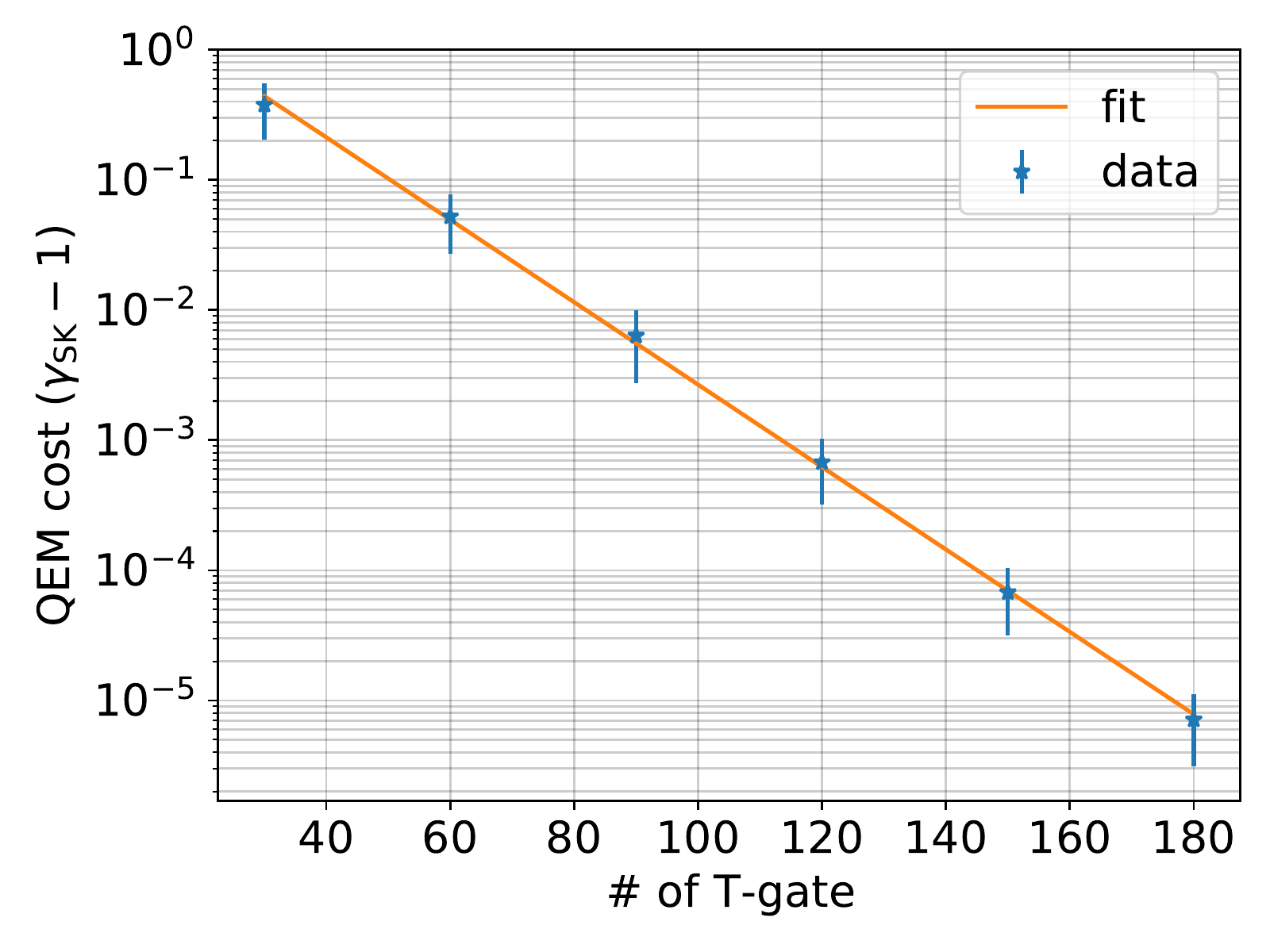}
        \label{fig:solovay_gamma_drop}
    }
    \end{tabular}
    \caption{Approximation error and QEM cost of improved Solovay-Kitaev method calculated for Haar-random unitary operations. Each color corresponds to the maximum count of $T$-gates. (a) Histogram of approximation errors due to the Solovay-Kitaev decomposition. (b) Histogram of QEM costs. (c) Histogram of QEM costs for approximation errors as a function of the number of allowed $T$-gates.}
\end{figure}

Next, we studied the performance of QEM when the Solovay-Kitaev decomposition is used. Since the actual QEM cost $\gamma_{\rm SK}$ depends on the target unitary operator, we drew a sample of unitary operations $U$ from a Haar-measure random distribution $\mu_{\rm H}$. Then, we decomposed a unitary gate in the form of $U = R_Z(\theta_1) \sqrt{X} R_Z(\theta_2) \sqrt{X} R_Z(\theta_3)$, where $\sqrt{X} = HSH$ is a Clifford operation. We used the improved Solovay-Kitaev algorithm from Ref.\,\cite{ross2014optimal}. This algorithm enables us to approximate an arbitrary Pauli-$Z$ rotation $R_Z(\theta) = \exp(i\frac{\theta}{2} Z)$ with an operator $\tilde{U}$ which is described as a sequence of Clifford operations and $T$-gates. We set the maximum count of $T$-gates for each decomposition for three Pauli-$Z$ rotations to check the trade-off relation between the $T$-gate count and the approximation accuracy.
Fig.\,\ref{fig:solovay_opnorm} shows the histogram of errors evaluated with operator norm $|| U - \tilde{U}||$. 
As expected, there was an exponential decrease in the approximation errors.

Next, we calculated the QEM cost by using Eq.\,(\ref{Eq:sv_decomp}). 
Fig.\,\ref{fig:solovay_gamma} shows the histogram of QEM costs $\gamma_{\rm SK}$, and Fig.\,\ref{fig:solovay_gamma_drop} plots the QEM cost versus the number of allowed $T$-gates. 
We can see that $\gamma_{\rm SK}-1$ exponentially decreases according to the number of $T$-gates, and its variance also decreases exponentially.
We fitted the QEM cost $\gamma_{\rm SK}$ with Eq.\,(\ref{Eq:SVcost}), and obtained $\beta_1 = 3.9(5)$ and $\beta_2 = 0.072(1)$.

\subsubsection{Performance analysis}
\begin{figure}[t!]
    \centering
    \includegraphics[width=0.8\columnwidth]{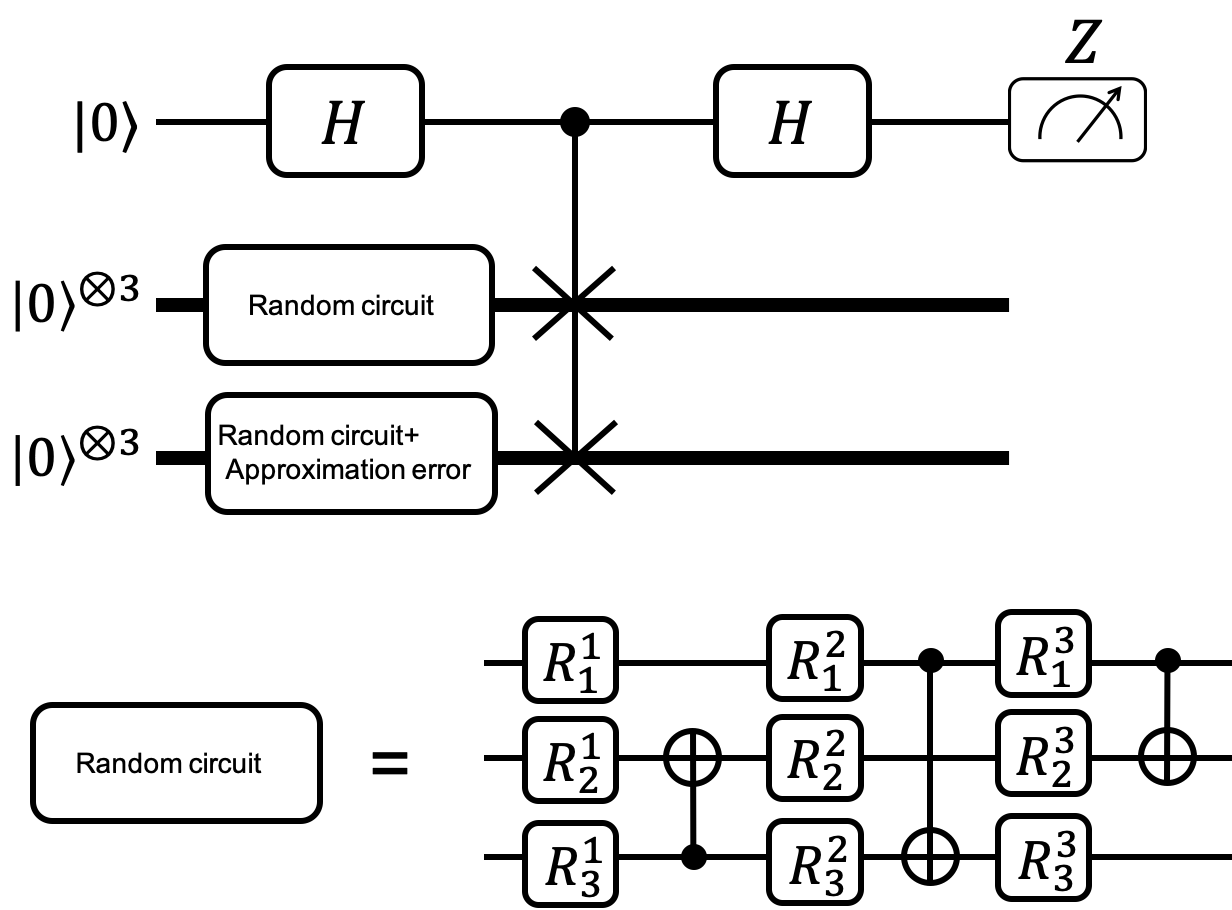}
    \caption{Schematic figure of simulated 7-qubit SWAP test circuit. We calculated the overlap of randomly generated states and the approximation of the generated state by using the Solovay-Kitaev algorithm. The random circuit was composed of three layers, each of which consisted of random single-qubit rotation gates and CNOT gates acting on two randomly chosen qubits. }
    \label{fig:swaptest}
\end{figure}

Next, we evaluated the performance of QEM for approximation errors due to the Solovay-Kitaev decomposition in a simulation of a SWAP test circuit with 7 qubits. A SWAP test circuit evaluates the overlap of two input states $\rho$ and $\sigma$ as $\mathrm{Tr}[\rho \sigma]$ by measuring ancilla qubits~\cite{ekert2002direct}. 
We set one of the input states to the ideal state and the other to the state affected by approximation errors. 
A schematic diagram is shown in Fig.\,\ref{fig:swaptest}. 

The ideal state was generated by using random quantum circuits composed of three layers. In each layer, random single-qubit unitary operations were simultaneously applied; then a CNOT gate acted on two randomly chosen qubits. The same random quantum circuit was applied to the approximate state by applying the Solovay-Kitaev decomposition to each single-qubit rotation. In this case, if there are no approximation errors, we necessarily obtain +1 as measurement outcomes since the input states are the same; hence the expectation value is also $+1$ for the Pauli-$Z$ operator of the ancilla qubit. On the other hand, the expectation value becomes smaller than unity when the inner product is reduced by approximation errors. Since approximation errors cannot be treated in the framework of stabilizer simulation, we simulated the quantum circuits by directly updating the state vector after each gate. 

\begin{figure*}
    \centering
    \subfigure[]{
        \includegraphics[width=\textwidth]{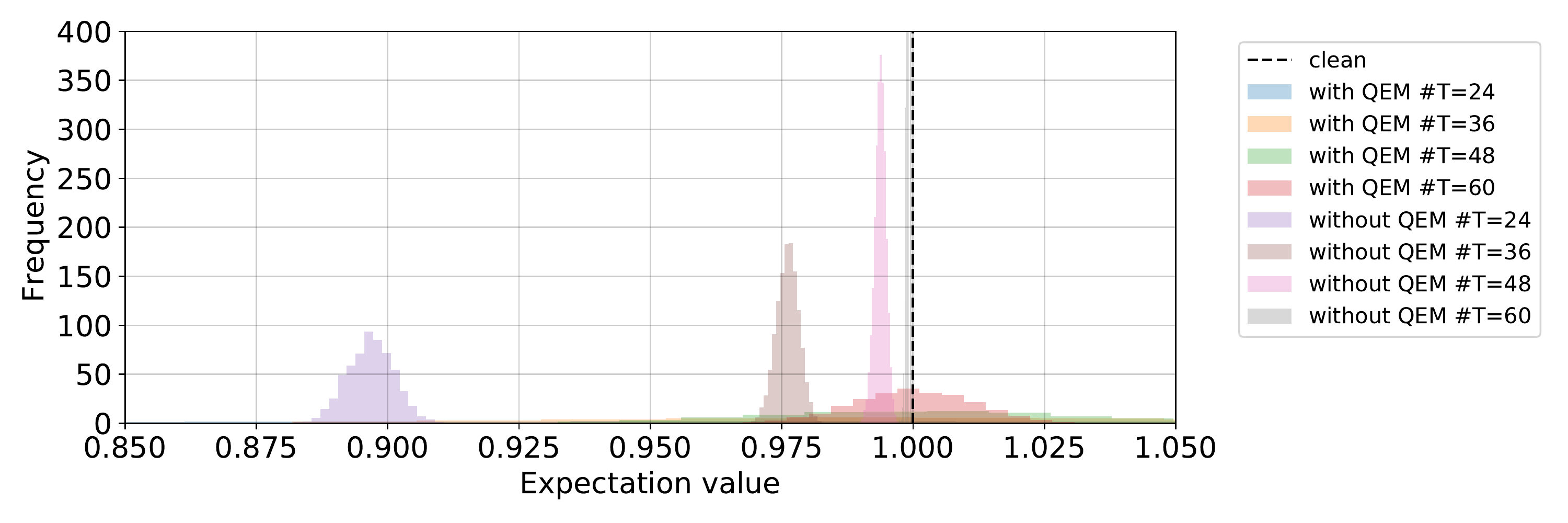}
        \label{fig:demonstrate_sk}
    } \\
    \begin{tabular}{cc}
    \subfigure[]{
        \includegraphics[width=7.5cm]{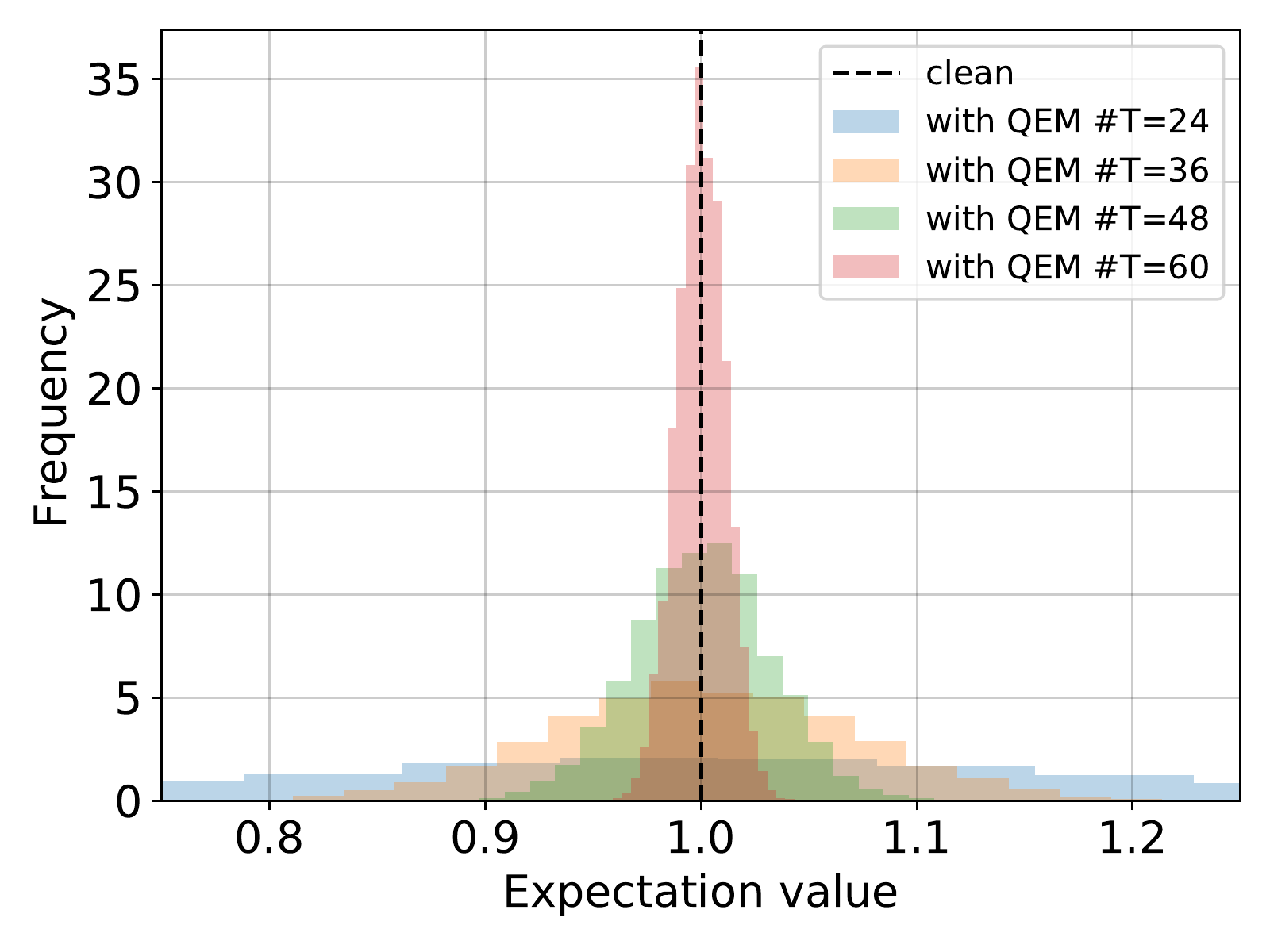}
        \label{fig:demonstrate_sk_mitigated}
    } & 
    \subfigure[]{
        \includegraphics[width=7.5cm]{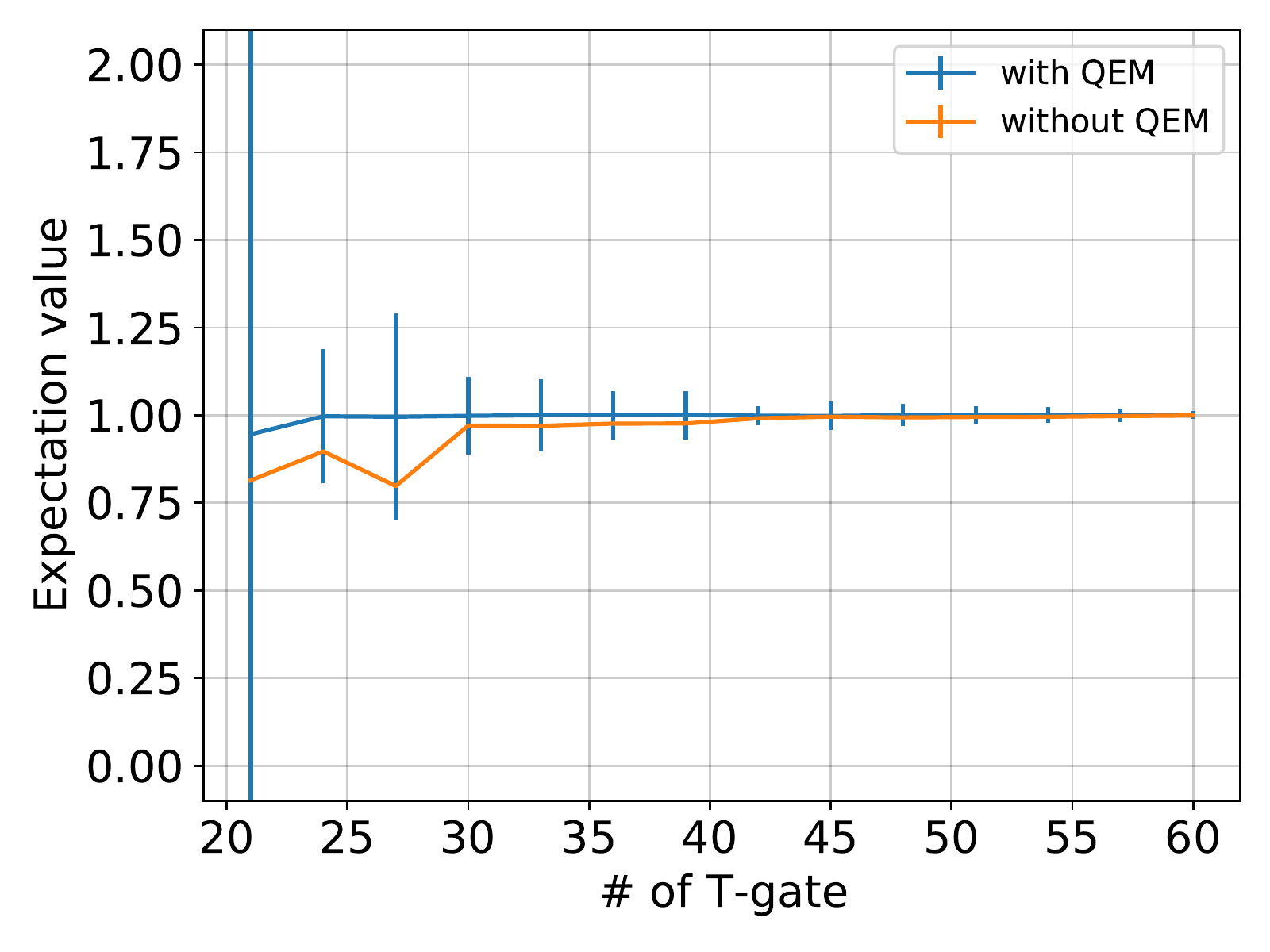}
        \label{fig:demonstrate_sk_decay}
    }
    \end{tabular}
    \caption{Histogram of expectation values of SWAP test with approximation errors. (a) Histogram of expectation values. (b) The same figure zoomed in around the ideal expectation value. (c) Sample averages and standard deviation as a function of the allowed number of $T$-gates.}
\end{figure*}

We numerically performed a series of $10^4$ experiments and computed the expectation values from $10^4$ single-shot measurements. The number of allowed $T$-gates in each Solovay-Kitaev decomposition for single-qubit unitary operations was varied from $24$ to $60$. 
Fig.\,\ref{fig:demonstrate_sk} shows the results, while Fig.\,\ref{fig:demonstrate_sk_mitigated} shows the data around the ideal expectation value. Moreover, Fig.\,\ref{fig:demonstrate_sk_decay} shows the mean value of $10^4$ samples for each logical error probability together with its standard deviation (the error bar). Note that standard deviation with $21$ $T$-gates is $4.65$.
We can see that our QEM technique successfully removed bias from the expectation value in the noisy cases. Compared with the QEM cost for decoding errors, we obtained a larger QEM cost in this case. This is consistent with the results reported in Ref.~\cite{endo2018practical} that indicates the QEM cost for unitary errors tends to be larger than that for stochastic errors. 

This problem may be alleviated by performing several Solovay-Kitaev decompositions with the same accuracy, constructing randomizing approximation errors, and removing the coherent component of the noise.
Note that with a sufficiently large sample size, our QEM technique enables the effective number of $T$-gates to be increased by inserting additional Clifford gates and Pauli channels and conducting repeated sampling, with negligible additional hardware requirements.

\subsection{Quantum error mitigation with estimation errors}
Probabilistic error cancellation assumes that the noise maps to be canceled are known in advance. While we can determine the approximation error of the Solovay-Kitaev decomposition within the numerical precision, it is hard to exactly characterize the noise maps of the decoding errors because GST is affected by finite sampling, as discussed in Sec.\,\ref{sec:effect_estimation_error}. In this section, we numerically evaluated the performance of our framework in the case of finite estimation errors. 

For the benchmarks, we chose the same quantum circuit, noise model, and observable as in Sec.\,\ref{sec:logical_error_performance_analysis}. We evaluated the expectation value for a 100-qubit noisy Clifford circuit that is unity if there is no noise. A non-uniform depolarizing channel where Pauli-$X,Y,Z$ occurs with probabilities $(p_x, p_y, p_z)$ was inserted after each gate. We used the probabilities obtained from the simulation of surface codes shown in Table.\,\ref{tab:logical_probs}. The point of difference from the previous simulations is that these probabilities were over- or under-estimated as $((1+r)p_x, (1+r)p_y, (1+r)p_z)$ ($r>-1$) in probabilistic error cancellation. Here, $r=0$ corresponds to an exact characterization and perfect error mitigation, while $r=-1$ corresponds to FTQC without error mitigation. As in the discussion in Sec.\,\ref{sec:effect_estimation_error}, we expect the distribution is such that its mean value is the same as in the simulation of the mitigated noise model $(|r|p_x, |r|p_y, |r|p_z)$ and its variance is approximately $\Gamma$, which is the overhead of QEM determined by the estimated noise model $((1+r)p_x, (1+r)p_y, (1+r)p_z)$.
\begin{figure*}
    \centering
    \subfigure[]{
        \includegraphics[width=\textwidth]{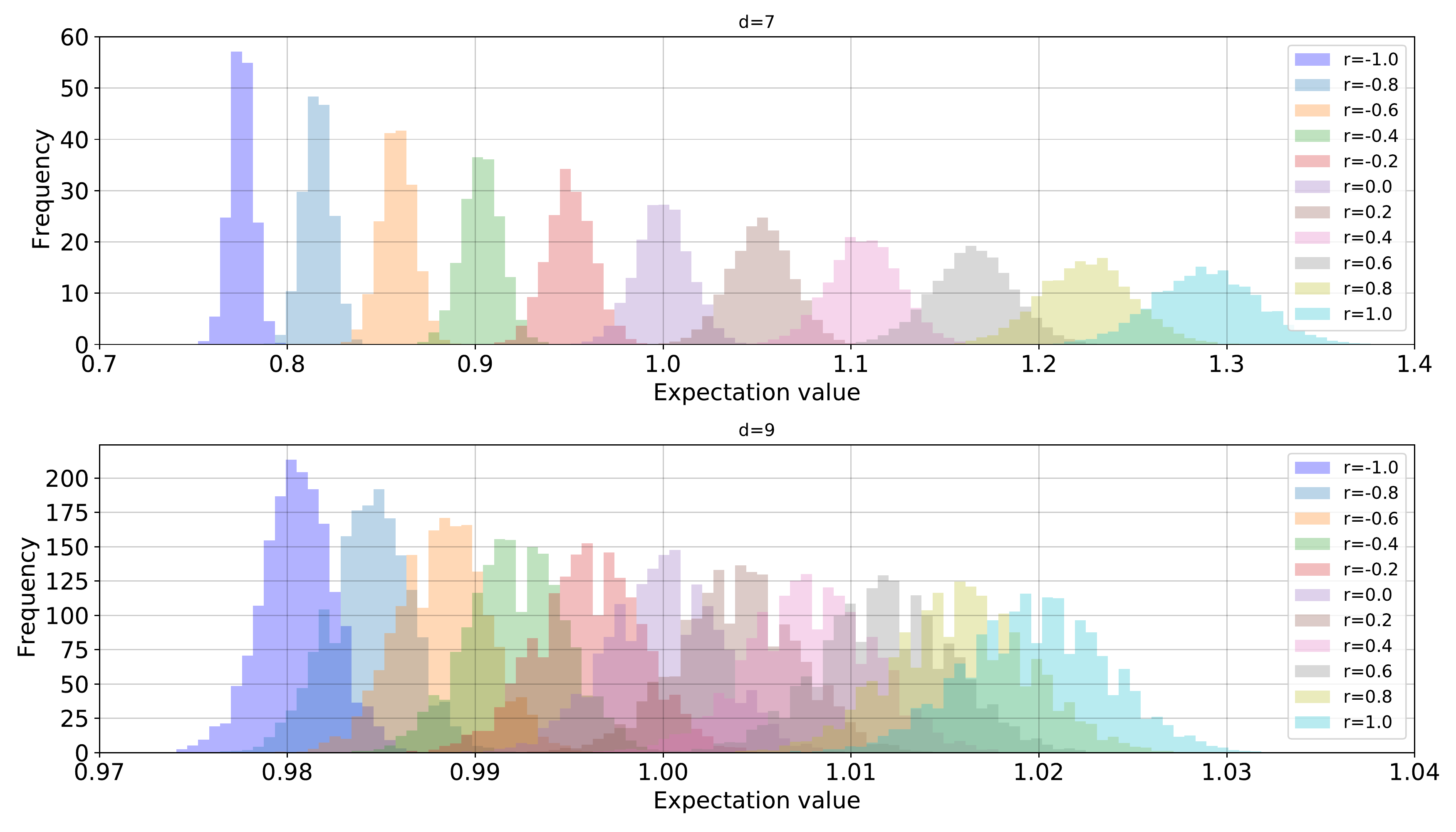}
        \label{fig:demonstrate_estimation_error}
    } \\
    \begin{tabular}{cc}
    \subfigure[]{
        \includegraphics[width=7.5cm]{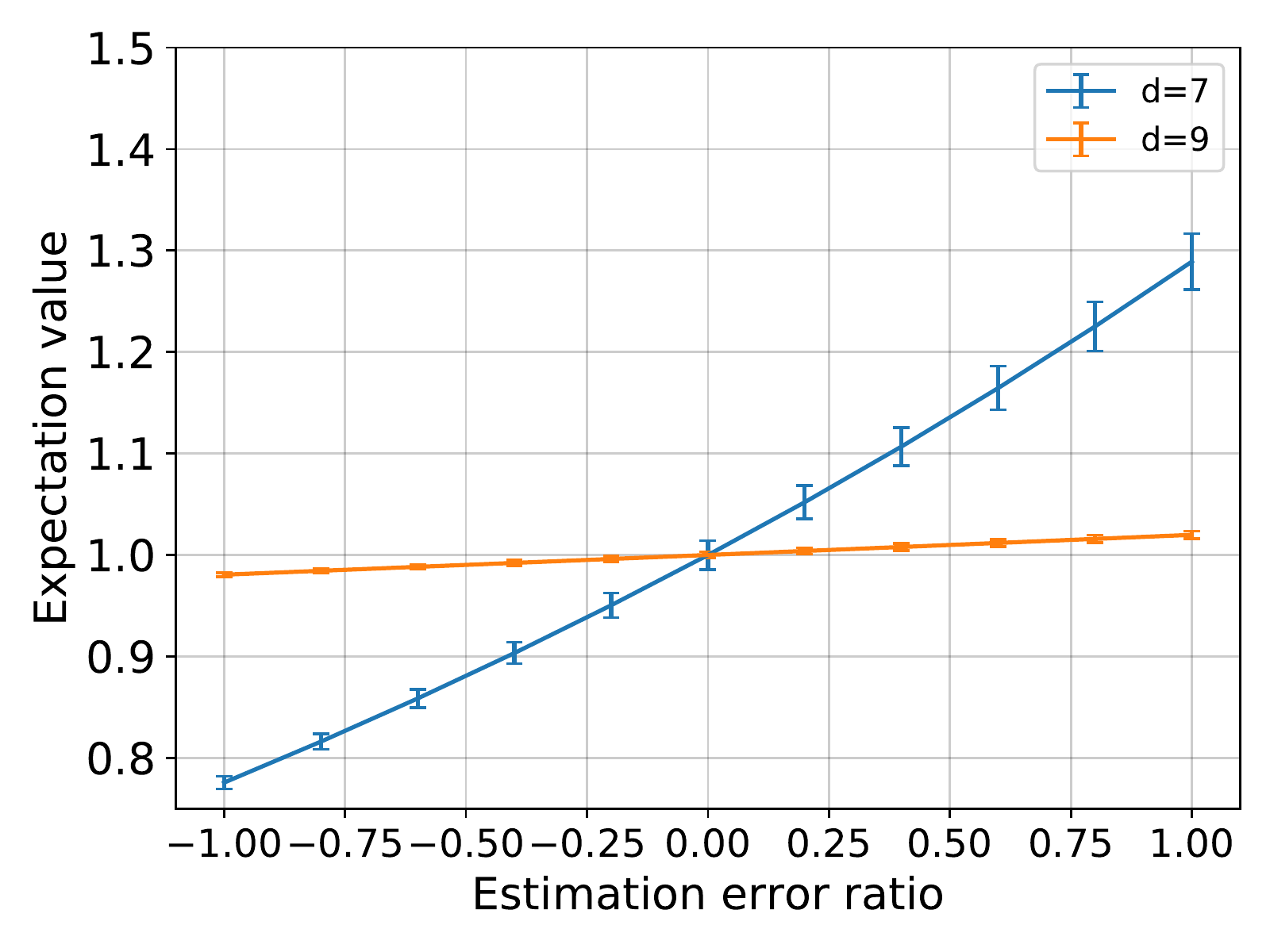}
        \label{fig:demonstrate_estimation_error_decay}
    }
    \end{tabular}
    \caption{Histogram of expectation values with estimation errors. (a) Histogram of expectation values. (b) Sample averages and standard deviation as a function of estimation accuracy.}
\end{figure*}
The histogram in the case of finite estimation errors parametrized by $r$ for $d=7$ and $d=9$ are plotted in Fig.\,\ref{fig:demonstrate_estimation_error}, and the mean values and standard deviations are plotted as a function of $r$ in Fig.\,\ref{fig:demonstrate_estimation_error_decay}. 
We can see that the residual bias increases exponentially with $|r|$. 
With infinite samples, QEM is beneficial when $r$ is sufficiently smaller than unity. Comparing the cases of under-estimation $r<0$ and over-estimation $r>0$ with the same absolute value $|r|$, we find that over-estimation ($r>0$) has a larger variance than that of under-estimation ($r<0$), while there is similar amount of bias in the expectation values. Since QEM prefers under-estimation to over-estimation, we conclude that characterization methods with a weighted penalty may lead to a further improvement in QEM.

\subsection{Practical utility of quantum error mitigation for FTQC}
While we have shown our method effectively decreases the hardware requirement of FTQC with the examples of random Clifford and swap-test circuits, the practical availability of QEM in the region where FTQC is used for useful applications with a quantum advantage is also vital. In this section, we discuss the enhancement of computation accuracy in this regime with our protocol.

We estimated how many quantum logical Clifford operations $N_G$ are required in the practical region from the existing resource estimation. Note that there are other noise sources, such as imperfect $T$-gate preparation, that can be counted as the overhead of logical operations in distillation processes, and thus we counted the effects of them as decoding errors. We considered two scenarios for evaluation; an optimistic one and a realistic one. An optimistic scenario is the case of lightweight applications mainly for showing a quantum advantage, i.e., an application with the minimum possible $N_G$ that cannot be simulated with the existing classical computers. We referred to the analysis of Refs.\,\cite{arute2019quantum,pednault2019leveraging} to estimate the maximum problem size tractable with the existing classical computers. According to them, we expect that quantum circuits with depth 100 and 100 logical qubits are sufficient to go beyond the limitation of classical simulation and that we can achieve this with $N_G \sim 10^4$.
The other scenario is the ground-energy estimation of spin models and chemical molecules since the quantum simulation is expected as one of the most resource-efficient applications whose quantum advantage is well studied. We picked an expected number of gates from recent the state-of-the-art resource estimation in Refs.\,\cite{kivlichan2020improved} and \cite{babbush2018encoding} that discuss the cost for calculating ground-state energy. These papers utilize quantum simulation with Trotterization~\cite{suzuki1991general} and qubitization~\cite{low2019hamiltonian} as a subroutine, respectively. Trotterization is a method to simulate quantum systems by approximating Hamiltonian dynamics with Trotter decomposition~\cite{suzuki1991general}, and qubitization~\cite{low2019hamiltonian} is a recently proposed method that constructs the state after time-evolution by repetitive applications of Grover-like iterations. 
According to the Tables.\,1 and 2 in Ref.\,\cite{kivlichan2020improved} and the Table.\,IV in Ref.\,\cite{babbush2018encoding}, approximately $10^{8}$ $T$-gates are required to simulate a Hubbard model that is hard to simulate with classical computers. 
Note that while these algorithms use phase estimation sampling to obtain the ground energy as binary digits and are not a procedure to estimate expectation values, we can still apply QEM to these algorithms with small overheads. This is because these problems can be translated into a series of decision problems. See Appendix.\,\ref{sec:apply_qem_to_bqp} for a detailed explanation. Considering there is an overhead of executing magic-state distillation and so on, we expect that $N_G = 10^{10}$ is required as a pessimistic estimation. 

\begin{figure*}
    \centering
    \subfigure[]{
        \includegraphics[width=7.5cm]{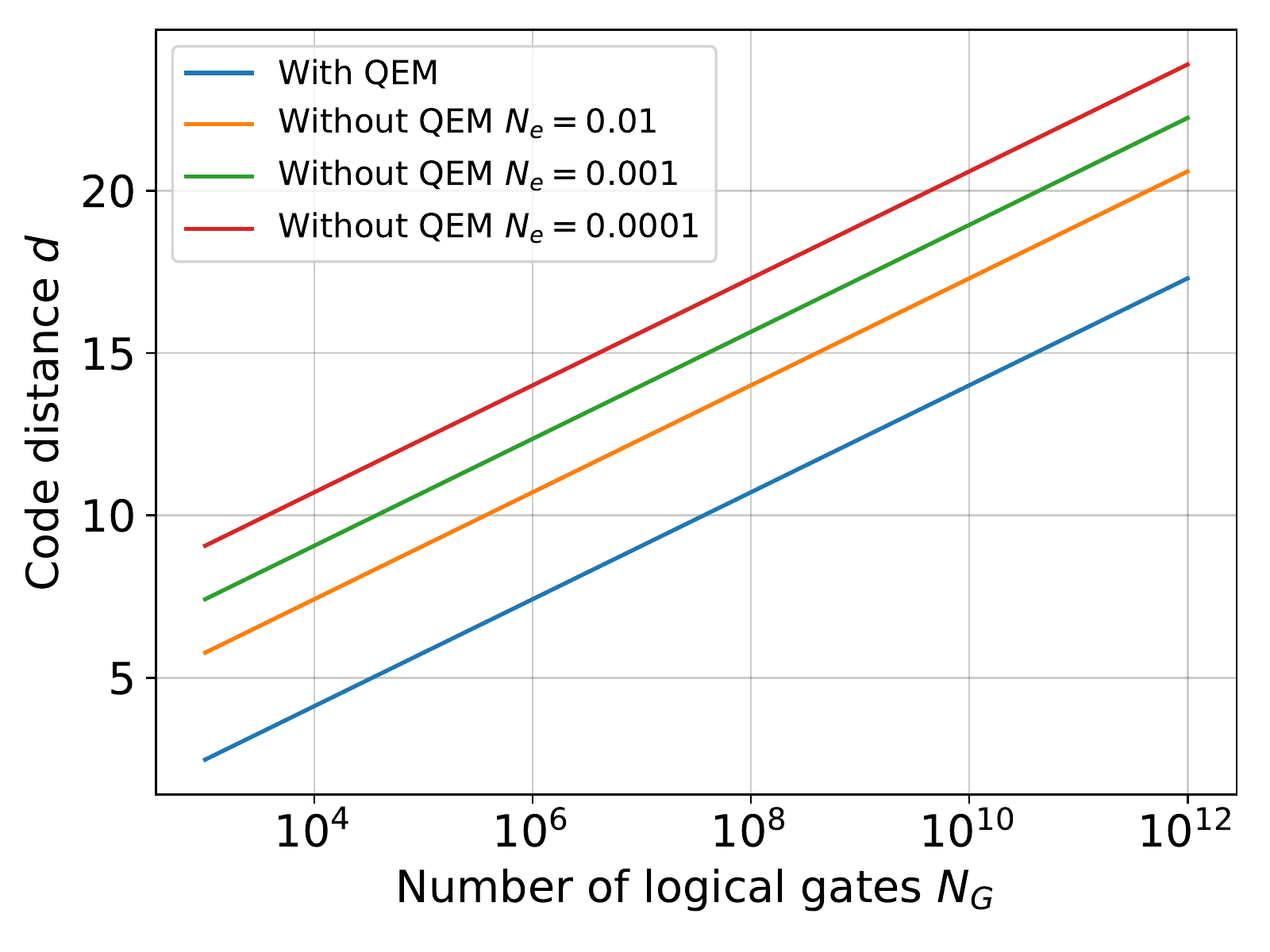}
        \label{fig:resource_estimation_large_application}
    }
    \subfigure[]{
        \includegraphics[width=7.5cm]{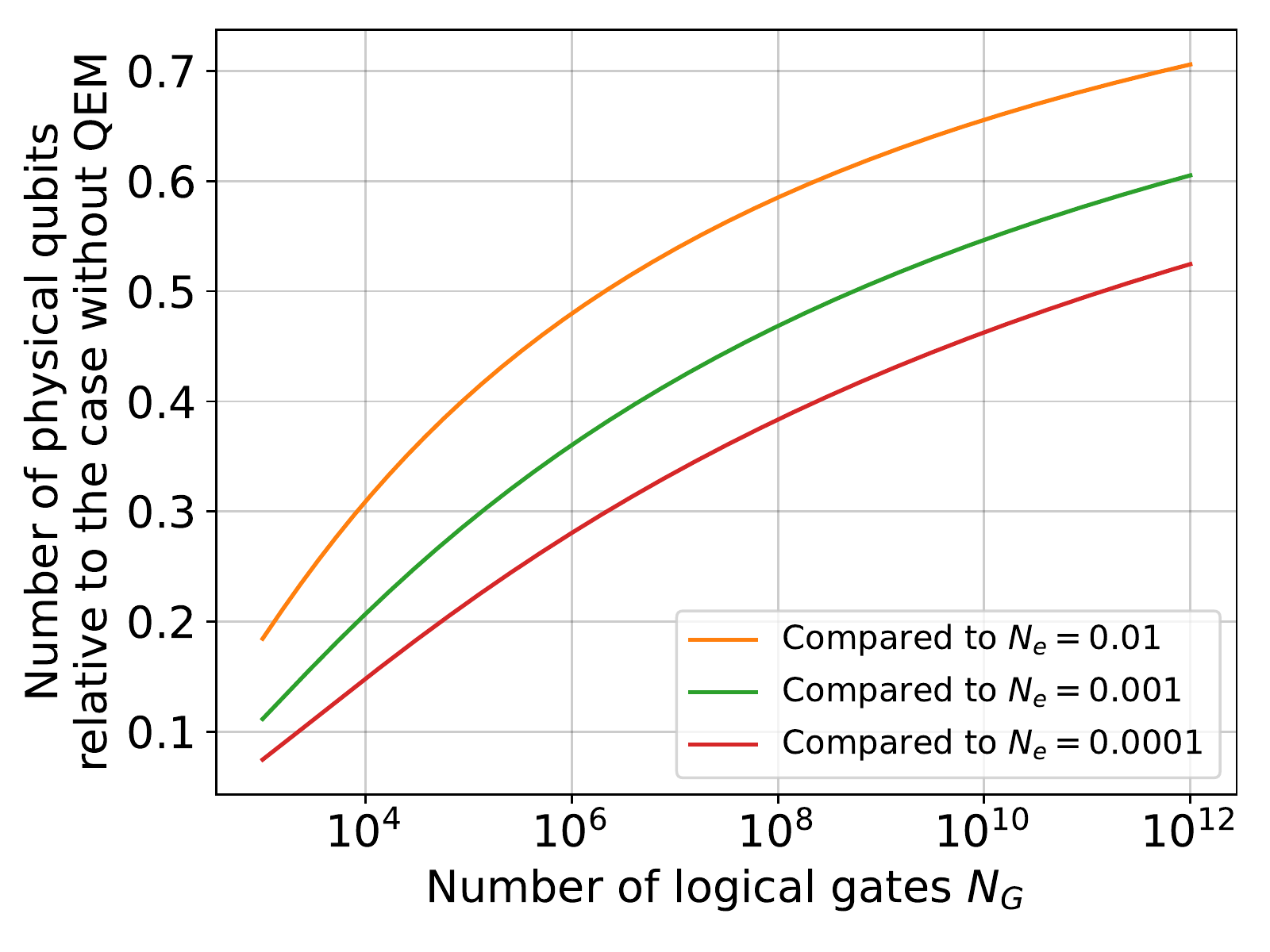}
        \label{fig:resource_estimation_large_application_reduce_qubit}
    }
    \caption{(a) The relation between required number of gates for executing algorithm $N_G$ and required code distance for encoding the information of qubit $d$. The relation is plotted for several numbers of allowed logical error counts during FTQC $N_{\rm e}$. (b) The ratio between the number of required physical qubits per logical qubit with and without QEM.}
    \label{fig:resource_estimation_large_application_both}
\end{figure*}

Next, we considered how QEM can reduce the effect of decoding errors of logical operations during FTQC. Without QEM, a logical error probability $p_{\rm L}$ must be much smaller than the inverse of the number of logical operations $N_G$. In other words, the mean number of logical errors $N_G p_{\rm L}$ must satisfy $N_G p_{\rm L}=O(1)$. Suppose the allowed mean number of errors without QEM for satisfying the required accuracy as $N_{\rm e}$, which becomes small as required accuracy becomes strict. On the other hand, with QEM, the bias of expectation values caused by logical errors whose mean number is below unity can be mitigated with $e^{O(1)}$, i.e., about $e^4 \sim 55$, sampling overheads according to Eq.\,(\ref{Eq:tradeoff}). Thus, the required logical error rates are relaxed from $p_{\rm L} \sim N_{\rm e} / N_G$ to $p_{\rm L} \sim 1 / N_G$. The effective error rates of elementary logical operations such as logical Clifford gates decrease exponentially with the code distance. When we focus on the advantage of QEM in the logical space, we can estimate the relations between the problem size $N_G$ and code distances $d$ in terms of the required mean number of errors without QEM $N_{\rm e}$ as shown in Fig.\,\ref{fig:resource_estimation_large_application}.
In this figure, the required code distance for reliable accuracy is plotted as a function of the number of logical gates required for executing an algorithm for several cases of $N_{\rm e}$. For the calculation, we used Eq.\,(\ref{Eq:threshold}) with $p/p_{\rm th} = 0.1$. The number of physical qubits per logical qubit scales as $O(d^2)$ in the case of two-dimensional topological codes. We plot how the number of physical qubits are reduced with QEM $d_{\rm mit}^2 / d_{\rm nmit}^2$ in Fig.\,\ref{fig:resource_estimation_large_application_reduce_qubit}, where $d_{\rm mit}$ and $d_{\rm nmit}$ are the required code distance with and without QEM, respectively. 
While the improvement of the code distance $d$ is constant, the impact of resource reduction depends on the expected technologies and the size of the problems of interest.

When we consider applications in the early FTQC era with $N_G \sim 10^4$, according to Fig.\,\ref{fig:resource_estimation_large_application}, the required code distance is reduced from about $9$ to $4$ if $N_{\rm e} = 10^{-3}$ and the required number of qubits is reduced to $21\%$. This advantage becomes large when the required accuracy $N_{\rm e}$ becomes strict. Even when we take the cost for distillation and lattice surgery into account, FTQC with $100$ logical qubits with the code distance $4$ is estimated to require about $10^4$ physical qubits with QEM, which can be the promising milestone to show the computational supremacy on the logical space. While this advantage becomes comparably small when we see promising long-term applications in $N_G \sim 10^{10}$, the code distance is still reduced from $19$ to $14$, which suppresses the number of physical qubits to $55\%$. Thus, we can conclude that in both cases, our proposal is expected to drastically alleviate the hardware requirement in practice.

It should be noted that the reduction of the code distance is vital not only for the reduction of the number of physical qubits but also for relaxing the requirement of error decoding architectures. To estimate occurred Pauli errors on physical qubits during FTQC, we need classical peripherals for decoding with sufficiently small latency for stabilizer measurement cycles. However, recent results show the tractable size of realistic implementation can decode up to about code distance $11$~\cite{holmes2020nisq+,ueno2021qecool,das2021lilliput}. While this value may be improved in the future, it is obvious that the performance of decoding units is another restriction of FTQC.  When we assume tractable code distance is limited to $11$, we can use at most $10^5$ logical gates with $N_{\rm e} = 10^{-3}$, which is just around the limitation of the classical simulation. On the other hand, the use of QEM can increase the available logical gates to $10^8$. Thus, our proposal can be the key to push the performance of FTQC from a classically simulatable region to the quantum supremacy regime.

While we only discussed the reduction of code distance, we can similarly reduce the effective required number of $T$-gates. While the effective increase of the resource is also constant as in the case of code distance, this can be used not only for reducing the total number of generated $T$-gates but also for mitigating the bias of generation throughput of magic states during FTQC protocol. Since magic-state generation succeeds probabilistically, the number of available magic states per time unit statistically fluctuates and cannot be estimated in advance. We expect our method can exempt these kinds of difficult run-time scheduling and make long-time execution of FTQC reliable. Furthermore, in the case of distributed FTQC, logical CNOT gates between distant nodes require entanglement distribution and distillation, which also typically succeed probabilistically. Thus, QEM can be used for reducing a wide range of difficulties in FTQC.

\section{Discussion}
\label{sec:conclusion}
We have described a method to effectively decrease errors in FTQC by performing QEM in the logical space. In the case of decoding errors due to insufficient code distances and magic-state distillation, we can perform QEM with small modification on quantum circuits. In particular, owing to the Pauli frame, we can perform QEM without implementing any physical operations if decoding errors are stochastic Pauli maps, while QEM operations may induce additional errors when we implement it physically on general quantum circuits. In regard to the approximation errors due to the Solovay-Kitaev decomposition, we cannot always use the Pauli frame because QEM employs not only Pauli operations but also Clifford operations. Since Clifford operations can be efficiently performed in FTQC and the number of decoding processes for error correction is much larger than the gate count in Solovay-Kitaev algorithms, this overhead is negligible. We have verified the trade-off between costs of QEM and the code distance and the number of $T$-gates. Furthermore, we have estimated the sampling cost of gate set tomography to obtain the decoding noise map to the required accuracy, and clarified that our approach enables quantum computing corresponding to more than the achievable code distance. We have numerically compared the computation with and without QEM on FTQC with the same sampling number and have shown the advantage of QEM even under the existence of finite estimation errors. We also have estimated the required resources with and without QEM in the early-FTQC era, and have shown that the required number of physical qubits can be suppressed up to tens of percent.

It should be noted that this is the first result to clearly show QEM can dramatically improve the computation accuracy for useful applications under realistic assumptions as we have shown by using the example of quantum simulation on FTQC. This is because the computational advantage of typical NISQ algorithms such as variational algorithms are empirically assumed and the required runtime of the algorithm has not been revealed; yet the accuracy of FTQC algorithms can be estimated reliably depending on the complexity of the problem. Accordingly, the usefulness of quantum error mitigation can be clearly discussed in the FTQC scenario.

Another important aspect of providing the theory and implementation of QEM for FTQC rather than NISQ computing is as follows. While it is known that QEM is the most effective when the mean number of errors during computation is in the order of unity~\cite{endo2018practical}, this criterion is not necessarily satisfied in the NISQ regime depending on the problem size. On the other hand, because we are allowed to tune code distances, magic-state distillation levels, and the number of $T$-gates per Solovay-Kitaev decomposition in FTQC, it is highly likely that we can satisfy this criterion. Thus, the main drawback of QEM, the exponential growth of the sampling overheads, can be circumvented; therefore we can find the highly practical regimes where QEM can help enhance the computation accuracy. Accordingly, we can conclude that the theory of QEM on FTQC is more versatile than that for NISQ computing.

As mentioned in the main text, promising applications of our method are quantum phase estimation algorithms and Hamiltonian simulation algorithms for investigating quantum many-body dynamics. There are algorithmic errors in the Trotter decomposition~\cite{lloyd1996universal} and recently proposed methods such as Taylor series~\cite{berry2015simulating} and quantum signal processing~\cite{low2019hamiltonian,low2017optimal}. In Refs.~\cite{endo2019mitigating,vazquez2020enhancing}, it is shown that such algorithmic errors can be mitigated by employing extrapolation. Since algorithmic errors can be controlled by changing the simulation accuracy, this technique can also be naturally incorporated in an FTQC scenario. Thus, the dominant errors in FTQC can be compensated via QEM.

The first generation of FTQC may not be sufficiently large for naively solving large and useful problems. While the architecture of distributed quantum computing is the most straightforward approach to increase the total number of qubits, it requires interconnections between quantum nodes, which induces additional overheads for entanglement distillation. Thus, we sometimes cannot use the sufficient number of distilled entanglements for distributed FTQC. In this context, techniques developed in NISQ era~\cite{mitarai2019constructing,mitarai2020overhead} for solving larger problems with small NISQ computers may also be useful in the middle-term FTQC. Our work is the first proposal that makes the best of a technique tailored for NISQ devices in the context of FTQC.

Finally, we should discuss the difference between our scheme and a similar work that combines QEM with the quantum error correction proposed by \,\textcite{mcclean2020decoding}. Their method considers implementing quantum error correction for NISQ devices via classical post-processing in the case that experimentalists cannot implement stabilizer measurements because of the limited connectivity and large error rates of NISQ devices. Although their method enables the state to be projected to the code space via quantum subspace expansion~\cite{mcclean2017hybrid}, logical errors cannot be fully eliminated. On the other hand, our scheme assumes FTQC can be performed but the number of qubits and $T$-gates cannot be increased infinitely. The remarkable advantage of our method is that we can fully eliminate the decoding errors and approximation errors by using a greater number of measurements at negligible hardware overhead, given the good characterization of the noise model. 

\section*{Note added}
After we uploaded this work to arXiv, three relevant works appeared that have a similar concept that quantum error mitigation is incorporated in fault-tolerant quantum computing to relax the hardware requirement at the cost of sampling overheads~\cite{xiong2020sampling,piveteau2021error,lostaglio2021error}. Ref.\,\cite{xiong2020sampling} shows quantum error mitigation for encoded qubits but they focus on concatenated codes rather than topological codes. Ref.\,\cite{piveteau2021error} uses quantum error mitigation for implementing $T$-gate without magic-state distillation and shows efficient characterization methods for the errors of $T$-gate under the assumption that logical Clifford operations are perfect. Ref.\,\cite{piveteau2021error} also discusses on how to relax the costs for implementing $T$-gates by using the concepts of robustness of magic. 

Compared to these works, we emphasize that our framework consider a different scenario where logical Clifford operations are imperfect, and thus FTQC suffers from decoding errors of logical Clifford procedure, logical noise on prepared magic states, and insufficient magic-state supply. This difference makes our framework versatile in the early FTQC era. While we provided consistent analysis of gate set tomography with noisy logical Clifford gates, we have refined the treatment of the noise of magic state preparation in gate set tomography, motivated by Refs.\,\cite{piveteau2021error,lostaglio2021error}.

\section*{Acknowledgement}
This work is supported by PRESTO, JST, Grant No.\,JPMJPR1916; ERATO, JST, Grant No.\,JPMJER1601; CREST, JST, Grant No.\,JPMJCR1771; MEXT Q-LEAP Grant No.\,JPMXS0120319794 and JPMXS0118068682; Moonshot R\&D, JST, Grant No.\,JPMJMS2061. We would like to thank Takanori Sugiyama for a fruitful discussion on gate set tomography. We acknowledge useful discussions with Zhenyu Cai, Xiao Yuan, Rui Asaoka, and Kaoru Yamamoto.

\appendix
\section{Pauli transfer matrix}
Any quantum map can be represented as a matrix called a Pauli transfer matrix (PTM). Suppose we perform a quantum process $\mathcal{E}$ on $n$ qubits that maps an $n$-qubit density matrix to another density matrix. We denote the set of $n$-qubit Pauli operators as $\mathcal{P}^{(n)} = \{I,X,Y,Z\}^{\otimes n}$. The set of $n$-qubit Pauli operators forms a basis of $4^n \times 4^n$ matrices and the elements are mutually orthonormal, $\frac{1}{d} {\rm Tr}[P_i P_j] =  \delta_{ij}$, for $d=2^n$.
The PTM representation of process $\mathcal{E}$ is defined with the Pauli basis as follows:
\begin{align}
\mathcal{M}(\mathcal{E})_{ij} = \frac{1}{d}{\rm Tr}[P_j \mathcal{E}(P_i)],
\end{align}
where $P_i$ is the $i$-th element of $\mathcal{P}^{(n)}$. Note that since any physical process maps a self-adjoint operator to another self-adjoint operator, all the elements of the Pauli transfer matrix are real values.
The density matrix can also be represented as a column vector with the Pauli basis:
\begin{align}
\ket{\rho}\rangle_i = {\rm Tr}[P_i \rho].
\end{align}
Note that an element of the vector corresponding to $P_i = I$ is the trace of $\rho$. A measurement of an observable $O$ can be mapped to a row vector,
\begin{align}
\langle \langle O|_j= \frac{1}{d} \mathrm{Tr}[O P_j ].
\end{align}
Note that the PTM representation satisfies 
\begin{align}
\mathrm{Tr}[O \mathcal{E}(\rho)] = \langle \langle O| \mathcal{M}(\mathcal{E}) \ket{\rho}\rangle.
\end{align}

There are several important properties of the Pauli transfer matrix representation. Given a composite map $\mathcal{E}=\mathcal{E}_A \circ \mathcal{E}_B$, the PTM representation is $\mathcal{M}(\mathcal{E}_B \circ \mathcal{E}_A) = \mathcal{M}(\mathcal{E}_B) \mathcal{M}(\mathcal{E}_A)$. Due to the linearity of the PTM representation, for the map $\mathcal{E}': \rho \mapsto \sum_k q_k \mathcal{E}_k(\rho)$, we have  $\mathcal{M}(\mathcal{E}') = \sum_k q_k \mathcal{M}(\mathcal{E}_k)$. When no confusion is possible, we will simply represent the Pauli transfer matrix $\mathcal{M}(\mathcal{E})$ as $\mathcal{E}$.

\section{Coefficients for quasi-probability decomposition}
\label{sec:coef}
To perform probabilistic error cancellation on an arbitrary $m$-qubit noise map $\mathcal{N}$, we need to decompose the inverse of the noise map $\mathcal{N}^{-1}$ into a linear combination of physical quantum processes.
According to Ref.\,\cite{endo2018practical}, any TPCP-map can be represented as a linear combination of Clifford operations and Pauli channels. This is because the set of Pauli transfer matrices of $m$-qubit Clifford operations and Pauli channels forms a basis of $m$-qubit Pauli transfer matrices. 
Here, we will introduce the following $16$ operators: 
\begin{equation}
\begin{aligned}
\mathcal{B} = \left \{ I, \sigma_j, \frac{I+i\sigma_j}{\sqrt{2}}, \frac{I + \sigma_j}{2}, \frac{\sigma_{j} + \sigma_{j+1}}{\sqrt{2}}, \frac{\sigma_{j} + i\sigma_{j+1}}{2} \right\},
\end{aligned}
\end{equation}
where $j \in \{1,2,3\}$, $(\sigma_1, \sigma_2, \sigma_3) = (X,Y,Z)$ and $\sigma_{4} = \sigma_{1}$. The Pauli transfer matrices of $\mathcal{B}^{\otimes m}$ comprise a complete basis of $n$-qubit Pauli transfer matrices; i.e., any quantum map can be represented as a linear combination of Clifford operations and Pauli channels. 
Since the application of non-Clifford operations requires complicated processes such as magic-state injection, distillation, and teleportation, this property is vital to performing probabilistic error cancellation on an arbitrary noise map in FTQC.

Specifically, when the noise can be modeled as stochastic Pauli errors, we can cancel it only with Pauli operations. This is because the set of Pauli transfer matrices of Pauli operations forms a basis of diagonal $n$-qubit Pauli transfer matrices.
Since we can perform logical Pauli operations only by updating the Pauli frame, which is stored in a classical memory, in FTQC, we can cancel stochastic logical Pauli noise without acting on the actual quantum device.
Suppose that the noise model is described by a Pauli transfer matrix acting on $m$ qubits:
\begin{equation}
\begin{aligned}
\mathcal{N}_{\rm Pauli} = (1-p_{\rm err}) \mathcal{I} + \sum_{g \neq I^{\otimes m}} p_g \mathcal{P}_g,
\end{aligned}
\end{equation}
where $\mathcal{I}$ is the identity map, $\mathcal{P}_g$ is the Pauli transfer matrix of Pauli operator $g$, $p_g~(g \in \{I,X,Y,Z\}^{\otimes m})$ is the probability at which the Pauli error $g$ occurs, and $p_{\rm err} = \sum_{g \neq I^{\otimes m}} p_g$.
This noise can be canceled with the following map:
\begin{equation}
\begin{aligned}
\mathcal{N}_{\rm Pauli}^{-1} = \sum_g \eta_g \mathcal{P}_g,
\end{aligned}
\end{equation}
where 
\begin{equation}
\begin{aligned}
\eta_g &= 4^{-n} \sum_{g'} \frac{c(g, g')}{\sum_{g''} p_{g''} c(g', g'')}.
\end{aligned}
\end{equation}
Note that $c(g,g')$ is a function of two Pauli operators such that $c(g,g')=1$ if $gg'=g'g$ and $c(g,g')=-1$ otherwise.
In the case of single-qubit Pauli noise, the coefficients and QEM cost $\gamma_Q$ can be explicitly expressed as follows.
\begin{widetext}
\begin{equation}
\begin{aligned}
\eta_{I}&=\frac{1}{4}\bigg(1+ \frac{1}{1-2(p_Y+p_Z)}+\frac{1}{1-2(p_Z+p_X)}+\frac{1}{1-2(p_X+p_Y)}\bigg) \\ 
\eta_X&=\frac{1}{4}\bigg(1+ \frac{1}{1-2(p_Y+p_Z)}-\frac{1}{1-2(p_Z+p_X)}-\frac{1}{1-2(p_X+p_Y)}\bigg) \\
\eta_Y&=\frac{1}{4}\bigg(1-\frac{1}{1-2(p_Y+p_Z)}+\frac{1}{1-2(p_Z+p_X)}-\frac{1}{1-2(p_X+p_Y)}\bigg) \\
\eta_Z&= \frac{1}{4}\bigg(1-\frac{1}{1-2(p_Y+p_Z)}-\frac{1}{1-2(p_Z+p_X)}+\frac{1}{1-2(p_X+p_Y)}\bigg) \\
\gamma_Q &= \sum_{g \in \{I, X, Y, Z\}} |\eta_g| = \frac{1}{2}\bigg(-1+\frac{1}{1-2(p_Y+p_Z)}+\frac{1}{1-2(p_Z+p_X)}+\frac{1}{1-2(p_X+p_Y)} \bigg)
\label{QEM_cost_Pauli_1q}
\end{aligned}
\end{equation}
\end{widetext}

Next, we show that the first-order approximation of the QEM cost for stochastic Pauli noise is Eq.\,(\ref{Eq:costandlogicalerror}).
Consider an unphysical map,
\begin{equation}
\begin{aligned}
\mathcal{N}_{\rm Pauli}' \equiv (1+ p_{\rm err})\mathcal{I}-\sum_{g \neq I^{\otimes m}} p_{g } \mathcal{P}_g.
\end{aligned}
\end{equation}
We can easily show
\begin{equation}
\begin{aligned}
\|\mathcal{N}_{\rm Pauli}' \mathcal{N}_{\rm Pauli} -\mathcal{I} \| \leq 4 p_{\rm err} ^2, 
\end{aligned}
\end{equation}
where $\| \cdot \|$ denotes the operator norm for the Pauli transfer matrix. Here, we can see that $\mathcal{N}_{\rm Pauli}$ is an approximation of the inverse map $\mathcal{N}_{\rm Pauli}^{-1}$ up to first order, and we have $\mathcal{N}_{\rm Pauli}'=\mathcal{N}_{\rm Pauli}^{-1}+O(p_{\rm err}^2)$. Accordingly, the QEM cost can be approximated as 
\begin{align}
\gamma_{Q} \approx p_{\rm err }+ \sum_g p_g = 1 + 2 p_{\rm err}
\end{align}
when $p_{\rm err} \ll 1$.

\section{Probabilistic error cancellation with gate set tomography}
\label{sec:PECwithgateset}
In this section, we explain how probabilistic error cancellation can be implemented using the result of gate set tomography. Furthermore, we also show this method is compatible with the Pauli frame.

\subsection{Gate set tomography}
Suppose that our goal is to characterize the gate set $\{\mathcal{G}_1, \mathcal{G}_2, ..., \mathcal{G}_{N_s} \},$ which involves at most $N$ qubits. To implement gate set tomography, we measure 
\begin{align}
\tilde{\mathcal{G}}_{ij}= \langle \langle O_i | \mathcal{G} |\rho_j \rangle \rangle, 
\end{align}
where $\mathcal{G}$ is one of the gates in the gate set and $\langle \langle O_i |$ and $|\rho_j \rangle \rangle$ are linearly independent $4^N$ observables and states. Note that the measurement results are generally noisy because of state preparation and measurement (SPAM) errors.  Let $O_i^{0}$ and $\rho_j^{(0)}$ be error-free observables and states. Here, we set $O_i^{(0)} \in \{I, X, Y, Z \}^{\otimes N}$ and $\rho_j^{(0)} \in \{\ket{0}, \ket{1}, \ket{+}, \ket{+_y} \}^{\otimes N}$ with $\ket{+}=\frac{1}{\sqrt{2}}(\ket{0}+\ket{1})$ and $\ket{+_y}=\frac{1}{\sqrt{2}}(\ket{0}+i \ket{1})$. Note that $\langle \langle I |$ corresponds to a trivial measurement whose outcome is always unity. By inserting the identity operator $\mathcal{I}=\sum_k |O_k^{(0)} \rangle \rangle \langle \langle O_k^{(0)} |$, we obtain
\begin{equation}
\tilde{\mathcal{G}}_{ij}= \sum_{k k'} A^{({\rm out})}_{ik} \mathcal{G}_{k k'} A^{({\rm in})}_{k' j},
\end{equation}
where $A^{({\rm out})}_{ik}=\langle \langle O_i| O^{(0)}_k \rangle \rangle$, $A^{({\rm in})}_{k' j}=\langle \langle O_{k'}^{(0)}|\rho_j \rangle \rangle$, and $\mathcal{G}_{k k'}=\langle \langle O_k^{(0)}| \mathcal{G} | O_{k'}^{(0)} \rangle \rangle$. Thus, we have
\begin{align}
\tilde{\mathcal{G}} = A^{({\rm out})} \mathcal{G} A^{({\rm in})}.
\end{align}
Note that $A^{({\rm out})}$ and $A^{({\rm in})}$ are affected by SPAM errors, which cannot be experimentally measured because they cannot be separated from each other. 

In addition, we replace $\mathcal{G}$ with the identity operation to obtain
\begin{align}
g= A^{({\rm out})} A^{({\rm in})}. 
\end{align}

In a typical scenario of gate set tomography, the estimation of the process is represented as
\begin{equation}
\begin{aligned}
\mathcal{G}^{\rm est}&=B g^{-1} \tilde{\mathcal{G}} B^{-1} \\
&= B A^{({\rm in}) -1} \mathcal{G} A^{({\rm in})} B^{-1}
\end{aligned}
\end{equation}
with $B$ being an arbitrarily chosen invertible matrix. Denoting the error-free $A^{({\rm in})}$ matrix as $A^{({\rm in}) (0)}$, a feasible choice is to set $B=A^{({\rm in}) (0)}$ when the initialization error is small. We estimate the initial state and measurement as follows:
\begin{equation}
\begin{aligned}
| \rho_j^{\rm est} \rangle \rangle &= B_{\bullet, j}= B A^{({\rm in}) -1} |\rho_j \rangle \rangle \\
\langle \langle O_i^{\rm est}|& = (g B^{-1})_{i, \bullet} = \langle \langle O_i |A^{({\rm in})} B^{-1},
\end{aligned}
\end{equation}
which can be computed from the $B$ and $g$ matrices. Now, by implementing the same procedure for $\mathcal{G}_k (k=1,2,...N_s)$ and assuming identical SPAM errors for each experiment, we can estimate $\mathcal{G}^{\rm est}_k= B A^{({\rm in}) -1} \mathcal{G} A^{({\rm in})} B^{-1}$. Although the estimated gate set, initial states, and measurements may differ from the true ones $\{|\rho_j \rangle \rangle, \langle \langle O_i |, \mathcal{G}_k \}$ due to SPAM errors, they give the correct expectation value for the gate-set sequence as follows:
\begin{equation}
\langle \langle O_i^{\rm est}| \prod_k \mathcal{G}^{\rm est}_k | \rho_j^{\rm est} \rangle \rangle = \langle \langle O_i| \prod_k \mathcal{G}_k | \rho_j \rangle \rangle.
\end{equation}
Throughout this paper, we will refer to the transformation due to SPAM errors and the $B$ matrix as the gauge and denote $G_a$. In the aforementioned case, $G_a=B A^{({\rm in}) -1}$.

Note that the choice of the gauge is crucial because it affects the set of required QEM operations; on the other hand, the QEM operations are restricted to Pauli operations because of the use of the Pauli frame. Therefore, we need to carefully choose the gauge so that only Pauli operations are used for QEM. We propose to estimate the gate set, initial states, and measurements as follows:
\begin{equation}
\begin{aligned}
\label{Eq:gst_definition}
\mathcal{G}_k^{\rm est}&= \tilde{\mathcal{G}} g^{-1}=A^{({\rm out})} \mathcal{G} A^{({\rm out}) -1} \\
|\rho_j^{\rm est} \rangle \rangle &= g_{\bullet,j }=A^{({\rm out})} |\rho_j \rangle \rangle \\
\langle \langle O_i ^{\rm est}| &= \langle \langle O_i^{(0)} |= \langle \langle O_i|A^{({\rm out}) -1}.
\end{aligned}
\end{equation}
The above formalism corresponds to the case with gauge $G_a=A^{({\rm out})}$. Later, we will see that this choice of gauge is compatible with QEM with the Pauli frame in the presence of stochastic Pauli errors. 

\subsection{Probabilistic error cancellation}
Here, we discuss how to derive quasi-probabilities in order to implement probabilistic error cancellation based on the results of gate set tomography. In addition, we show that this method is compatible with the Pauli frame. Let us assume that we have obtained the estimations of the gate set, initial states, and measurements as follows: 
\begin{equation}
\begin{aligned}
\mathcal{G}^{\rm est}_k&=G_a \mathcal{G}_k G_a^{-1} \\
|\rho_j^{\rm est} \rangle \rangle &= G_a | \rho_j \rangle \rangle \\
\langle \langle O_i^{\rm est}|&=\langle \langle O_i | G_a^{-1}.
\end{aligned}
\end{equation}
Let us also assume that gate set tomography is applied to the basis operations for QEM and estimates are obtained as $\mathcal{B}_l^{\rm est}=G_a \mathcal{B}_l G_a^{-1}$. Here, we will denote the ideal operation as $\mathcal{U}_k$. We will attempt  to invert the estimated noise process, i.e., $\mathcal{N}^{{\rm est} -1}_k= \mathcal{U}_k \mathcal{G}_k^{{\rm est} -1}= \sum_l q_l \mathcal{B}_l^{\rm est}$. In reality, what we implement via probabilistic error cancellation corresponds to $\mathcal{N}_k^{\rm{mit} -1}= \sum_l q_l \mathcal{B}_l = G_a^{-1} \mathcal{U}_k G_a \mathcal{G}^{-1}_k$. Therefore, the error-mitigated gate can be expressed as $\mathcal{G}^{\rm{mit}}_k=\mathcal{N}_k^{\rm{mit} -1} \mathcal{G}_k= G_a^{-1} \mathcal{U}_k G_a$. Similarly, we try to construct the ideal initial states and measurements, i.e., $|\rho^{(0)}_j \rangle \rangle=\sum_{l'} q_{l'} \mathcal{B}_{l'}^{\rm est} |\rho_j^{\rm est} \rangle \rangle$ and $\langle \langle O_i^{(0)}|=\sum_{l''} q_{l''} \langle \langle O_i^{\rm est}| \mathcal{B}_{l''}^{\rm est}$, which actually correspond to $|\rho^{\rm{mit}}_j \rangle \rangle= G_a |\rho^{(0)} \rangle \rangle$ and $\langle \langle O_i^{\rm{mit}}|=\langle \langle O^{(0)}_i |G_a^{-1} $, respectively. Accordingly, we obtain the error-mitigated expectation value for the sequence of quantum gates,
\begin{align}
\langle \langle O_i^{\rm{mit}}| \prod_k \mathcal{G}^{\rm{mit}}_k |\rho_j^{\rm{mit}} \rangle \rangle = \langle \langle O_i^{(0)}| \prod_k \mathcal{U}_k |\rho_j^{(0)} \rangle \rangle.
\end{align}

Now, let us discuss the compatibility of this method with the Pauli frame. When using the Pauli frame for QEM, only Pauli operations are allowed. This indicates that the solution of 
\begin{align}
\mathcal{N}^{{\rm est} -1}_k=  \sum_l q_l \mathcal{B}_l^{\rm est}
\end{align}
only contains Pauli operators. Here, we choose the gauge $G_a=A^{({\rm out})}$. Note that in the presence of stochastic Pauli measurement errors, $A^{({\rm out})}$ can also be described by a diagonal matrix corresponding to a stochastic Pauli error. We can easily check that the Pauli operations are invariant under this gauge and that $\mathcal{N}^{{\rm est} -1}_k$ is also described by stochastic Pauli noise when the gates suffer stochastic Pauli errors. Thus, only Pauli operators are required for QEM. Similarly, we can also show only Pauli operations are required to realize  $|\rho^{(0)}_j \rangle \rangle=\sum_{l'} q_{l'} \mathcal{B}_{l'}^{\rm est} |\rho_j^{\rm est} \rangle \rangle$ and $\langle \langle O_i^{(0)}|=\sum_{l''} q_{l''} \langle \langle O_i^{\rm est}| \mathcal{B}_{l''}^{\rm est}$ for the state preparation in the presence of stochastic Pauli errors. Therefore, this method is fully compatible with the Pauli frame.

\subsection{Efficiency of gate set tomography for decoding errors}
In this section, we discuss the efficiency of gate set tomography. Here, we assume that the noise of elementary logical operations is modeled as stochastic Pauli noise. Then, the noise map is of the form $\mathcal{N} = \sum_{g \in \{I, X, Y, Z\}^{\otimes N}} p_g g \rho g$, where $p_g = O(p_{\rm L})$. We will evaluate the number of required samplings to estimate $p_g$ within a precision of $(1\pm r)p_g$.

When we estimate each element of $\tilde{\mathcal{G}}$ and $g$ with the standard error $\delta$, it is enough to perform $N_{\rm GST} = O(\delta^{-2})$ samplings. Suppose we have obtained $\tilde{\mathcal{G}}$ and $g$ with a small statistical fluctuation, as $\tilde{\mathcal{G}} + \Delta \tilde{\mathcal{G}}$ and $g + \Delta g$, respectively, where each element of $\Delta \tilde{\mathcal{G}}$ and $\Delta g$ is in the order of $O(\delta)$. We can show that the gate set tomography estimation in Eq.\,(\ref{Eq:gst_definition}) can be performed with small standard errors. $|\rho_j^{\rm est} \rangle \rangle$ can be directly obtained from $g$, and $\langle \langle O_i^{\rm est}|)$ can be obtained without any statistical errors. Then, we obtain $\mathcal{G}^{\rm est}$ as $\mathcal{G}^{\rm est} = (\tilde{\mathcal{G}} + \Delta \tilde{\mathcal{G}})(g + \Delta g)^{-1} \sim \tilde{\mathcal{G}} g^{-1} + \Delta \tilde{\mathcal{G}} g^{-1} - \tilde{\mathcal{G}} g^{-1} \Delta g g^{-1}$. Thus, the elements of $\mathcal{G}^{\rm est}$ have the same order of standard error. The noise map of $\mathcal{G}^{\rm est}$ can be obtained as $\mathcal{N} = \mathcal{G}^{\rm est} (\mathcal{G}^{(0)})^{-1}$, where $\mathcal{G}^{(0)}$ is an error-free gate. Thus, the logical error probabilities can be also estimated with the standard error $O(\delta)$. Let $p_g^{\rm est}$ for $g\in\{I,X,Y,Z\}^{\otimes N}$ be the estimated logical error. Then, to achieve an estimation accuracy of the form $p_g^{\rm est} = (1+r)p_g$, its standard error must be smaller than $r p_g$. Thus, the number of samples required to achieve an accuracy factor $r$ in the above form is given by $N_{\rm GST} = O((r p_{\rm L})^{-2})$. Note that when we evaluate the noise map of magic-state preparation, we perform state tomography for the magic state twirled with Clifford gates, which also requires $O((r p_{\rm L})^{-2})$ samplings for the same accuracy.

Next, we will show that the required number of samplings can be improved to $N_{\rm GST} = O(r^{-2} p_{\rm L}^{-1})$ when the target gate of gate set tomography is a logical Clifford gate. When there is no noise, every element of $\tilde{\mathcal{G}}$ and $g$ is obtained as the results of Pauli measurements for a stabilizer state. Each element of $\tilde{\mathcal{G}}$ and $g$ is zero if the measured state is not an eigenstate of an observable, and $\pm 1$ otherwise. Since the noise is modeled as stochastic Pauli noise, these elements are zero even in the presence of noise. Thus, we do not need to perform sampling for such elements.
When the elements are $\pm 1$ without noise, the elements with noise become $\pm (1-2\mu)$, which is obtained as the mean value of a random variable $\hat{m}$ such that $\hat{m}=\pm 1$ with probability $1-\mu$ and $\hat{m}=\mp 1$ with probability $\mu$, where $\mu = O(p_{\rm L})$. Supposing that $N_{\rm GST}$ samplings are performed on this distribution, $\pm (1-2\mu)$ can be estimated with the standard error $O \left(\sqrt{p_{\rm L} N_{\rm GST}^{-1}}\right)$. Therefore, to estimate each element of $\tilde{\mathcal{G}}$ and $g$ with the standard error $\delta$, it is enough to perform $N_{\rm GST} = O(\delta^{-2} p_{\rm L})$ samplings. This means when we need to estimate logical error probabilities with the accuracy $\delta = r p_g$, we only need to perform  $N_{\rm GST} = O(r^{-2} p_{\rm L}^{-1})$ samplings.

\section{Surface code and lattice surgery}
\label{sec:surface_code_and_lattice_surgery}
\begin{figure}
    \centering
    \includegraphics[width=7cm]{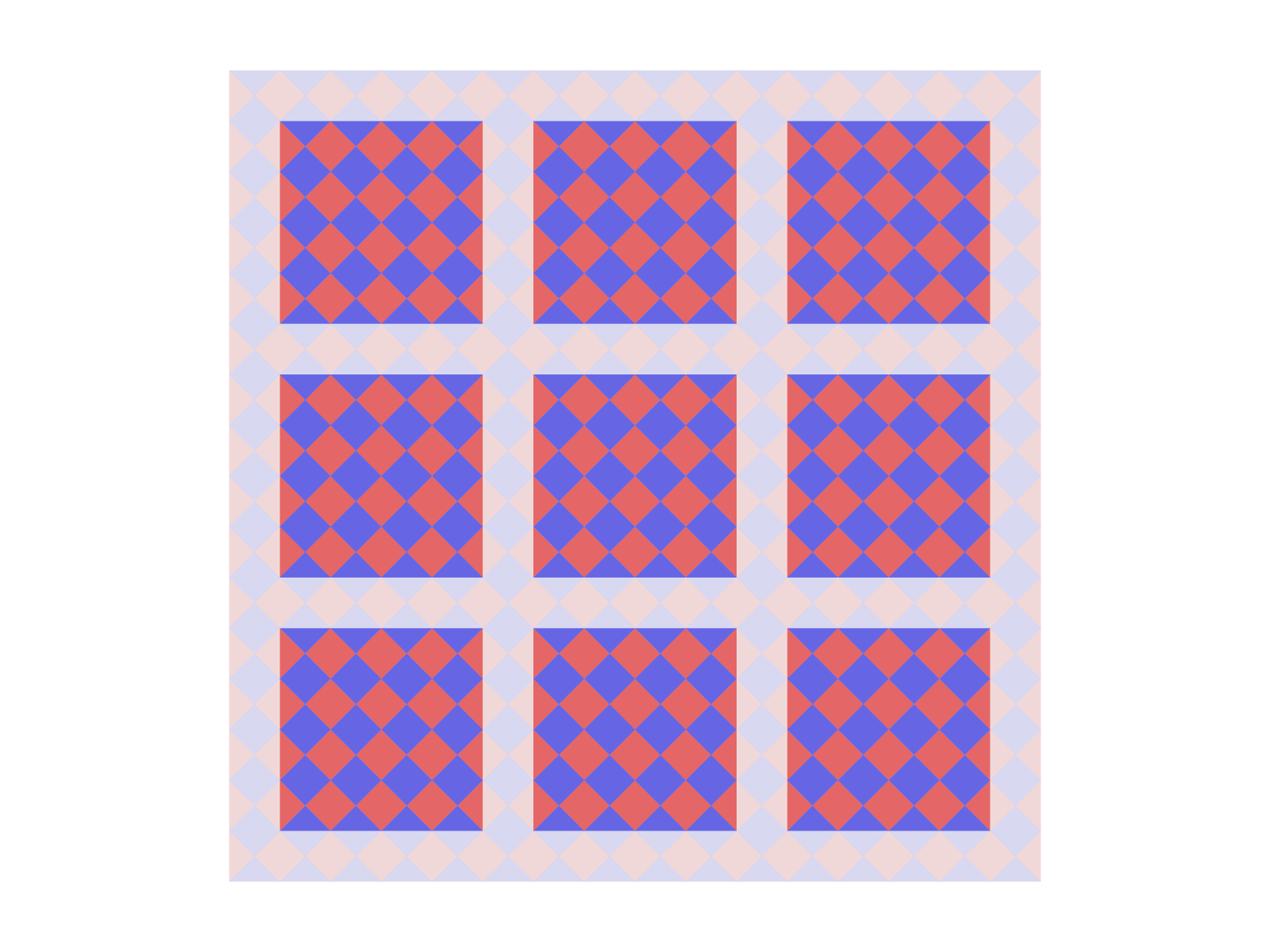}
    \caption{Schematic diagram of nine logical qubits for $d=5$ surface code. The vertices of the blue and red squares correspond to physical qubits. Each boldly colored patch consisting of blue and red squares represents a single logical qubit. In each logical qubit, the bold red (blue) squares represent a Pauli-$Z$ (Pauli-$X$) stabilizer operator acting on their vertices.}
    \label{fig:surfacecode}
\end{figure}

While the scope of our proposal is not limited to a specific architecture of FTQC, as an example, let us consider FTQC with surface codes and lattice surgery.
Surface code \cite{dennis2002topological,bravyi1998quantum} is one of the most promising quantum error-correcting codes for integrated devices such as superconducting qubits. 
This is because surface code has a large threshold value, its stabilizer measurements can be done in a short and constant depth, and it requires physical qubits that are allocated in a two-dimensional grid and interact only with the nearest neighboring ones. 
An array of logical qubits is shown in Fig.\,\ref{fig:surfacecode}. 
There are nine boldly colored patches in the figure, each of which corresponds to a logical qubit. The data qubits are allocated on the vertices of the boldly colored red and blue squares. The red (blue) squares in each patch correspond to a stabilizer operator which acts on their vertices as a Pauli-$Z$ (Pauli-$X$) operator. The width of each boldly colored patch $d$ is equivalent to the code distance. Thus, we use $O(d^2)$ physical qubits per logical qubit for surface codes with code distance $d$. Note that while we can reduce the number of physical qubits by using rotated surface codes~\cite{horsman2012surface}, we used the surface codes shown in Fig.\,\ref{fig:surfacecode} for the sake of a simple numerical simulation.

Meanwhile, lattice surgery~\cite{horsman2012surface,fowler2018low} is a method to increase the number of logical qubits and perform logical two-qubit gates with physical qubits in a planer topology. 
When we prepare logical qubits as patches, we can perform a Hadamard gate transversally. Although a logical CNOT-gate is also a transversal in surface codes, we cannot perform it fault-tolerantly in planer topology. Instead, we implement multi-qubit Pauli measurements by merging and splitting patches corresponding to logical qubits. By using multi-qubit Pauli measurements and feed-forward operations, we can indirectly perform logical two-qubit Clifford gates fault-tolerantly. For non-transversal operations, we can perform them by injecting, distilling, and performing gate teleportation with magic states, $\ket{A}$ and $\ket{Y}=SH\ket{0}$ where $S = \exp \left(i\frac{\pi}{4} Z \right)$. Using these magic states and the technique of gate teleportation, we can indirectly apply non-transversal operations such as $S$-gate and $T$-gate. While we can perform an $S$-gate using $\ket{Y}$ without consuming it, a magic state $\ket{A}$ is consumed whenever a $T$-gate is performed. Therefore, the number of required magic states for $T$-gate is a dominant factor in the execution time.

Although the above-mentioned strategy of FTQC was used in the main text, there are several other possible strategies choosing codes and logical operations that improve efficiency and feasibility.
For instance, we can also construct logical qubits as defect pairs in a single large patch and perform two-qubit Clifford gates with braiding~\cite{fowler2012surface}. We can also use concatenated Steane codes or color codes if we can perform CNOT operations more flexibly. 
An $S$-gate can be achieved with code deformation~\cite{brown2017poking} instead of magic-state injection and distillation. We can choose $CCZ$-gate as a magic state instead of a $T$-gate~\cite{gidney2019efficient}. We can estimate the recovery operations by using algorithms that achieve small latencies through the use of threshold degradation~\cite{delfosse2017almost,holmes2020nisq+}. In any case, our method is general, and we expect that it is practical.

\newcommand{\dket}[1]{\ket{#1}\rangle}
\newcommand{\dbra}[1]{\langle\bra{#1}}
\newcommand{\dbraket}[1]{\langle \braket{#1} \rangle}

\newcommand{\rhodev}[1]{\dket{\rho_{\rm dev}^{(#1)}}}
\newcommand{\Udev}[1]{\mathcal{U}_{\rm dev}^{(#1)}}
\newcommand{\PF}[1]{\mathcal{P}_{\rm PF}^{(#1)}}
\newcommand{\Prec}[1]{\mathcal{P}_{\rm rec}^{(#1)}}
\newcommand{\Uideal}[1]{\tilde{\mathcal{U}}^{(#1)}}

\newcommand{\hist}[1]{\bm{h}(#1)}
\newcommand{\ph}[1]{p_{\hist{#1}}}
\newcommand{\qh}[1]{q_{\hist{#1}}}
\newcommand{\Udevh}[1]{\mathcal{U}_{\rm dev}^{(\hist{#1})}}
\newcommand{\PFh}[1]{\mathcal{P}_{\rm PF}^{(\hist{#1})}}
\newcommand{\costh}[1]{\gamma_Q^{(\hist{#1})}}
\newcommand{\parityh}[1]{s^{(\hist{#1})}}

\section{Concrete process for logical operations}
\label{sec:concrete_logical_operations}
In this section, we explicitly show how our framework works by updating the quantum state of experimental devices and the Pauli frame, i.e., a classical memory for storing and updating Pauli operators for obtaining error-corrected results via post-processing of measurement outcomes. Note that there is latency in collecting sufficient information to correct quantum states since there are errors in syndrome measurements and we cannot instantaneously perform feedback operations for error correction; the required recovery Pauli operations need to be continuously updated in the Pauli frame classically because unitary Clifford operators are performed for computation while storing the outcomes of the syndrome measurements. Finally, we obtain the error-corrected results from the outcomes of post-processing measurements based on the state of the Pauli frame.

First, we describe a typical construction of FTQC without QEM as studied in Refs.\,\cite{fowler2012towards,fowler2018low}. We reformulate it with the superoperator representation so that we can introduce notations that will simplify the description of our framework.  
Then, we explain the FTQC framework incorporating both QEC and QEM. In this framework, the decoding errors are assumed to be not negligible but can be mitigated by probabilistic error cancellation in the logical space.

In FTQC, the process of making syndrome measurements on all the logical qubits is called a {\it code cycle}. For reliable decoding, we need to wait for several cycles depending on the code distance before processing the next logical operation. Here, we will call the unit of latency for the slowest logical operation (i.e. the maximum number of cycles that are required before being ready for the next logical operation) a {\it step} and assume that in each logical operation, every logical qubit waits until all the logical qubits become ready for the next logical operation. 
This leads to a simple definition of the states of the quantum devices and the classical memory at a certain step. We should emphasize that this unit is introduced only for the sake of illustration and that our scheme can be applied to asynchronous processing of logical operations.

\subsection{FTQC architecture without QEM}
At the $t$-th step of FTQC without QEM, we have to update two components: a quantum device of which the state is $\rhodev{t}$ and the Pauli frame $\PF{t}$. Let us denote the Pauli transfer matrix of an actual quantum process that suffers noise until the $t$-th step as $\Udev{t}$ and an ideal initial state as $\dket{\rho_0}$. Then, the state at the $t$-th step can be represented as $\rhodev{t} = \Udev{t} \dket{\rho_0}$. Note that $\Udev{t}$ is the superoperator of a trace-preserving and completely positive map, but it is not a unitary process, since the actual process requires several intermediate measurements.
Denote the ideal operations until the $t$-th step as $\Uideal{t}$.
Here, we aim to have
\begin{align}
\PF{t} \Udev{t} = \Uideal{t}
\end{align}
for an arbitrary step $t$.
Note that since several successive syndrome values are required to calculate $\PF{t}$, the Pauli frame at the $t$-th step turns out with a latency in cycles that is at least proportional to the code distance~\cite{fowler2012towards}.

We will focus on a simplified universal set of logical operations, including preparation of logical $\dket{0_{\rm L}}$ and $\dket{A_{\rm L}}$, logical Clifford operations including logical Pauli operations, logical single-qubit Pauli-$Z$ measurements, and logical gate teleportation for performing a logical $T$-gate. While these logical operations can be divided into more basic logical operations such as merge-and-split operations in lattice surgery~\cite{horsman2012surface}, we will use this set for simplicity. In the following, we illustrate how the physical states and the Pauli frame are updated in each case.

\subsubsection{Preparation of logical states}
There are two types of initialization in FTQC: preparation of a logical $\dket{0_{\rm L}}$ state and a logical magic state $\dket{A_{\rm L}}$ implemented in the surface code. Here, we describe an operation to add a clean logical qubit:
\begin{equation}
\begin{aligned}
\mathcal{Q}_0 \dket{\rho} &:= \dket{\rho} \otimes \dket{0_{\rm L}} \\
\mathcal{Q}_A \dket{\rho} &:= \dket{\rho} \otimes \dket{A_{\rm L}}
\end{aligned}
\end{equation}

Let us first explain the procedure for preparing $\ket{0_{\rm L}}$. 
We join $n$ physical qubits to a system, where $n$ is the number of qubits used to construct a logical qubit. The initial state of the joined data qubits can be any random state. We measure all the joined data qubits in the Pauli-$Z$ basis. If there is no error in this measurement, we obtain a computational basis $\ket{\bm{x}}$, where $\bm{x} \in \{0,1\}^n$ is $n$-bit outcomes. We suppose a Pauli operator $P_{\bm{x}} = \bigotimes_i X^{x_i}$; then we find $P_{\bm{x}} \ket{\bm{x}} = \ket{0}^{\otimes n}$. 
The state $\ket{0}^{\otimes n}$ is the +1 eigenstate of all the $Z$-stabilizer operators and the logical Pauli-$Z$ operator, but not an eigenstate of any $X$-stabilizer operator. We perform $X$-stabilizer measurements to project $\ket{0}^{\otimes n}$ to the code space. If the state is projected to the $+1$ eigenspace for all the $X$-stabilizer operators, the state becomes the $+1$ eigenstate of all the stabilizer operators and the logical Pauli-$Z$ operator, which is the definition of $\ket{0_{\rm L}}$. If any of them is $-1$, we can find a Pauli-$Z$ operator $P_{\bm{z}}$ which anti-commutes with all the $X$-stabilizer operators with $-1$ outcome and commutes with the other stabilizer operators. Accordingly, we have that $P_{\bm{z}} P_{\bm{x}} \ket{\psi}$, where $\ket{\psi}$ is the state after $X$-stabilizer measurements on $\ket{\bm{x}}$, is equal to $\ket{0}_{\rm L}$. 
The procedure to add $\ket{0_{\rm L}}$ is as follows: suppose that $\Udev{t+1}$ is a sequence to join $n$ qubits; measure all of them in the Pauli-$Z$ basis, and perform $X$-stabilizer measurements. The Pauli frame $\PF{t+1}$ is the tensor product of $\PF{t}$ and the superoperator of $P_{\bm{z}} P_{\bm{x}}$.
In practice, Pauli errors may occur during the above measurements. Since the error rates are expected to be below the threshold value, they can be reliably detected in the succeeding stabilizer measurements and corrected by updating $P_{\bm{z}} P_{\bm{x}}$. See Sec.\,\ref{Sec:logical_clifford} for the update of the Pauli frame with latency.

The other initialization is the preparation of the logical magic state $\ket{A_{\rm L}}$, which is, as detailed later, used for performing non-Clifford gates fault-tolerantly via gate teleportation. In order to use magic states for gate teleportation, the infidelity of the logical magic state must be comparable to or smaller than the required logical error rate. On the other hand, since $\ket{A_{\rm L}}$ is not an eigenstate of the logical Pauli-$Z$ operator or logical Pauli-$X$ operator, we cannot use the same method we used above for preparing $\ket{0_{\rm L}}$. 
Instead, a logical magic state $\ket{A_{\rm L}}$ with sufficient fidelity for gate teleportation can be constructed as follows: I) create a noisy magic state $\ket{A_{\rm L}}$ whose code distance is a small constant $d_s$, II) expand the code distance from $d_s$ to $d$ fault-tolerantly, and III) create a clean magic state from several noisy magic states with magic-state distillation. If the fidelity of noisy magic states is sufficiently large and a magic state is sufficiently distilled, we can obtain a clean magic state together with its Pauli frame. See Refs.\,\cite{li2015magic,fowler2012towards,fowler2018low} for details on these procedures.

Thus, we can perform an update $\Uideal{t+1} = \mathcal{Q} \Uideal{t}$, where $\mathcal{Q} \in \{\mathcal{Q}_0, \mathcal{Q}_A\}$, by updating $\Udev{t}$ and $\PF{t}$ as 
\begin{equation}
\begin{aligned}
\Udev{t+1} &= (\mathcal{I} \otimes \PF{Q}) \mathcal{Q} \Udev{t} \\
\PF{t+1} &= \PF{t} \otimes \PF{Q}
\end{aligned}
\end{equation}
where $\PF{Q}$ turns out with a latency because of noisy syndrome measurements.

\subsubsection{Logical Pauli Operation}
Logical Pauli operations are special cases of logical Clifford operations and can be performed much more easily than general Clifford operations because of the Pauli frame. When we perform a logical Pauli operation $\mathcal{P}^{(t)}$ at the $t$-th step, we need to construct $\PF{t+1}$ and $\Udev{t+1}$ such that $\PF{t+1} \Udev{t+1} = \Uideal{t+1} = \mathcal{P}^{(t)} \Uideal{t}$.
Since a logical Pauli operation is a product of physical Pauli operations, there are two ways to achieve this. One is to update only the Pauli frame as follows:
\begin{equation}
\begin{aligned}
\label{Eq:update}
\PF{t+1} &= \mathcal{P}^{(t)} \PF{t} \\
\Udev{t+1} &= \Udev{t}
\end{aligned}
\end{equation}
Note that this is an instantaneous and noiseless operation because it is processed only on the classical computer. We refer to this scheme as a Pauli operation by {\it software update}.

The other is to update the physical device instead of the Pauli frame.
\begin{align}
\PF{t+1} &= \PF{t} \\
\Udev{t+1} &= \mathcal{P}^{(t)} \Udev{t}
\end{align}
Compared with the software update, this operation requires operation of the physical quantum devices, and hence, it may involve an error. We call this scheme a {\it hardware update}. Note that this procedure is a transversal single-qubit Pauli operation, so we expect that it negligibly increases the physical error rate per code cycle. 
Although the hardware update seems to have no advantage in a typical architecture of FTQC, logical Pauli operations by hardware update are required in the Pauli twirling of logical errors in FTQC, as described in Appendix.\,\ref{sec:decoding_error_approx}.

\subsubsection{Logical Clifford Operation}
\label{Sec:logical_clifford}
Unlike logical Pauli operations, Clifford operations require physical operations to be performed on quantum devices. Here, we denote the Pauli transfer matrix of physical errors at the $t$-th step as $\Prec{t}$. Note that since quantum states are projected to the code space with syndrome measurements, we expect that $\Prec{t}$ can be approximated as a Pauli error. For the case that we cannot assume $\Prec{t}$ to be a Pauli error, we can twirl it with stochastic Pauli operations by hardware update and project it to the stochastic Pauli errors. See Appendix.\,\ref{sec:decoding_error_approx} for details.

Pauli errors are detected via syndrome measurements. Furthermore, the recovery Pauli operation $\Prec{t}$ can be estimated with a high success probability and the Pauli frame can be updated in the same way as Eq.\,(\ref{Eq:update}). If there are no measurement errors, the recovery operation for a certain cycle can be estimated only from the syndrome values during the same cycle. However, when the syndrome measurements suffer physical errors, we need to idle the logical qubits for $d$ cycles to collect sufficient syndrome measurement outcomes before starting the next Clifford operation to guarantee exponential decay of the logical error rate~\cite{fowler2012towards}. Therefore, the recovery operations turn out with a latency of at least $d$ cycles. In practice, the latency is larger than $d$ cycles since we also need post-processing for signal discrimination and of the decoding algorithms on classical peripherals.

Suppose that we perform a logical Clifford operation $\mathcal{C}^{(t)}$ and that physical Pauli errors $\Prec{t}$ happen during the Clifford operation, which appear after the decoding. Here, we can ignore recovery failures when the code distance is sufficiently large. We aim to obtain $\PF{t+1}$ and $\Udev{t+1}$ that satisfy $\PF{t+1} \Udev{t+1} = \Uideal{t+1} = \mathcal{C}^{(t)} \Uideal{t}$. 
A logical Clifford operation updates the state of the quantum devices as follows:
\begin{align}
\Udev{t+1} = \Prec{t} \mathcal{C}^{(t)} \Udev{t}.
\end{align}
In addition, the Pauli frame is updated as follows:
\begin{align}
\PF{t+1} = \mathcal{C}^{(t)} \PF{t} (\mathcal{C}^{(t)})^{-1} \Prec{t}.
\end{align}
Since a logical Clifford operation is also a Clifford operation on physical qubits in the case of stabilizer codes, $\PF{t+1}$ is a Pauli operator. 

In practice, $\Prec{t}$ is expected to have a large latency since it requires post-processing to estimate the recovery operation. Thus, another Pauli error might happen during the idling operations performed while waiting for the estimation of $\Prec{t}$. Nevertheless, the next logical Clifford operation or logical Pauli measurement can be performed before $\Prec{t}$ is estimated. As an example, consider the case in which $\mathcal{C}^{(t+1)}$ is processed before the last Pauli frame is updated. The Pauli frame of the next step is represented as 
\begin{widetext}
\begin{align}
\PF{t+2} &= \Prec{t+1} \mathcal{C}^{(t+1)} \PF{t+1} (\mathcal{C}^{(t+1)})^{-1} \nonumber \\
&= (\mathcal{C}^{(t+1)} \mathcal{C}^{(t)} \PF{t} (\mathcal{C}^{(t)})^{-1} (\mathcal{C}^{(t+1)})^{-1} ) (\mathcal{C}^{(t+1)} \Prec{t} (\mathcal{C}^{(t+1)})^{-1}) \Prec{t+1} .
\end{align}
\end{widetext}
Since Pauli actions commute, we can update the part $\mathcal{C}^{(t+1)} \Prec{t} (\mathcal{C}^{(t+1)})^{-1}$ even in the next step. We can use a similar technique when the next operation is a logical Pauli measurement, which is explained in the next section.
Therefore, as long as all the successive operations are Clifford operations or Pauli measurements, we can postpone the application of the recovery operators and updates of the Pauli frame. This technique is essential not only for improving the throughput of logical operations in FTQC but also for avoiding exponential growth in latency~\cite{holmes2020nisq+}. 

Note that in the lattice surgery scheme, logical CNOTs are performed by preparing a logical $\dket{0_{\rm L}}$ state, making logical Pauli measurements, and adaptive Pauli operations~\cite{fowler2018low}. While none of them is a Clifford operation, we can consider them on the whole to constitute a unitary process. Adaptive Pauli operations can be postponed for the same reason as the updates of the Pauli frame.

\subsubsection{Single-qubit logical Pauli measurement}
We can perform a destructive single-qubit logical Pauli-$Z$ measurement by making physical Pauli-$Z$ measurements on all the data qubits of the target logical qubit. Note that logical Pauli measurements with another Pauli basis can be performed by swapping $Z$ and $X$ or combining the logical Pauli-$Z$ measurement with single-qubit Logical Clifford operations. 
The outcome of a logical measurement is calculated as the parity of the measurement outcomes of the data qubits. In other words, let $n$ be the number of data qubits of the target logical qubit, and $\Pi_{\bm{x}} = \prod_i (I+ (-1)^{x_i} Z_i)/2$ for $\bm{x} \in \{0,1\}^n$ be an operator that projects all the data qubits in a physical computational basis $\ket{\bm{x}}$. 
A single-bit outcome of a single-qubit logical measurement is calculated as a parity of the outcomes of physical measurements $\bm{x}$. We denote this function as $f: \{0,1\}^n \rightarrow \{0,1\}$. Therefore, we aim to estimate $\bm{x}$ and obtain $f(\bm{x})$ fault tolerantly.

When we perform physical Pauli-$Z$ measurements, physical Pauli errors $\Prec{t}$ occur on the data qubits. Also, as explained in the section on logical Clifford operations, the Pauli frame $\PF{t}$ does not have the same timing as the logical measurement. Even in this case, the recovery operation can be applied after the physical Pauli-$Z$ measurements have been performed on the data qubits, since 
\begin{align}
\langle \bra{\Pi_{\bm{x}}} \Prec{t} \Udev{t}
&= \langle \bra{\Pi_{\bm{x}\oplus \bm{y}}} \PF{t} \Udev{t} \nonumber \\
&= \langle \bra{\Pi_{\bm{x}\oplus \bm{y}}} \mathcal{\tilde{U}}^{(t)} 
\end{align}
where $\oplus$ represents an element-wise ${\rm XOR}$, and $\bm{y} \in\{0,1\}^n$ is a binary vector such that $y_i = 1$ if $\PF{t} \Prec{t}$ acts on the $i$-th data qubits as a Pauli-$X$ or $Y$ operator and $y_i=0$ otherwise. We denote this function as $\bm{y} = {\rm mask}(\mathcal{P}_{\rm PF}^{(t)}\mathcal{P}_{\rm rec}^{(t)})$.

Therefore, we obtain $\bm{x}$ as an outcome of the measurement on the uncorrected state $\dket{\rho_{\rm dev}^{(t)}}$. Notice that $\bm{y}$ appears when the decoding process catches up with the code cycle of the logical measurements, and we can retrieve the error-corrected logical outcome as $f(\bm{x} \oplus {\rm mask}(\PF{t} \Prec{t}))$. The update rules can be written as
\begin{align}
\mathcal{U}_{\rm dev}^{(t+1)} &= \frac{1}{p_{\bm{x}}} \langle \bra{\Pi_{\bm{x}}} \mathcal{P}_{\rm rec}^{(t)} \mathcal{U}_{\rm dev}^{(t)} \\
\mathcal{P}_{\rm PF}^{(t+1)} &= {\rm Dis}_M[\mathcal{P}_{\rm PF}^{(t)}],
\end{align}
where $p_{\bm{x}}$ is the probability with which we obtain $\bm{x}$ and ${\rm Dis}_M[\cdot]$ is an operation to discard the Pauli frame corresponding to the measured logical qubits. In contrast to the logical initialization, the update for the logical measurement will reduce the space of the target logical qubit.

Several things should be noted in regard to the decoding algorithm at the measurement timing. The Pauli frame can be represented in the form $(\bigotimes_i X^{x_i} Z^{z_i})$ for $x_i, z_i \in \{0,1\}$. We call $\{x_i\}$ and $\{z_i\}$ as the $X$-part and $Z$-part of the Pauli frame, respectively. Since all the data qubits are directly measured in this logical operation, syndrome measurements are not performed in this code cycle. Instead, we calculate the values of the $Z$-stabilizer syndrome measurements without measurement errors as parities of $\bm{x}$ because we directly measure the data qubits without resorting to ancilla qubits. Since there is no effective measurement error in the $Z$-stabilizer syndrome measurements at the cycle of the logical measurement, an estimation of the $X$-part of $\PF{t} \Prec{t}$ can be converted into an instance of a graph problem called minimum-weight perfect matching~\cite{fowler2012towards}. On the other hand, the information required to construct a perfect matching problem for estimating the $Z$-part of the Pauli frame is lost by making direct $Z$-basis measurements. Nevertheless, this does not affect the computation since only the $X$-part of the Pauli frame is relevant to determining $\bm{y}={\rm mask}(\PF{t}\Prec{t})$.

\subsubsection{Gate teleportation with magic state}
For a universal FTQC, we need to perform non-Clifford operations on logical qubits encoded in surface code. To this end, in a typical scenario, we create a logical qubit prepared in $\dket{A_{\rm L}}$ and perform $T$-gate on the target system by consuming the magic state $\dket{A_{\rm L}}$.

First, we describe gate teleportation without noise. Let $\dket{\rho} \otimes \dket{A}$ be the tensor product of the target system and a magic state. It is known that the following equation holds for an arbitrary $\rho$~\cite{fowler2012towards,trout2015magic}:
\begin{align}
(\mathcal{S}^{f(\bm{x})} \otimes \dbra{\Pi_{\bm{x}}}) \Lambda \mathcal{Q}_A \dket{\rho} &=  (\mathcal{S}^{f(\bm{x})} \otimes \dbra{\Pi_{\bm{x}}}) \Lambda (\dket{\rho} \otimes \dket{A_{\rm L}}) \nonumber \\
&= \mathcal{T} \ket{\rho}\rangle \dbraket{ \Pi_{\bm{x}} | +_{\rm L}},
\end{align}
where $\Lambda$ is the Pauli transfer matrix of a logical CNOT-gate acting on $\dket{\rho}$ as a control and $\dket{A_{\rm L}}$ as a target, $\mathcal{T}$ is the Pauli transfer matrix of the $T$-gate, and $\mathcal{S}^{f(\bm{x})}$ is the Pauli transfer matrix of the adaptive $S$-gate that is applied to the state if $f(\bm{x})=1$ (see the last section for the definition of the function $f$).

Using the above fact, we can construct the following update rules for the target system:
\begin{widetext}
\begin{align}
\label{Eq:teleport_without_qem}
\Udev{t+1} &= \frac{1}{p_{\bm{x}}} (\mathcal{S}^{f(\bm{x}\oplus \bm{y})} \otimes \dbra{\Pi_{\bm{x}}}) \Prec{t} \Lambda (\mathcal{I} \otimes \PF{Q}) \mathcal{Q}_A \Udev{t} \\
\PF{t+1} &= \mathcal{S}^{f(\bm{x}\oplus \bm{y})} {\rm Dis}_{Q}[\Lambda (\PF{t} \otimes \PF{Q}) \Lambda^{-1} \Prec{t}] \mathcal{S}^{-f(\bm{x}\oplus \bm{y})},
\end{align} 
\end{widetext}
where $\Prec{t}$ represents physical detectable errors, 
\begin{align}
p_{\bm{x}} &= \dbraket{\Pi_{\bm{x} \oplus \bm{y}} | +_{\rm L} }, \\
\bm{y} &=  {\rm mask}({\rm Dis}_{\rho}[\Lambda (\PF{t} \otimes \PF{Q}) \Lambda^{-1} \PF{t}]), 
\end{align}
and ${\rm Dis}_{Q}[\cdot]$ (${\rm Dis}_{\rho}[\cdot]$) is an operation to discard the Pauli frame of the magic state (target state). Then, we can verify that the above update rule obeys Eq.\,(\ref{Eq:update}), as follows.
\begin{widetext}
\begin{align}
\PF{t+1} \Udev{t+1} 
&= \frac{1}{p_{\bm{x}}} \mathcal{S}^{f(\bm{x}\oplus \bm{y})} {\rm Dis}_{Q}[ \Lambda (\PF{t} \otimes \PF{Q}) \Lambda^{-1} \Prec{t}] \mathcal{S}^{-f(\bm{x}\oplus \bm{y})} (\mathcal{S}^{f(\bm{x}\oplus \bm{y})} \otimes \dbra{\Pi_{\bm{x}}}) \Prec{t} \Lambda (\mathcal{I} \otimes \PF{Q}) \mathcal{Q}_A \Udev{t} \nonumber \\
&= \frac{1}{p_{\bm{x}}} (\mathcal{S}^{f(\bm{x}\oplus \bm{y})} \otimes \dbra{\Pi_{\bm{x}}}) {\rm Dis}_{Q}[ \Lambda (\PF{t} \otimes \PF{Q}) \Lambda^{-1} \Prec{t} ] \Prec{t} \Lambda (\mathcal{I} \otimes \PF{Q}) \mathcal{Q}_A \Udev{t} \nonumber \\
&= \frac{1}{p_{\bm{x}}} (\mathcal{S}^{f(\bm{x}\oplus \bm{y})} \otimes \dbra{\Pi_{\bm{x} \oplus \bm{y}}}) (\Lambda (\PF{t} \otimes \PF{Q}) \Lambda^{-1} \Prec{t}) \Prec{t} \Lambda (\mathcal{I} \otimes \PF{Q}) \mathcal{Q}_A \Udev{t} \nonumber \\
&= \frac{1}{p_{\bm{x}}} (\mathcal{S}^{f(\bm{x}\oplus \bm{y})} \otimes \dbra{\Pi_{\bm{x} \oplus \bm{y}}}) \Lambda \mathcal{Q}_A \PF{t} \Udev{t} \nonumber \\
&= \mathcal{T} \Uideal{t} = \Uideal{t+1}.
\end{align}
\end{widetext}
Note that we have used the following property of the discard operation for an arbitrary Pauli operation $\mathcal{P}$:
\begin{align}
\dbra{\Pi_{\bm{x}}} &= \dbra{\Pi_{\bm{x} \oplus {\rm mask}(\mathcal{P})}} \mathcal{P} \\
\mathcal{P} &= {\rm Dis}_{Q}[\mathcal{P}] {\rm Dis}_{\rho}[\mathcal{P}]
\end{align}
Note as well that $\bm{y}$ has latency due to the  decoding process and the $(\mathcal{S}^{f(\bm{x})}$ and $\dbra{\Pi_{\bm{x}}})$ operations cannot be performed simultaneously. Thus, there is an additional delay in determining $f(\bm{x} \oplus \bm{y})$, which is required for determining whether we perform $\mathcal{S}$ or not.

\subsection{FTQC architecture with QEM}
The failure probability of decoding and residual errors on the prepared magic states are not negligible when the code distance of quantum error correction is not enough and when we cannot perform sufficient magic-state distillation. Here, let us consider the case that a decoding error $\mathcal{N}$ happens after each elementary logical operation. We will assume that $\mathcal{N}$ is a stochastic logical Pauli channel that is characterized in advance.
While these assumptions do not strictly hold in practice, we nonetheless expect that they hold with negligible errors in typical quantum devices. See Appendix.\,\ref{sec:decoding_error_approx} for a justification of this assumption. Note that even if there is a finite discrepancy because of this approximation, our scheme can decrease bias in the expectation values as long as the discrepancy is small.
In this section, we show that probabilistic error cancellation can be integrated into the FTQC architecture. More concretely, we show that each logical operation in the previous section can be modified so that probabilistic error cancellation can remove the residual noise of QEC only with additional logical Pauli operations by software update.

Unlike in the NISQ scenario, FTQC operations are probabilistic, since intermediate measurements are involved in the gate teleportation of the magic states, and subsequent operations are adaptatively chosen corresponding to the measurement outcomes. We need to determine QEM operations accordingly because decoding of the noise processes may change depending on the measurement outcomes. Here, let us denote the set of outcomes of intermediate measurements and QEM operations up to the $t$-th step as $\hist{t}$, with the corresponding Pauli frame and the state of the hardware denoted as $\PFh{t}$ and $\Udevh{t}$. Note that we do not independently define the measurement outcomes and QEM operations since they affect each other. 
The probability that $\hist{t}$ can be expressed as $\ph{t} \qh{t}$, where $\ph{t}$ is the probability with which a certain measurement outcome of intermediate measurements is observed and $\qh{t}$ is the probability with which a certain QEM operation is performed. 
Note that, these probabilities are functions of $\hist{t}$. Furthermore, we denote the parity of the QEM operation (the product of the parities of the generated QEM operations) and the QEM cost at the $t$-th step as $\parityh{t}$ and $\costh{t}$. 

In our framework, we can construct a consistent FTQC architecture incorporating QEM by satisfying the following equation:

\begin{align}
\label{Eq:ftqc_qem_keep}
\sum_{\hist{t}} \qh{t} \ph{t} \costh{t} \parityh{t} \PFh{t} \Udevh{t} = \Uideal{t}.
\end{align}

Here, we explain why satisfying the above equation is sufficient. If we can construct $\Udevh{t}$ as a product of physical processes, we can sample the density matrix $\dket{\rho^{(\hist{t})}} = \PFh{t} \Udevh{t} \dket{\rho_0}$ with probability $\qh{t} \ph{t}$.
Suppose the spectral decomposition of a given observable $O$ is $O = \sum_i O_i \Pi_i$ ($O_i \in \mathbb{R}$). Sampling from the final state with POVM elements $\{\Pi_i\}$ with weight $\costh{t} \parityh{t} O_i$ gives 
\begin{widetext}
\begin{equation}
\begin{aligned}
\sum_{\hist{t}, i} \qh{t} \ph{t} \costh{t} \parityh{t} O_i \dbraket{\Pi_i | \rho^{(\hist{t})}}
&= \dbra{O}  \sum_{\hist{t}} \qh{t} \ph{t} \costh{t} \parityh{t} \PFh{t} \Udevh{t} \dket{\rho_0} \\ 
&= \dbraket{O | \Uideal{t} | \rho_0}, 
\label{Eq: unbiased}
\end{aligned}
\end{equation}
\end{widetext}
which shows that QEM can recover an unbiased expectation value of the observables when Eq.\,(\ref{Eq:ftqc_qem_keep}) holds.  

\subsubsection{Logical Pauli operation}
Since logical Pauli operations by software update are instantaneous and noiseless, we do not need to perform error mitigation on them.

\subsubsection{Logical Clifford operation}
Suppose that we want to apply a Clifford operation $\mathcal{C}$ at the $(t+1)$-th step; here, $\mathcal{C}$ is followed by not only detected physical Pauli errors $\Prec{t+1}$ but also logical Pauli noise $\mathcal{N}$ reflecting a decoding failure with a non-negligible probability; i.e., the quantum device is updated as 
\begin{equation}
\begin{aligned}
\Udevh{t+1} &= \Prec{t} \mathcal{N} \mathcal{C} \Udevh{t}.
\end{aligned}
\end{equation}
While $\Prec{t}$ is revealed with latency and canceled by updating the Pauli frame, $\mathcal{N}$ is not canceled and affects the expectation values without QEM. Here, We show that we can cancel $\mathcal{N}$ by applying QEM with a probabilistic update of the Pauli frame. 
Our goal is to find a set of update rules $(\costh{t}, \parityh{t}, \PFh{t}, \Udevh{t}) \mapsto (\costh{t+1}, \parityh{t+1}, \PFh{t+1}, \Udevh{t+1})$ which satisfies Eq.\,(\ref{Eq:ftqc_qem_keep}) for 
\begin{equation}
\begin{aligned}
\Uideal{t+1} = \mathcal{C} \Uideal{t}.
\end{aligned}
\end{equation}

Since we know the stochastic logical Pauli noise $\mathcal{N}$ in advance, we can calculate the inverse of the noise map $\mathcal{N}^{-1} = \sum_i \eta_i \mathcal{P}_i$. Next, we decompose the non-zero coefficients $\eta_i$ into $\eta_i = \gamma_Q \mathrm{sgn}(\eta_i) q_i$, where $\gamma_Q = \sum_i |\eta_i|$ and $q_i = |\eta_i|/\gamma_Q$. 
We randomly choose $i$ with probability $q_i$, and the index $i$ is appended to give $\hist{t+1}$. We find that $\qh{t+1} = q_i \qh{t}$ and $\ph{t+1} = \ph{t}$.
Then, we update the state of the Pauli frame and the classical coefficients with the following update rules:
\begin{align}
\PFh{t+1} &= \mathcal{C} \PFh{t} \mathcal{C}^{-1} \mathcal{P}_i \Prec{t}  \\
\parityh{t+1} &= \mathrm{sgn}(\eta_i) \parityh{t} \\
\costh{t+1} &= \gamma_Q \costh{t}
\end{align}
Now, we can verify that Eq.\,(\ref{Eq:ftqc_qem_keep}) is satisfied in the $(t+1)$-th step if it is satisfied in the $t$-th step.
\begin{widetext}
\begin{align}
\sum_{\hist{t+1}} \qh{t+1} \ph{t+1} \costh{t+1} \parityh{t+1} \PFh{t+1} \Udevh{t+1} 
&= \sum_{\hist{t}} \qh{t} \ph{t} \costh{t} \parityh{t} \mathcal{C} \PFh{t} \mathcal{C}^{-1} \big(\sum_i \eta_i \mathcal{P}_i \big)  \mathcal{N} \mathcal{C} \Udevh{t} \nonumber \\
&= \mathcal{C} \sum_{\hist{t}} q_{\hist{t}} p_{\hist{t}} \costh{h} \parityh{t} \PFh{t} \Udevh{t} \nonumber \\
&= \mathcal{C} \mathcal{\tilde{U}}^{(t)} = \mathcal{\tilde{U}}^{(t+1)}.
\end{align}
\end{widetext}

As we will see later, we can construct a similar update rule for probabilistic error cancellation for a single qubit logical Pauli measurement; therefore, we can inductively show that a decomposition specified with the update rule satisfies Eq. (\ref{Eq:ftqc_qem_keep}).

\subsubsection{Logical initialization and single-qubit logical Pauli measurement}
When code distances and magic-state distillation processes are insufficient, errors in the logical state preparation are not negligible. The noise maps for logical $\ket{0}$ state preparation can be assumed to be stochastic logical Pauli noise. While the noise map on magic-state preparation due to insufficient distillation may not be approximated as a stochastic Pauli noise, it can be twirled by error-mitigated logical Clifford operations. Thus, these errors are assumed to be inserted stochastic logical Pauli noise maps just after initialization. We can consider that these preparations are ideal, and instead, there exists a virtual noisy idling operation just after initialization. We can mitigate these errors with the same update rules as in the case of logical Clifford operations by treating these errors as decoding errors. 
Similarly, the errors of single-qubit logical Pauli measurement can be considered to be a probabilistic logical bit-flip just before the logical measurement; as such, they can be processed in the same manner.

\subsubsection{Gate teleportation with magic state}
To perform gate teleportation on the $T$-gate, we can try the following process:
\begin{align}
(\mathcal{I} \otimes \dbra{\Pi_{\bm{x}}}) \Lambda \mathcal{Q}_A
\end{align}
and then perform $\mathcal{S}^{f(\bm{x})}$ depending on the measurement outcomes to indirectly perform $\mathcal{T}$ on the target system. However, in practice, what is performed until the measurement is made is 
\begin{align}
(\mathcal{I} \otimes \dbra{\Pi_{\bm{x}}}) \mathcal{N} \Prec{t} \Lambda (\mathcal{I} \otimes \PF{Q}) \mathcal{Q}_A
\end{align}
where $\Prec{t}$ means the physical errors caused by the logical magic-state preparation, logical CNOT, and logical measurements. Compared with the case without QEM Eq.\,(\ref{Eq:teleport_without_qem}), there is additional logical Pauli noise $\mathcal{N}$ that is caused by the failure of the decoding to estimate $\Prec{t}$. 

Since this procedure involves an intermediate measurement, we obtain the outcome $\bm{x}$ randomly. The procedure for the inverse decomposition is the same as that of a logical Clifford operation: we calculate the inverse of the noise map $\mathcal{N}^{-1} = \sum_i \eta_i \mathcal{P}_i$ and decompose each non-zero coefficient $\eta_i$ into $\eta_i = \gamma_Q \mathrm{sgn}(\eta_i) q_i$, where $\gamma_Q = \sum_i |\eta_i|$ and $q_i = |\eta_i|/\gamma_Q$. 
Note that the measurement result $\bm{x}$ is added to $\hist{t}$ together with the choice of QEM operation $i$ to obtain $\hist{t+1}$. Next, $\qh{t}$ is updated to $\qh{t+1} = q_i \qh{t}$ and $\ph{t}$ is updated to $\ph{t+1} = p_{\bm{x}} \ph{t}$, where $p_{\bm{x}}$ is the probability that the measurement outcome is ${\bm{x}}$.
Accordingly, we can update the relevant elements with the following rules:
\begin{widetext}
\begin{align}
\Udevh{t+1} &= \frac{1}{p_{\bm{x}}} (\mathcal{S}^{f(\bm{x} \oplus \bm{y})} \otimes \dbra{\Pi_{\bm{x}}}) \mathcal{N} \Prec{t} \Lambda (\mathcal{I} \otimes \PF{Q}) \mathcal{Q}_A \Udevh{t} \\
\PFh{t+1} &= \mathcal{S}^{f(\bm{x} \oplus \bm{y})} {\rm Dis}_{A}[\mathcal{R}^{(\hist{t})}] \mathcal{S}^{-f(\bm{x} \oplus \bm{y})} \\
\parityh{t+1} &= \mathrm{sgn}(\eta_i) \parityh{t} \\
\costh{t+1} &= \gamma_Q \costh{t},
\end{align}
\end{widetext}
where
\begin{align}
\mathcal{R}^{(\hist{t})} &= \Lambda (\PFh{t} \otimes \PF{Q}) \Lambda^{-1} \mathcal{P}_i \Prec{t}, \\
\bm{y} &= {\rm mask}({\rm Dis}_{\rho}[\mathcal{R}^{(\hist{t})}]).
\end{align}
Note that an additional decoding error that depends on whether we perform $\mathcal{S}$ or not would occur after the delayed application of $\mathcal{S}^{f(\bm{x} \oplus \bm{y})}$. While this map is omitted for simplicity, we can perform probabilistic error cancellation on it in the same way. 

With the above update rules, we can verify that Eq.\,(\ref{Eq:ftqc_qem_keep}) is obeyed at the $(t+1)$-th step. The product of $\Udevh{t+1}$ and $\PFh{t+1}$ is evaluated as follows:
\begin{widetext}
\begin{equation}
\begin{aligned}
\PFh{t+1} \Udevh{t+1} 
&= \frac{1}{p_{\bm{x}}} \mathcal{S}^{f(\bm{x} \oplus \bm{y})} {\rm Dis}_{A}[\mathcal{R}^{(\hist{t})}] \mathcal{S}^{-f(\bm{x} \oplus \bm{y})} (\mathcal{S}^{f(\bm{x} \oplus \bm{y})} \otimes \dbra{\Pi_{\bm{x}}}) \mathcal{N} \Prec{t} \Lambda (\mathcal{I} \otimes \PF{Q}) \mathcal{Q}_A \Udevh{t} \\
&= \frac{1}{p_{\bm{x}}} (\mathcal{S}^{f(\bm{x} \oplus \bm{y})} \otimes \dbra{\Pi_{\bm{x}}}) {\rm Dis}_{A}[\mathcal{R}^{(\hist{t})}] \mathcal{N} \Prec{t} \Lambda (\mathcal{I} \otimes \PF{Q}) \mathcal{Q}_A \Udevh{t} \\
&= \frac{1}{p_{\bm{x}}} (\mathcal{S}^{f(\bm{x} \oplus \bm{y})} \otimes \dbra{\Pi_{\bm{x}\oplus \bm{y}}}) \mathcal{R}^{(\hist{t})} \mathcal{N} \Prec{t} \Lambda (\mathcal{I} \otimes \PF{Q}) \mathcal{Q}_A \Udevh{t} \\
&= \frac{1}{p_{\bm{x}}} (\mathcal{S}^{f(\bm{x} \oplus \bm{y})} \otimes \dbra{\Pi_{\bm{x}\oplus \bm{y}}}) \mathcal{P}_i \mathcal{N} \Lambda \mathcal{Q}_A \PFh{t} \Udevh{t}
\end{aligned}
\end{equation}
Therefore, the expectation value is evaluated as follows.
\begin{equation}
\begin{aligned}
& \sum_{\hist{t+1}} \qh{t+1} \ph{t+1} \costh{t+1} \parityh{t+1} \PFh{t+1} \Udevh{t+1} \\
&= \sum_{\hist{t}} \sum_{\bm{x}} \qh{t} \ph{t} \costh{t} \parityh{t} (\mathcal{S}^{f(\bm{x} \oplus \bm{y})} \otimes \dbra{\Pi_{\bm{x}\oplus \bm{y}}}) 
 (\sum_i \eta_i \mathcal{P}_i) \mathcal{N} \Lambda \mathcal{Q}_A \PFh{t} \Udevh{t} \\
&= \sum_{\hist{t}} \sum_{\bm{x}} \qh{t} \ph{t} \costh{t} \parityh{t} (\mathcal{S}^{f(\bm{x} \oplus \bm{y})} \otimes \dbra{\Pi_{\bm{x}\oplus \bm{y}}}) \Lambda \mathcal{Q}_A \PFh{t} \Udevh{t} \\
&= \mathcal{T} (\sum_{\bm{x}} \dbraket{\Pi_{\bm{x} \oplus \bm{y}} | +_{\rm L}}) \sum_{\hist{t}} p_{\bm{x}} \qh{t} \ph{t} \costh{t} \parityh{t} \PFh{t} \Udevh{t} \\
&= \mathcal{T} \Uideal{t} = \Uideal{t+1} 
\end{aligned}
\end{equation}
\end{widetext}
Therefore, we have verified that all the logical operations obey Eq.\,(\ref{Eq:ftqc_qem_keep}).

\section{On the noise model of the decoding errors}
\label{sec:decoding_error_approx}
In the main text, we assumed that the decoding errors for elementary logical operations can be modeled as Markovian and stochastic Pauli noise obeying Eq.\,(\ref{Eq:logicalerror}). Here, we justify this assumption. 
First, we will determine whether an actual noise model of syndrome-measurement cycles of surface codes can be treated as Markovian or not. When the syndrome measurements may output an incorrect value, we need $d$ consecutive syndrome values for reliably estimating the recovery operations. Since the quantum states are in the logical code space only after the recovery operations, the actual quantum states are not in the code space during FTQC. This makes it hard to evaluate the logical noise map for several cycles during FTQC, because the map does not take a logical state to another logical state, while we need to evaluate the logical noise map in advance in order to perform QEM on the code space.
To avoid this problem, we assume that the following noise model can well approximate the actual noise model: suppose that we can perform perfect syndrome measurements in the $l d$-th cycle, where $l=1,2,...$, and that we can perform recovery operations just after that. Then, the quantum state is in the logical code space at the $l d$-th cycle. In this case, we can define a logical error map $\mathcal{M}_{\rm dec}$ from the $(l-1)d$-th cycle to the $l d$-th cycle. Here, we assume that if we have a logical operation $\mathcal{U}$ requiring $\chi d$ cycles, the logical map including effective decoding errors can be approximated with $\mathcal{M}_{\rm dec}^{\chi} \mathcal{U}$ when the code distance $d$ is sufficiently large. 
If this assumption holds, we can cancel the noise map $\mathcal{M}_{\rm dec}^{\chi}$ by performing QEM on each logical operation. Although the actual $\chi$ depends on the logical operations, we will assume that $\chi=1$ for simplicity. 

\begin{figure*}
    \centering
    \includegraphics[width=14.5cm]{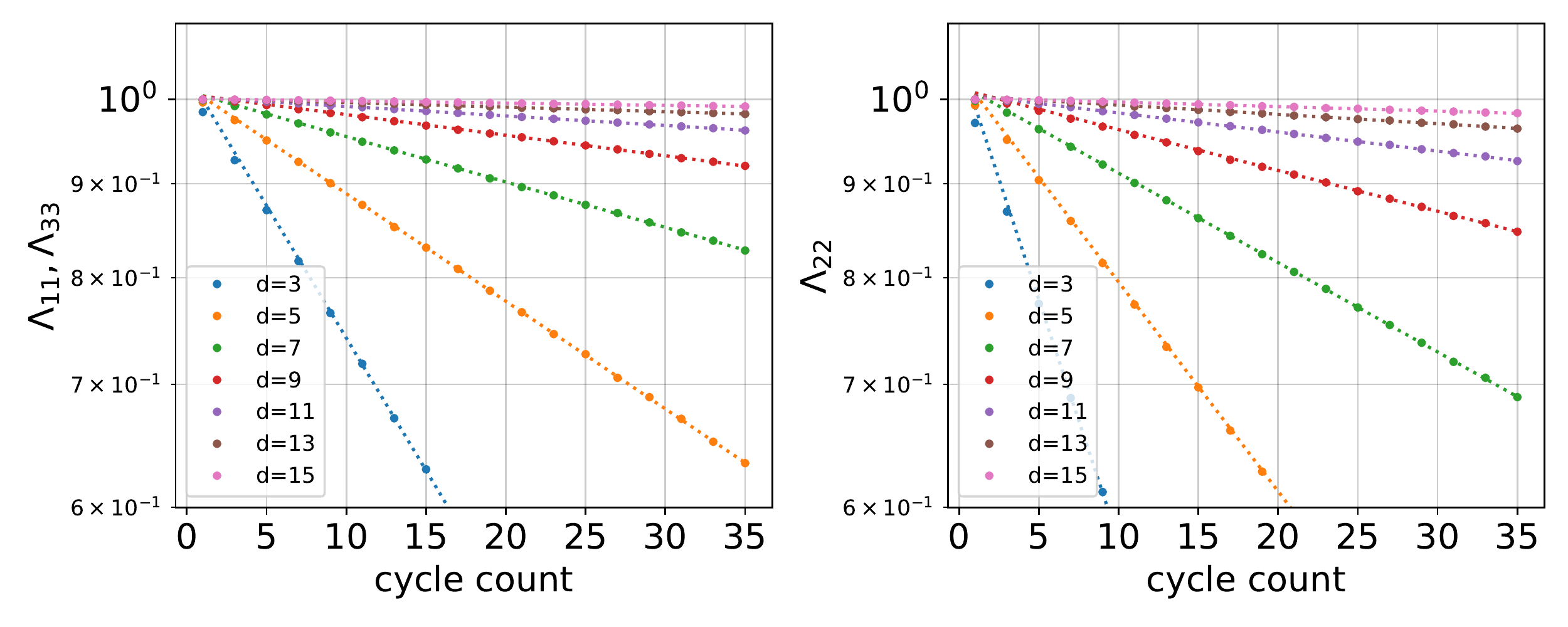}
    \caption{Diagonal elements of Pauli transfer matrix for a noise map of a single logical qubit during $c$ syndrome-measurement cycles plotted versus the number of code cycles. Each color corresponds to the code distance of a logical qubit. The circles are numerical results, and the dashed lines are fitting results with an exponentially decaying function.}
    \label{fig:decode_cycle}
\end{figure*}

The following numerical analysis shows that this assumption holds at least when the physical errors are stochastic Pauli noise. Let $\mathcal{M}_{\rm dec, c}$ be a noise map for $c$-cycle idling of a single logical qubit with code distance $d$, and let $\Lambda^{(c)}$ be the Pauli transfer matrix of $\mathcal{M}_{\rm dec, c}$. Here, $\mathcal{M}_{\rm dec, c}$ is a stochastic noise map since a stochastic Pauli error can only cause logical Pauli errors; thus, $\Lambda^{(c)}$ is a diagonal matrix. Our assumption can be rephrased as follows: There is an effective Pauli transfer matrix $\Lambda_{\rm eff}$ such that $\Lambda^{(c)} = \Lambda_{\rm eff}^c$ for sufficiently large $c$. Equivalently, we assume that each diagonal element decays exponentially to the number of cycles $c$. Since $\Lambda_{00}$ is always unity for stochastic Pauli errors, we are interested in the other diagonal elements.
Fig.\,\ref{fig:decode_cycle} plots the diagonal elements, except $\Lambda_{00}$, according to the number of cycles.
We utilize the same settings as in Sec.\,\ref{sec:logical_error_performance_analysis}, i.e., a depolarizing noise map with $p=0.01$. Note that $\Lambda_{33}$ and $\Lambda_{11}$ are equal, since the behavior of surface codes is symmetric for Pauli-$X$ and $Z$ errors. In the figure, the circles are the numerical results, and the dashed lines are fitting results with an exponentially decaying function. We used data points with $c>20$ for the fitting. The results agree well with our assumption, except for the region where the cycle count $c$ is around $1$. Thus, we can conclude that if a physical error is stochastic Pauli noise, we can define an effective logical noise map for each logical operation.

Next, we discuss the case in which the physical noise cannot be modeled as stochastic Pauli noise. In practice, the noise of calibrated quantum operations is expected to be almost stochastic Pauli, since the unitary component of the physical errors can be canceled by echoing. In addition, it is expected that the noise map per code cycle in the code space of surface codes is well approximated as stochastic Pauli noise at a large distance~\cite{bravyi2018correcting}. While the logical noise map on a prepared magic state also suffers the noise from magic-state injection and distillation, this noise can be twirled to stochastic Pauli noise with error-mitigated logical Clifford operations.
Nevertheless, there may be non-negligible coherence in quantum noise of logical Clifford gates. In this case, we can twirl the noise map per code cycle and remove the unitary component of the noise via logical Pauli operations with hardware update (See Appendix.\,\ref{sec:concrete_logical_operations} for the definition of hardware update). Below, we show that we can perform twirling on logical noise caused by logical Clifford operations if logical Pauli operations with physical operations, i.e., logical Pauli operations without updating Pauli frame, can be performed with negligible error rates. Note that the logical error rates for logical Pauli operations are expected to be sufficiently smaller than those for logical Clifford operations. This is because logical Pauli operations can be performed with transversal single-qubit operations that are completed in a single cycle, and the errors caused by these operations are negligible compared with those caused by two-qubit operations for stabilizer measurements.
Suppose that we perform a logical Clifford operation $\mathcal{C}$ and a logical noise map $\mathcal{M}_{\rm dec}^{\chi}$ follows it. Further suppose that we perform twirling noise $\mathcal{M}_{\rm dec}^{\chi}$ with a set of Pauli operators $\mathbb{S}$. The twirling process can be described as follows:
\begin{equation}
\begin{aligned}
\left(\frac{1}{|\mathbb{S}|}\sum_{\mathcal{P} \in \mathbb{S}} \mathcal{P} \mathcal{M}_{\rm dec}^{\chi} \mathcal{P}\right) \mathcal{C} = \frac{1}{|\mathbb{S}|} \sum_{\mathcal{P} \in \mathbb{S}} \mathcal{P} \mathcal{M}_{\rm dec}^{\chi} \mathcal{C} (\mathcal{C}^{\dagger} \mathcal{P} \mathcal{C}),
\end{aligned}
\end{equation}
where $\mathcal{P}$ is the superoperator of the logical Pauli operations. Since $\mathcal{C}^{\dagger} \mathcal{P} \mathcal{C}$ is a logical Pauli operation, $\mathcal{M}_{\rm dec}^{\chi}$ can be twirled simply with logical Pauli operations. The same arguments hold for Pauli measurements and feedback operations dependent on their outcomes. Since all the elemental logical operations except for magic-state injection are Clifford operations or Pauli channels, we can apply Pauli twirling to most of the quantum operations in FTQC. 
Note that while logical Pauli operations for computation can be done by updating the Pauli frame, the logical Pauli operations for twirling require a physical implementation on quantum devices. 
This is because when we attempt to perform logical Pauli operations via the Pauli frame for twirling, we need to update the Pauli frame according to the actual logical operation $\mathcal{M}_{\rm dec}^{\chi} \mathcal{C}$ (see Appendix.\,\ref{sec:concrete_logical_operations} for the formalism of the Pauli frame in superoperator representation). However, since $\mathcal{M}_{\rm dec}^{\chi}$ is not a stochastic Pauli noise, we cannot keep the Pauli frame as a Pauli operator; thus, we cannot continue tracking the frame as a Pauli operator. Thus, the Pauli twirling must be done physically.

\section {Details of the numerical analysis}
\label{sec:numercal_detail}
In the numerical simulations for evaluating the decoding errors, we used a uniform depolarizing noise model, in which noise occurs on each physical qubit independently and acts as follows:
\begin{equation}
\begin{aligned}
\mathcal{E}(\rho) = (1-p) \rho + \frac{p}{3}(X\rho X + Y\rho Y + Z \rho Z).
\label{Eq:depolarising}
\end{aligned}
\end{equation}
This error acts on data qubits at the beginning of each cycle and on ancillary qubits just before the measurement. As indicated in the main text, we assumed perfect syndrome measurements at the $0$-th and $d$-th cycles, which guarantees that the quantum states at these cycles are in logical space with recovery operations regardless of whether the decoding is successful or not. Then, we evaluated the logical error probabilities during these cycles.
To estimate the recovery operation, we used a minimum-weight perfect matching decoder~\cite{dennis2002topological}. This decoder reduces the decoding problem to an instance of the minimum-weight perfect matching problem. While this problem is NP-hard when there are Pauli-$Y$ errors, we can approximately solve it by using Edmonds' blossom algorithm~\cite{edmonds1965paths}. It is known that surface codes show threshold behavior even with this approximation. We used the implementation in Ref.\,\cite{kolmogorov2009blossom} for solving this problem. To estimate the logical error rate, we evaluated $10^5$ samples for each data point in Fig.\,\ref{fig:logical_error_zoom} and $10^6$ samples for the other figures.

In the performance evaluation of error mitigation for decoding errors, we assumed that the error channel for each logical gate is a non-uniform logical depolarizing channel obtained in the benchmark of the logical error probabilities in surface codes. Since there is no perfect syndrome measurement in practice, this assumption does not hold exactly. Nevertheless, this approximation is asymptotically correct, and thus, we used it to evaluate the performance of QEM in the case of logical errors.

For the simulation of the Clifford circuits, we used a stabilizer circuit simulator of which the memory allocations were optimized so that the updates for the actions of the Clifford operations become sequential. With this technique, the simulation of the stabilizer circuits that were dominated by Clifford operations rather than by Pauli measurements becomes hundreds of times faster than the existing stabilizer simulators~\cite{aaronson2004improved}.

For the Solovay-Kitaev algorithm, we used the method and implementation proposed in Ref.\,\cite{ross2014optimal}. While we need to limit the allowed number of $T$-gates, this method outputs a sequence of Clifford and $T$-gates according to the allowed error rate $\varepsilon$. Thus, we searched for the minimum error rate $\varepsilon^*$ with which the algorithm outputs a sequence wherein the number of $T$-gates is smaller than the allowed number of $T$-gates. 
This search involved using a simple bisection method, and it was repeated until the accuracy reached $10^{-14}$. For the simulation of the SWAP test circuits, we used Qulacs~\cite{suzuki2021qulacs}, which is a simulator for general noisy quantum circuits and is fast especially when a huge number of simulations have to be performed on small quantum circuits.

\section{Quantum error mitigation for decision problems}
\label{sec:apply_qem_to_bqp}

Several important algorithms in FTQC, such as prime factoring and calculation of the ground-state energy, are a procedure to obtain the computational results via phase estimation sampling~\cite{gidney2021factor,kivlichan2020improved} and are not a procedure to calculate expectation values of observables. 
Since typical QEM techniques are designed to reduce the bias in the expectation value caused by noise, it is not clear whether QEM can be applied to such sampling algorithms. In this section, we show that QEM can be utilized for mitigating errors in a wider range of problems than calculating expectation values. In particular, we show that several promising long-term algorithms, such as ground-state energy estimation via phase estimation sampling and Shor's factoring algorithm, can be decomposed into a series of decision problems, and show that QEM can be applied to each algorithm to solve the decision problem. To the best of our knowledge, this has not been mentioned in the context of QEM, while a similar concept has been known in the context of quasi-probability sampling for classically simulating quantum circuits\,\cite{bennink2017unbiased,hakkaku2021sampling}.

First, we show that ground-energy estimation with phase estimation sampling\,\cite{kivlichan2020improved,babbush2018encoding} can be decomposed into a series of decision problems with a bisection method. In a quantum phase estimation routine, we prepare an initial state, perform phase estimation based on quantum simulation. Then, we obtain a quantum state $\ket{\psi} = \sum_i \alpha_i \ket{\tilde{E}_i}\ket{\psi_i}$, where $\ket{\tilde{E}_i}$ is the energy of the $i$-th eigenstate or its relevant value in binary representation, $\ket{\psi_i}$ is an eigenstate of the given Hamiltonian, and $\alpha_i$ corresponds to the overlap between the $i$-th eigenstate and the initial state. The Pauli-$Z$ basis measurement is performed on the first register of $\ket{\psi}$, and $E_i$ is sampled with the probability $|\alpha_i|^2$.

\begin{figure}
    \centering
    \includegraphics[width=7.5cm]{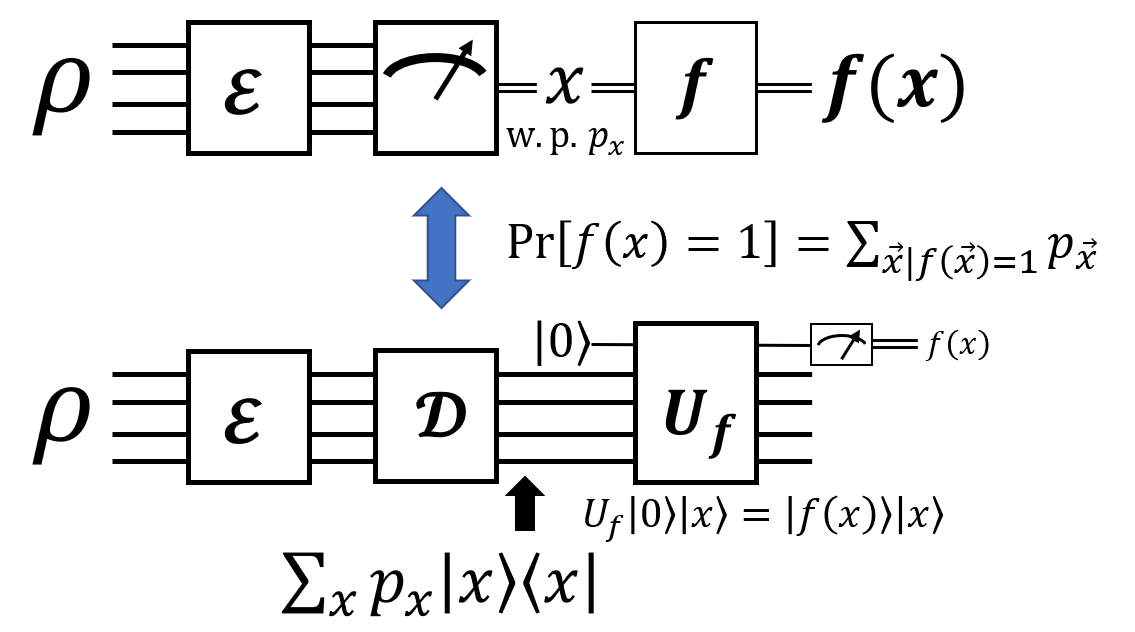}
    \caption{The diagram of circuit conversion from the original decision problem to fully quantum picture. Here, $\rho$ is a $n$-qubit initial state, $\mathcal{E}$ is a noisy process, $\mathcal{D}$ is a perfect dephasing noise map, $f$ is a classical binary function on $n$-bit string $x$, and $U_f$ is a quantum circuit that simulate $f$ as $U\ket{0}\ket{x} = \ket{f(x)}\ket{x}$. See the main text for details.}
    \label{fig:decision_circuit_convert}
\end{figure}

Now, we apply a bisection method to estimate the ground state energy. The subroutine outputs $1$ if the sampled energy is smaller than a given parameter $K$ and outputs $0$ otherwise. 
Suppose that the prepared initial state has an overlap with the ground state larger than $1/\textrm{poly}(n)$. If the ground energy is smaller than $K$, the subroutine outputs $1$ with probability more than $1/\textrm{poly}(n)$. Otherwise, the subroutine always outputs $0$. We denote the process before the Pauli-$Z$ measurements as $\mathcal{P}$, and the classical post-processing after the measurements to determine 0 or 1 as $f$. Then, we can conclude that the procedure as the sampling of a bit-string $x$ with the probability $\langle \langle \Pi_x | \mathcal{P} | 0 \rangle \rangle$, where $\Pi_x$ is the POVM element on the first register, and outputs a single bit $f(x)$ with the classical post-processing.
When the quantum process $\mathcal{P}$ suffers from errors, we can apply QEM to $\mathcal{P}$ to mitigate errors. While there may exists a complicated classical post-processing using a sampled bit-string after the quantum process $\mathcal{P}$, the combination of sampling $\Pi_x$ and classical post-processing $f$ can be interpreted as a quantum process with strong dephasing noise. The evaluation of decision bit $f(x)$ can also be interpreted as the evaluation of the Pauli-$Z$ operator. Therefore, we can interpret the whole subroutine as a fully quantum process for calculating the expectation value, and we can apply QEM to the subroutine. The process of circuit conversion is shown in Fig.\,\ref{fig:decision_circuit_convert}.

With this subroutine, we can solve the original ground-energy estimation with a bisection method. We assume the minimum and maximum possible energy $E_{\min}$ and $E_{\rm max}$ as variables, respectively. Then, the subroutine is called several times to check whether the ground energy is smaller than $E_{\rm mid} = (E_{\rm max} + E_{\rm min})/2$. If the ground energy is smaller than $(E_{\rm max} + E_{\rm min})/2$, the polynomial number of calls is enough to determine the inequality with high confidence. According to the output, we reduce a possible range of the ground energy $[E_{\rm min}, E_{\rm max}]$ by updating $E_{\rm max}$ or $E_{\rm min}$ with $E_{\rm mid}$. Since the range is halved in each iteration, we need $O(\log \epsilon)$ iterations to achieve $E_{\rm max} - E_{\rm min} < \epsilon$.

When the number of logical errors in each iteration is $O(1)$ on average, the sampling overhead due to the application of QEM is also constant. While this conversion enables the application of QEM, this requires $O(\log \epsilon)$-times more iterations for binary search compared with the original algorithm. Nevertheless, since these iterations are independent computational tasks and the total number of logical operations and hardware requirements per single execution do not change. Therefore, we can conclude that QEM can be applied not only for the evaluation of the expectation values but also for long-term applications.

For a fair comparison, the following fact should also be noted. In the case of prime factoring, we can apply the same decomposition to the algorithm by finding the minimum non-trivial factor of a given integer. However, when the noise model is stochastic and the sampling overhead of QEM is constant, i.e., the mean number of logical errors during FTQC is $O(1)$, we can obtain a correct answer for prime factoring with a small overhead without using QEM. Since prime factoring is in the NP class, we can efficiently check whether the submitted answer is correct or not. When the mean number of errors in FTQC is $O(1)$, i.e., the overhead of QEM is constant, a noiseless sampling occurs with a constant probability. This means we can obtain a correct non-trivial factor with constant sampling overheads even without QEM. Thus QEM is not effective for problems in the intersection of the BQP and NP class such as prime factoring. In other words, QEM is useful for problems satisfying the following two conditions; 1) problems are in the BQP class but not in the NP class, and 2) problems can be reduced to a series of decision problems. Useful algorithms for long-term applications, such as quantum simulation and the estimation of ground energy of spin models and molecules, satisfy these conditions.

\bibliography{bibQEC}

\begin{thebibliography}{70}%
\makeatletter
\providecommand \@ifxundefined [1]{%
 \@ifx{#1\undefined}
}%
\providecommand \@ifnum [1]{%
 \ifnum #1\expandafter \@firstoftwo
 \else \expandafter \@secondoftwo
 \fi
}%
\providecommand \@ifx [1]{%
 \ifx #1\expandafter \@firstoftwo
 \else \expandafter \@secondoftwo
 \fi
}%
\providecommand \natexlab [1]{#1}%
\providecommand \enquote  [1]{``#1''}%
\providecommand \bibnamefont  [1]{#1}%
\providecommand \bibfnamefont [1]{#1}%
\providecommand \citenamefont [1]{#1}%
\providecommand \href@noop [0]{\@secondoftwo}%
\providecommand \href [0]{\begingroup \@sanitize@url \@href}%
\providecommand \@href[1]{\@@startlink{#1}\@@href}%
\providecommand \@@href[1]{\endgroup#1\@@endlink}%
\providecommand \@sanitize@url [0]{\catcode `\\12\catcode `\$12\catcode
  `\&12\catcode `\#12\catcode `\^12\catcode `\_12\catcode `\%12\relax}%
\providecommand \@@startlink[1]{}%
\providecommand \@@endlink[0]{}%
\providecommand \url  [0]{\begingroup\@sanitize@url \@url }%
\providecommand \@url [1]{\endgroup\@href {#1}{\urlprefix }}%
\providecommand \urlprefix  [0]{URL }%
\providecommand \Eprint [0]{\href }%
\providecommand \doibase [0]{https://doi.org/}%
\providecommand \selectlanguage [0]{\@gobble}%
\providecommand \bibinfo  [0]{\@secondoftwo}%
\providecommand \bibfield  [0]{\@secondoftwo}%
\providecommand \translation [1]{[#1]}%
\providecommand \BibitemOpen [0]{}%
\providecommand \bibitemStop [0]{}%
\providecommand \bibitemNoStop [0]{.\EOS\space}%
\providecommand \EOS [0]{\spacefactor3000\relax}%
\providecommand \BibitemShut  [1]{\csname bibitem#1\endcsname}%
\let\auto@bib@innerbib\@empty
%</preamble>
\bibitem [{\citenamefont {Shor}(1999)}]{shor1999polynomial}%
  \BibitemOpen
  \bibfield  {author} {\bibinfo {author} {\bibfnamefont {P.~W.}\ \bibnamefont
  {Shor}},\ }\bibfield  {title} {\bibinfo {title} {Polynomial-time algorithms
  for prime factorization and discrete logarithms on a quantum computer},\
  }\href@noop {} {\bibfield  {journal} {\bibinfo  {journal} {SIAM review}\
  }\textbf {\bibinfo {volume} {41}},\ \bibinfo {pages} {303} (\bibinfo {year}
  {1999})}\BibitemShut {NoStop}%
\bibitem [{\citenamefont {Harrow}\ \emph {et~al.}(2009)\citenamefont {Harrow},
  \citenamefont {Hassidim},\ and\ \citenamefont {Lloyd}}]{harrow2009quantum}%
  \BibitemOpen
  \bibfield  {author} {\bibinfo {author} {\bibfnamefont {A.~W.}\ \bibnamefont
  {Harrow}}, \bibinfo {author} {\bibfnamefont {A.}~\bibnamefont {Hassidim}},\
  and\ \bibinfo {author} {\bibfnamefont {S.}~\bibnamefont {Lloyd}},\ }\bibfield
   {title} {\bibinfo {title} {Quantum algorithm for linear systems of
  equations},\ }\href@noop {} {\bibfield  {journal} {\bibinfo  {journal}
  {Physical review letters}\ }\textbf {\bibinfo {volume} {103}},\ \bibinfo
  {pages} {150502} (\bibinfo {year} {2009})}\BibitemShut {NoStop}%
\bibitem [{\citenamefont {Nielsen}\ and\ \citenamefont
  {Chuang}(2002)}]{nielsen2002quantum}%
  \BibitemOpen
  \bibfield  {author} {\bibinfo {author} {\bibfnamefont {M.~A.}\ \bibnamefont
  {Nielsen}}\ and\ \bibinfo {author} {\bibfnamefont {I.}~\bibnamefont
  {Chuang}},\ }\href@noop {} {\bibinfo {title} {Quantum computation and quantum
  information}} (\bibinfo {year} {2002})\BibitemShut {NoStop}%
\bibitem [{\citenamefont {Lidar}\ and\ \citenamefont
  {Brun}(2013)}]{lidar2013quantum}%
  \BibitemOpen
  \bibfield  {author} {\bibinfo {author} {\bibfnamefont {D.~A.}\ \bibnamefont
  {Lidar}}\ and\ \bibinfo {author} {\bibfnamefont {T.~A.}\ \bibnamefont
  {Brun}},\ }\href@noop {} {\emph {\bibinfo {title} {Quantum error
  correction}}}\ (\bibinfo  {publisher} {Cambridge University Press},\ \bibinfo
  {year} {2013})\BibitemShut {NoStop}%
\bibitem [{\citenamefont {Fowler}\ \emph
  {et~al.}(2012{\natexlab{a}})\citenamefont {Fowler}, \citenamefont
  {Mariantoni}, \citenamefont {Martinis},\ and\ \citenamefont
  {Cleland}}]{fowler2012surface}%
  \BibitemOpen
  \bibfield  {author} {\bibinfo {author} {\bibfnamefont {A.~G.}\ \bibnamefont
  {Fowler}}, \bibinfo {author} {\bibfnamefont {M.}~\bibnamefont {Mariantoni}},
  \bibinfo {author} {\bibfnamefont {J.~M.}\ \bibnamefont {Martinis}},\ and\
  \bibinfo {author} {\bibfnamefont {A.~N.}\ \bibnamefont {Cleland}},\
  }\bibfield  {title} {\bibinfo {title} {Surface codes: Towards practical
  large-scale quantum computation},\ }\href@noop {} {\bibfield  {journal}
  {\bibinfo  {journal} {Physical Review A}\ }\textbf {\bibinfo {volume} {86}},\
  \bibinfo {pages} {032324} (\bibinfo {year} {2012}{\natexlab{a}})}\BibitemShut
  {NoStop}%
\bibitem [{\citenamefont {Horsman}\ \emph {et~al.}(2012)\citenamefont
  {Horsman}, \citenamefont {Fowler}, \citenamefont {Devitt},\ and\
  \citenamefont {Van~Meter}}]{horsman2012surface}%
  \BibitemOpen
  \bibfield  {author} {\bibinfo {author} {\bibfnamefont {C.}~\bibnamefont
  {Horsman}}, \bibinfo {author} {\bibfnamefont {A.~G.}\ \bibnamefont {Fowler}},
  \bibinfo {author} {\bibfnamefont {S.}~\bibnamefont {Devitt}},\ and\ \bibinfo
  {author} {\bibfnamefont {R.}~\bibnamefont {Van~Meter}},\ }\bibfield  {title}
  {\bibinfo {title} {Surface code quantum computing by lattice surgery},\
  }\href@noop {} {\bibfield  {journal} {\bibinfo  {journal} {New Journal of
  Physics}\ }\textbf {\bibinfo {volume} {14}},\ \bibinfo {pages} {123011}
  (\bibinfo {year} {2012})}\BibitemShut {NoStop}%
\bibitem [{\citenamefont {Fowler}\ and\ \citenamefont
  {Gidney}(2018)}]{fowler2018low}%
  \BibitemOpen
  \bibfield  {author} {\bibinfo {author} {\bibfnamefont {A.~G.}\ \bibnamefont
  {Fowler}}\ and\ \bibinfo {author} {\bibfnamefont {C.}~\bibnamefont
  {Gidney}},\ }\bibfield  {title} {\bibinfo {title} {Low overhead quantum
  computation using lattice surgery},\ }\href@noop {} {\bibfield  {journal}
  {\bibinfo  {journal} {arXiv preprint arXiv:1808.06709}\ } (\bibinfo {year}
  {2018})}\BibitemShut {NoStop}%
\bibitem [{\citenamefont {Lloyd}(1996)}]{lloyd1996universal}%
  \BibitemOpen
  \bibfield  {author} {\bibinfo {author} {\bibfnamefont {S.}~\bibnamefont
  {Lloyd}},\ }\bibfield  {title} {\bibinfo {title} {Universal quantum
  simulators},\ }\href@noop {} {\bibfield  {journal} {\bibinfo  {journal}
  {Science}\ ,\ \bibinfo {pages} {1073}} (\bibinfo {year} {1996})}\BibitemShut
  {NoStop}%
\bibitem [{\citenamefont {Kivlichan}\ \emph {et~al.}(2020)\citenamefont
  {Kivlichan}, \citenamefont {Gidney}, \citenamefont {Berry}, \citenamefont
  {Wiebe}, \citenamefont {McClean}, \citenamefont {Sun}, \citenamefont {Jiang},
  \citenamefont {Rubin}, \citenamefont {Fowler}, \citenamefont {Aspuru-Guzik}
  \emph {et~al.}}]{kivlichan2020improved}%
  \BibitemOpen
  \bibfield  {author} {\bibinfo {author} {\bibfnamefont {I.~D.}\ \bibnamefont
  {Kivlichan}}, \bibinfo {author} {\bibfnamefont {C.}~\bibnamefont {Gidney}},
  \bibinfo {author} {\bibfnamefont {D.~W.}\ \bibnamefont {Berry}}, \bibinfo
  {author} {\bibfnamefont {N.}~\bibnamefont {Wiebe}}, \bibinfo {author}
  {\bibfnamefont {J.}~\bibnamefont {McClean}}, \bibinfo {author} {\bibfnamefont
  {W.}~\bibnamefont {Sun}}, \bibinfo {author} {\bibfnamefont {Z.}~\bibnamefont
  {Jiang}}, \bibinfo {author} {\bibfnamefont {N.}~\bibnamefont {Rubin}},
  \bibinfo {author} {\bibfnamefont {A.}~\bibnamefont {Fowler}}, \bibinfo
  {author} {\bibfnamefont {A.}~\bibnamefont {Aspuru-Guzik}}, \emph {et~al.},\
  }\bibfield  {title} {\bibinfo {title} {Improved fault-tolerant quantum
  simulation of condensed-phase correlated electrons via trotterization},\
  }\href@noop {} {\bibfield  {journal} {\bibinfo  {journal} {Quantum}\ }\textbf
  {\bibinfo {volume} {4}},\ \bibinfo {pages} {296} (\bibinfo {year}
  {2020})}\BibitemShut {NoStop}%
\bibitem [{\citenamefont {Babbush}\ \emph {et~al.}(2018)\citenamefont
  {Babbush}, \citenamefont {Gidney}, \citenamefont {Berry}, \citenamefont
  {Wiebe}, \citenamefont {McClean}, \citenamefont {Paler}, \citenamefont
  {Fowler},\ and\ \citenamefont {Neven}}]{babbush2018encoding}%
  \BibitemOpen
  \bibfield  {author} {\bibinfo {author} {\bibfnamefont {R.}~\bibnamefont
  {Babbush}}, \bibinfo {author} {\bibfnamefont {C.}~\bibnamefont {Gidney}},
  \bibinfo {author} {\bibfnamefont {D.~W.}\ \bibnamefont {Berry}}, \bibinfo
  {author} {\bibfnamefont {N.}~\bibnamefont {Wiebe}}, \bibinfo {author}
  {\bibfnamefont {J.}~\bibnamefont {McClean}}, \bibinfo {author} {\bibfnamefont
  {A.}~\bibnamefont {Paler}}, \bibinfo {author} {\bibfnamefont
  {A.}~\bibnamefont {Fowler}},\ and\ \bibinfo {author} {\bibfnamefont
  {H.}~\bibnamefont {Neven}},\ }\bibfield  {title} {\bibinfo {title} {Encoding
  electronic spectra in quantum circuits with linear t complexity},\
  }\href@noop {} {\bibfield  {journal} {\bibinfo  {journal} {Physical Review
  X}\ }\textbf {\bibinfo {volume} {8}},\ \bibinfo {pages} {041015} (\bibinfo
  {year} {2018})}\BibitemShut {NoStop}%
\bibitem [{\citenamefont {Holmes}\ \emph {et~al.}(2020)\citenamefont {Holmes},
  \citenamefont {Jokar}, \citenamefont {Pasandi}, \citenamefont {Ding},
  \citenamefont {Pedram},\ and\ \citenamefont {Chong}}]{holmes2020nisq+}%
  \BibitemOpen
  \bibfield  {author} {\bibinfo {author} {\bibfnamefont {A.}~\bibnamefont
  {Holmes}}, \bibinfo {author} {\bibfnamefont {M.~R.}\ \bibnamefont {Jokar}},
  \bibinfo {author} {\bibfnamefont {G.}~\bibnamefont {Pasandi}}, \bibinfo
  {author} {\bibfnamefont {Y.}~\bibnamefont {Ding}}, \bibinfo {author}
  {\bibfnamefont {M.}~\bibnamefont {Pedram}},\ and\ \bibinfo {author}
  {\bibfnamefont {F.~T.}\ \bibnamefont {Chong}},\ }\bibfield  {title} {\bibinfo
  {title} {Nisq+: Boosting quantum computing power by approximating quantum
  error correction},\ }\href@noop {} {\bibfield  {journal} {\bibinfo  {journal}
  {arXiv preprint arXiv:2004.04794}\ } (\bibinfo {year} {2020})}\BibitemShut
  {NoStop}%
\bibitem [{\citenamefont {Ueno}\ \emph {et~al.}(2021)\citenamefont {Ueno},
  \citenamefont {Kondo}, \citenamefont {Tanaka}, \citenamefont {Suzuki},\ and\
  \citenamefont {Tabuchi}}]{ueno2021qecool}%
  \BibitemOpen
  \bibfield  {author} {\bibinfo {author} {\bibfnamefont {Y.}~\bibnamefont
  {Ueno}}, \bibinfo {author} {\bibfnamefont {M.}~\bibnamefont {Kondo}},
  \bibinfo {author} {\bibfnamefont {M.}~\bibnamefont {Tanaka}}, \bibinfo
  {author} {\bibfnamefont {Y.}~\bibnamefont {Suzuki}},\ and\ \bibinfo {author}
  {\bibfnamefont {Y.}~\bibnamefont {Tabuchi}},\ }\bibfield  {title} {\bibinfo
  {title} {Qecool: On-line quantum error correction with a superconducting
  decoder for surface code},\ }\href@noop {} {\bibfield  {journal} {\bibinfo
  {journal} {arXiv preprint arXiv:2103.07526}\ } (\bibinfo {year}
  {2021})}\BibitemShut {NoStop}%
\bibitem [{\citenamefont {Das}\ \emph {et~al.}(2021)\citenamefont {Das},
  \citenamefont {Locharla},\ and\ \citenamefont {Jones}}]{das2021lilliput}%
  \BibitemOpen
  \bibfield  {author} {\bibinfo {author} {\bibfnamefont {P.}~\bibnamefont
  {Das}}, \bibinfo {author} {\bibfnamefont {A.}~\bibnamefont {Locharla}},\ and\
  \bibinfo {author} {\bibfnamefont {C.}~\bibnamefont {Jones}},\ }\bibfield
  {title} {\bibinfo {title} {Lilliput: A lightweight low-latency lookup-table
  based decoder for near-term quantum error correction},\ }\href@noop {}
  {\bibfield  {journal} {\bibinfo  {journal} {arXiv preprint arXiv:2108.06569}\
  } (\bibinfo {year} {2021})}\BibitemShut {NoStop}%
\bibitem [{\citenamefont {Arute}\ \emph {et~al.}(2019)\citenamefont {Arute},
  \citenamefont {Arya}, \citenamefont {Babbush}, \citenamefont {Bacon},
  \citenamefont {Bardin}, \citenamefont {Barends}, \citenamefont {Biswas},
  \citenamefont {Boixo}, \citenamefont {Brandao}, \citenamefont {Buell} \emph
  {et~al.}}]{arute2019quantum}%
  \BibitemOpen
  \bibfield  {author} {\bibinfo {author} {\bibfnamefont {F.}~\bibnamefont
  {Arute}}, \bibinfo {author} {\bibfnamefont {K.}~\bibnamefont {Arya}},
  \bibinfo {author} {\bibfnamefont {R.}~\bibnamefont {Babbush}}, \bibinfo
  {author} {\bibfnamefont {D.}~\bibnamefont {Bacon}}, \bibinfo {author}
  {\bibfnamefont {J.~C.}\ \bibnamefont {Bardin}}, \bibinfo {author}
  {\bibfnamefont {R.}~\bibnamefont {Barends}}, \bibinfo {author} {\bibfnamefont
  {R.}~\bibnamefont {Biswas}}, \bibinfo {author} {\bibfnamefont
  {S.}~\bibnamefont {Boixo}}, \bibinfo {author} {\bibfnamefont {F.~G.}\
  \bibnamefont {Brandao}}, \bibinfo {author} {\bibfnamefont {D.~A.}\
  \bibnamefont {Buell}}, \emph {et~al.},\ }\bibfield  {title} {\bibinfo {title}
  {Quantum supremacy using a programmable superconducting processor},\
  }\href@noop {} {\bibfield  {journal} {\bibinfo  {journal} {Nature}\ }\textbf
  {\bibinfo {volume} {574}},\ \bibinfo {pages} {505} (\bibinfo {year}
  {2019})}\BibitemShut {NoStop}%
\bibitem [{\citenamefont {Peruzzo}\ \emph {et~al.}(2014)\citenamefont
  {Peruzzo}, \citenamefont {McClean}, \citenamefont {Shadbolt}, \citenamefont
  {Yung}, \citenamefont {Zhou}, \citenamefont {Love}, \citenamefont
  {Aspuru-Guzik},\ and\ \citenamefont {O’brien}}]{peruzzo2014variational}%
  \BibitemOpen
  \bibfield  {author} {\bibinfo {author} {\bibfnamefont {A.}~\bibnamefont
  {Peruzzo}}, \bibinfo {author} {\bibfnamefont {J.}~\bibnamefont {McClean}},
  \bibinfo {author} {\bibfnamefont {P.}~\bibnamefont {Shadbolt}}, \bibinfo
  {author} {\bibfnamefont {M.-H.}\ \bibnamefont {Yung}}, \bibinfo {author}
  {\bibfnamefont {X.-Q.}\ \bibnamefont {Zhou}}, \bibinfo {author}
  {\bibfnamefont {P.~J.}\ \bibnamefont {Love}}, \bibinfo {author}
  {\bibfnamefont {A.}~\bibnamefont {Aspuru-Guzik}},\ and\ \bibinfo {author}
  {\bibfnamefont {J.~L.}\ \bibnamefont {O’brien}},\ }\bibfield  {title}
  {\bibinfo {title} {A variational eigenvalue solver on a photonic quantum
  processor},\ }\href@noop {} {\bibfield  {journal} {\bibinfo  {journal}
  {Nature communications}\ }\textbf {\bibinfo {volume} {5}},\ \bibinfo {pages}
  {1} (\bibinfo {year} {2014})}\BibitemShut {NoStop}%
\bibitem [{\citenamefont {Kandala}\ \emph {et~al.}(2017)\citenamefont
  {Kandala}, \citenamefont {Mezzacapo}, \citenamefont {Temme}, \citenamefont
  {Takita}, \citenamefont {Brink}, \citenamefont {Chow},\ and\ \citenamefont
  {Gambetta}}]{kandala2017hardware}%
  \BibitemOpen
  \bibfield  {author} {\bibinfo {author} {\bibfnamefont {A.}~\bibnamefont
  {Kandala}}, \bibinfo {author} {\bibfnamefont {A.}~\bibnamefont {Mezzacapo}},
  \bibinfo {author} {\bibfnamefont {K.}~\bibnamefont {Temme}}, \bibinfo
  {author} {\bibfnamefont {M.}~\bibnamefont {Takita}}, \bibinfo {author}
  {\bibfnamefont {M.}~\bibnamefont {Brink}}, \bibinfo {author} {\bibfnamefont
  {J.~M.}\ \bibnamefont {Chow}},\ and\ \bibinfo {author} {\bibfnamefont
  {J.~M.}\ \bibnamefont {Gambetta}},\ }\bibfield  {title} {\bibinfo {title}
  {Hardware-efficient variational quantum eigensolver for small molecules and
  quantum magnets},\ }\href@noop {} {\bibfield  {journal} {\bibinfo  {journal}
  {Nature}\ }\textbf {\bibinfo {volume} {549}},\ \bibinfo {pages} {242}
  (\bibinfo {year} {2017})}\BibitemShut {NoStop}%
\bibitem [{\citenamefont {McClean}\ \emph {et~al.}(2016)\citenamefont
  {McClean}, \citenamefont {Romero}, \citenamefont {Babbush},\ and\
  \citenamefont {Aspuru-Guzik}}]{mcclean2016theory}%
  \BibitemOpen
  \bibfield  {author} {\bibinfo {author} {\bibfnamefont {J.~R.}\ \bibnamefont
  {McClean}}, \bibinfo {author} {\bibfnamefont {J.}~\bibnamefont {Romero}},
  \bibinfo {author} {\bibfnamefont {R.}~\bibnamefont {Babbush}},\ and\ \bibinfo
  {author} {\bibfnamefont {A.}~\bibnamefont {Aspuru-Guzik}},\ }\bibfield
  {title} {\bibinfo {title} {The theory of variational hybrid quantum-classical
  algorithms},\ }\href@noop {} {\bibfield  {journal} {\bibinfo  {journal} {New
  Journal of Physics}\ }\textbf {\bibinfo {volume} {18}},\ \bibinfo {pages}
  {023023} (\bibinfo {year} {2016})}\BibitemShut {NoStop}%
\bibitem [{not()}]{note_blue_line}%
  \BibitemOpen
  \href@noop {} {}\bibinfo {note} {For drawing the blue line, we assumed that
  the ratio between physical error rates and the threshold is $0.1$ and the
  quantum algorithms are run with $O(n^3)$ time complexity for the problem size
  $n$ with the number of physical qubits being fixed.}\BibitemShut {Stop}%
\bibitem [{\citenamefont {Temme}\ \emph {et~al.}(2017)\citenamefont {Temme},
  \citenamefont {Bravyi},\ and\ \citenamefont {Gambetta}}]{temme2017error}%
  \BibitemOpen
  \bibfield  {author} {\bibinfo {author} {\bibfnamefont {K.}~\bibnamefont
  {Temme}}, \bibinfo {author} {\bibfnamefont {S.}~\bibnamefont {Bravyi}},\ and\
  \bibinfo {author} {\bibfnamefont {J.~M.}\ \bibnamefont {Gambetta}},\
  }\bibfield  {title} {\bibinfo {title} {Error mitigation for short-depth
  quantum circuits},\ }\href@noop {} {\bibfield  {journal} {\bibinfo  {journal}
  {Physical review letters}\ }\textbf {\bibinfo {volume} {119}},\ \bibinfo
  {pages} {180509} (\bibinfo {year} {2017})}\BibitemShut {NoStop}%
\bibitem [{\citenamefont {Endo}\ \emph {et~al.}(2018)\citenamefont {Endo},
  \citenamefont {Benjamin},\ and\ \citenamefont {Li}}]{endo2018practical}%
  \BibitemOpen
  \bibfield  {author} {\bibinfo {author} {\bibfnamefont {S.}~\bibnamefont
  {Endo}}, \bibinfo {author} {\bibfnamefont {S.~C.}\ \bibnamefont {Benjamin}},\
  and\ \bibinfo {author} {\bibfnamefont {Y.}~\bibnamefont {Li}},\ }\bibfield
  {title} {\bibinfo {title} {Practical quantum error mitigation for near-future
  applications},\ }\href@noop {} {\bibfield  {journal} {\bibinfo  {journal}
  {Physical Review X}\ }\textbf {\bibinfo {volume} {8}},\ \bibinfo {pages}
  {031027} (\bibinfo {year} {2018})}\BibitemShut {NoStop}%
\bibitem [{\citenamefont {Takagi}\ \emph {et~al.}(2021)\citenamefont {Takagi},
  \citenamefont {Endo}, \citenamefont {Minagawa},\ and\ \citenamefont
  {Gu}}]{takagi2021fundamental}%
  \BibitemOpen
  \bibfield  {author} {\bibinfo {author} {\bibfnamefont {R.}~\bibnamefont
  {Takagi}}, \bibinfo {author} {\bibfnamefont {S.}~\bibnamefont {Endo}},
  \bibinfo {author} {\bibfnamefont {S.}~\bibnamefont {Minagawa}},\ and\
  \bibinfo {author} {\bibfnamefont {M.}~\bibnamefont {Gu}},\ }\bibfield
  {title} {\bibinfo {title} {Fundamental limitations of quantum error
  mitigation},\ }\href@noop {} {\bibfield  {journal} {\bibinfo  {journal}
  {arXiv preprint arXiv:2109.04457}\ } (\bibinfo {year} {2021})}\BibitemShut
  {NoStop}%
\bibitem [{\citenamefont {Wang}\ \emph {et~al.}(2021)\citenamefont {Wang},
  \citenamefont {Czarnik}, \citenamefont {Arrasmith}, \citenamefont {Cerezo},
  \citenamefont {Cincio},\ and\ \citenamefont {Coles}}]{wang2021can}%
  \BibitemOpen
  \bibfield  {author} {\bibinfo {author} {\bibfnamefont {S.}~\bibnamefont
  {Wang}}, \bibinfo {author} {\bibfnamefont {P.}~\bibnamefont {Czarnik}},
  \bibinfo {author} {\bibfnamefont {A.}~\bibnamefont {Arrasmith}}, \bibinfo
  {author} {\bibfnamefont {M.}~\bibnamefont {Cerezo}}, \bibinfo {author}
  {\bibfnamefont {L.}~\bibnamefont {Cincio}},\ and\ \bibinfo {author}
  {\bibfnamefont {P.~J.}\ \bibnamefont {Coles}},\ }\bibfield  {title} {\bibinfo
  {title} {Can error mitigation improve trainability of noisy variational
  quantum algorithms?},\ }\href@noop {} {\bibfield  {journal} {\bibinfo
  {journal} {arXiv preprint arXiv:2109.01051}\ } (\bibinfo {year}
  {2021})}\BibitemShut {NoStop}%
\bibitem [{\citenamefont {Kitaev}(1997)}]{kitaev1997quantum}%
  \BibitemOpen
  \bibfield  {author} {\bibinfo {author} {\bibfnamefont {A.~Y.}\ \bibnamefont
  {Kitaev}},\ }\bibfield  {title} {\bibinfo {title} {Quantum computations:
  algorithms and error correction},\ }\href@noop {} {\bibfield  {journal}
  {\bibinfo  {journal} {Russian Mathematical Surveys}\ }\textbf {\bibinfo
  {volume} {52}},\ \bibinfo {pages} {1191} (\bibinfo {year}
  {1997})}\BibitemShut {NoStop}%
\bibitem [{\citenamefont {Dawson}\ and\ \citenamefont
  {Nielsen}(2005)}]{dawson2005solovay}%
  \BibitemOpen
  \bibfield  {author} {\bibinfo {author} {\bibfnamefont {C.~M.}\ \bibnamefont
  {Dawson}}\ and\ \bibinfo {author} {\bibfnamefont {M.~A.}\ \bibnamefont
  {Nielsen}},\ }\bibfield  {title} {\bibinfo {title} {The solovay-kitaev
  algorithm},\ }\href@noop {} {\bibfield  {journal} {\bibinfo  {journal} {arXiv
  preprint quant-ph/0505030}\ } (\bibinfo {year} {2005})}\BibitemShut {NoStop}%
\bibitem [{\citenamefont {Blume-Kohout}\ \emph {et~al.}(2013)\citenamefont
  {Blume-Kohout}, \citenamefont {Gamble}, \citenamefont {Nielsen},
  \citenamefont {Mizrahi}, \citenamefont {Sterk},\ and\ \citenamefont
  {Maunz}}]{blume2013robust}%
  \BibitemOpen
  \bibfield  {author} {\bibinfo {author} {\bibfnamefont {R.}~\bibnamefont
  {Blume-Kohout}}, \bibinfo {author} {\bibfnamefont {J.~K.}\ \bibnamefont
  {Gamble}}, \bibinfo {author} {\bibfnamefont {E.}~\bibnamefont {Nielsen}},
  \bibinfo {author} {\bibfnamefont {J.}~\bibnamefont {Mizrahi}}, \bibinfo
  {author} {\bibfnamefont {J.~D.}\ \bibnamefont {Sterk}},\ and\ \bibinfo
  {author} {\bibfnamefont {P.}~\bibnamefont {Maunz}},\ }\bibfield  {title}
  {\bibinfo {title} {Robust, self-consistent, closed-form tomography of quantum
  logic gates on a trapped ion qubit},\ }\href@noop {} {\bibfield  {journal}
  {\bibinfo  {journal} {arXiv preprint arXiv:1310.4492}\ } (\bibinfo {year}
  {2013})}\BibitemShut {NoStop}%
\bibitem [{\citenamefont {Greenbaum}(2015)}]{greenbaum2015introduction}%
  \BibitemOpen
  \bibfield  {author} {\bibinfo {author} {\bibfnamefont {D.}~\bibnamefont
  {Greenbaum}},\ }\bibfield  {title} {\bibinfo {title} {Introduction to quantum
  gate set tomography},\ }\href@noop {} {\bibfield  {journal} {\bibinfo
  {journal} {arXiv preprint arXiv:1509.02921}\ } (\bibinfo {year}
  {2015})}\BibitemShut {NoStop}%
\bibitem [{\citenamefont {Kitaev}(1995)}]{kitaev1995quantum}%
  \BibitemOpen
  \bibfield  {author} {\bibinfo {author} {\bibfnamefont {A.~Y.}\ \bibnamefont
  {Kitaev}},\ }\bibfield  {title} {\bibinfo {title} {Quantum measurements and
  the abelian stabilizer problem},\ }\href@noop {} {\bibfield  {journal}
  {\bibinfo  {journal} {arXiv preprint quant-ph/9511026}\ } (\bibinfo {year}
  {1995})}\BibitemShut {NoStop}%
\bibitem [{\citenamefont {Gidney}\ and\ \citenamefont
  {Eker{\aa}}(2021)}]{gidney2021factor}%
  \BibitemOpen
  \bibfield  {author} {\bibinfo {author} {\bibfnamefont {C.}~\bibnamefont
  {Gidney}}\ and\ \bibinfo {author} {\bibfnamefont {M.}~\bibnamefont
  {Eker{\aa}}},\ }\bibfield  {title} {\bibinfo {title} {How to factor 2048 bit
  rsa integers in 8 hours using 20 million noisy qubits},\ }\href@noop {}
  {\bibfield  {journal} {\bibinfo  {journal} {Quantum}\ }\textbf {\bibinfo
  {volume} {5}},\ \bibinfo {pages} {433} (\bibinfo {year} {2021})}\BibitemShut
  {NoStop}%
\bibitem [{\citenamefont {Li}\ and\ \citenamefont
  {Benjamin}(2017)}]{li2017efficient}%
  \BibitemOpen
  \bibfield  {author} {\bibinfo {author} {\bibfnamefont {Y.}~\bibnamefont
  {Li}}\ and\ \bibinfo {author} {\bibfnamefont {S.~C.}\ \bibnamefont
  {Benjamin}},\ }\bibfield  {title} {\bibinfo {title} {Efficient variational
  quantum simulator incorporating active error minimization},\ }\href@noop {}
  {\bibfield  {journal} {\bibinfo  {journal} {Physical Review X}\ }\textbf
  {\bibinfo {volume} {7}},\ \bibinfo {pages} {021050} (\bibinfo {year}
  {2017})}\BibitemShut {NoStop}%
\bibitem [{\citenamefont {Gottesman}(1997)}]{gottesman1997stabilizer}%
  \BibitemOpen
  \bibfield  {author} {\bibinfo {author} {\bibfnamefont {D.}~\bibnamefont
  {Gottesman}},\ }\bibfield  {title} {\bibinfo {title} {Stabilizer codes and
  quantum error correction},\ }\href@noop {} {\bibfield  {journal} {\bibinfo
  {journal} {arXiv preprint quant-ph/9705052}\ } (\bibinfo {year}
  {1997})}\BibitemShut {NoStop}%
\bibitem [{\citenamefont {Eastin}\ and\ \citenamefont
  {Knill}(2009)}]{eastin2009restrictions}%
  \BibitemOpen
  \bibfield  {author} {\bibinfo {author} {\bibfnamefont {B.}~\bibnamefont
  {Eastin}}\ and\ \bibinfo {author} {\bibfnamefont {E.}~\bibnamefont {Knill}},\
  }\bibfield  {title} {\bibinfo {title} {Restrictions on transversal encoded
  quantum gate sets},\ }\href@noop {} {\bibfield  {journal} {\bibinfo
  {journal} {Physical review letters}\ }\textbf {\bibinfo {volume} {102}},\
  \bibinfo {pages} {110502} (\bibinfo {year} {2009})}\BibitemShut {NoStop}%
\bibitem [{\citenamefont {Riesebos}\ \emph {et~al.}(2017)\citenamefont
  {Riesebos}, \citenamefont {Fu}, \citenamefont {Varsamopoulos}, \citenamefont
  {Almudever},\ and\ \citenamefont {Bertels}}]{riesebos2017pauli}%
  \BibitemOpen
  \bibfield  {author} {\bibinfo {author} {\bibfnamefont {L.}~\bibnamefont
  {Riesebos}}, \bibinfo {author} {\bibfnamefont {X.}~\bibnamefont {Fu}},
  \bibinfo {author} {\bibfnamefont {S.}~\bibnamefont {Varsamopoulos}}, \bibinfo
  {author} {\bibfnamefont {C.~G.}\ \bibnamefont {Almudever}},\ and\ \bibinfo
  {author} {\bibfnamefont {K.}~\bibnamefont {Bertels}},\ }\bibfield  {title}
  {\bibinfo {title} {Pauli frames for quantum computer architectures},\ }in\
  \href@noop {} {\emph {\bibinfo {booktitle} {Proceedings of the 54th Annual
  Design Automation Conference 2017}}}\ (\bibinfo {year} {2017})\ pp.\ \bibinfo
  {pages} {1--6}\BibitemShut {NoStop}%
\bibitem [{\citenamefont {Bravyi}\ \emph {et~al.}(2018)\citenamefont {Bravyi},
  \citenamefont {Englbrecht}, \citenamefont {K{\"o}nig},\ and\ \citenamefont
  {Peard}}]{bravyi2018correcting}%
  \BibitemOpen
  \bibfield  {author} {\bibinfo {author} {\bibfnamefont {S.}~\bibnamefont
  {Bravyi}}, \bibinfo {author} {\bibfnamefont {M.}~\bibnamefont {Englbrecht}},
  \bibinfo {author} {\bibfnamefont {R.}~\bibnamefont {K{\"o}nig}},\ and\
  \bibinfo {author} {\bibfnamefont {N.}~\bibnamefont {Peard}},\ }\bibfield
  {title} {\bibinfo {title} {Correcting coherent errors with surface codes},\
  }\href@noop {} {\bibfield  {journal} {\bibinfo  {journal} {npj Quantum
  Information}\ }\textbf {\bibinfo {volume} {4}},\ \bibinfo {pages} {1}
  (\bibinfo {year} {2018})}\BibitemShut {NoStop}%
\bibitem [{\citenamefont {Jones}\ \emph {et~al.}(2012)\citenamefont {Jones},
  \citenamefont {Van~Meter}, \citenamefont {Fowler}, \citenamefont {McMahon},
  \citenamefont {Kim}, \citenamefont {Ladd},\ and\ \citenamefont
  {Yamamoto}}]{jones2012layered}%
  \BibitemOpen
  \bibfield  {author} {\bibinfo {author} {\bibfnamefont {N.~C.}\ \bibnamefont
  {Jones}}, \bibinfo {author} {\bibfnamefont {R.}~\bibnamefont {Van~Meter}},
  \bibinfo {author} {\bibfnamefont {A.~G.}\ \bibnamefont {Fowler}}, \bibinfo
  {author} {\bibfnamefont {P.~L.}\ \bibnamefont {McMahon}}, \bibinfo {author}
  {\bibfnamefont {J.}~\bibnamefont {Kim}}, \bibinfo {author} {\bibfnamefont
  {T.~D.}\ \bibnamefont {Ladd}},\ and\ \bibinfo {author} {\bibfnamefont
  {Y.}~\bibnamefont {Yamamoto}},\ }\bibfield  {title} {\bibinfo {title}
  {Layered architecture for quantum computing},\ }\href@noop {} {\bibfield
  {journal} {\bibinfo  {journal} {Physical Review X}\ }\textbf {\bibinfo
  {volume} {2}},\ \bibinfo {pages} {031007} (\bibinfo {year}
  {2012})}\BibitemShut {NoStop}%
\bibitem [{\citenamefont {Fowler}\ \emph
  {et~al.}(2012{\natexlab{b}})\citenamefont {Fowler}, \citenamefont
  {Whiteside},\ and\ \citenamefont {Hollenberg}}]{fowler2012towards}%
  \BibitemOpen
  \bibfield  {author} {\bibinfo {author} {\bibfnamefont {A.~G.}\ \bibnamefont
  {Fowler}}, \bibinfo {author} {\bibfnamefont {A.~C.}\ \bibnamefont
  {Whiteside}},\ and\ \bibinfo {author} {\bibfnamefont {L.~C.}\ \bibnamefont
  {Hollenberg}},\ }\bibfield  {title} {\bibinfo {title} {Towards practical
  classical processing for the surface code},\ }\href@noop {} {\bibfield
  {journal} {\bibinfo  {journal} {Physical review letters}\ }\textbf {\bibinfo
  {volume} {108}},\ \bibinfo {pages} {180501} (\bibinfo {year}
  {2012}{\natexlab{b}})}\BibitemShut {NoStop}%
\bibitem [{\citenamefont {Ross}\ and\ \citenamefont
  {Selinger}(2014)}]{ross2014optimal}%
  \BibitemOpen
  \bibfield  {author} {\bibinfo {author} {\bibfnamefont {N.~J.}\ \bibnamefont
  {Ross}}\ and\ \bibinfo {author} {\bibfnamefont {P.}~\bibnamefont
  {Selinger}},\ }\bibfield  {title} {\bibinfo {title} {Optimal ancilla-free
  {Clifford+T} approximation of z-rotations},\ }\href@noop {} {\bibfield
  {journal} {\bibinfo  {journal} {arXiv preprint arXiv:1403.2975}\ } (\bibinfo
  {year} {2014})}\BibitemShut {NoStop}%
\bibitem [{\citenamefont {Campbell}(2019)}]{campbell2019random}%
  \BibitemOpen
  \bibfield  {author} {\bibinfo {author} {\bibfnamefont {E.}~\bibnamefont
  {Campbell}},\ }\bibfield  {title} {\bibinfo {title} {Random compiler for fast
  hamiltonian simulation},\ }\href@noop {} {\bibfield  {journal} {\bibinfo
  {journal} {Physical review letters}\ }\textbf {\bibinfo {volume} {123}},\
  \bibinfo {pages} {070503} (\bibinfo {year} {2019})}\BibitemShut {NoStop}%
\bibitem [{\citenamefont {Nielsen}\ \emph {et~al.}(2020)\citenamefont
  {Nielsen}, \citenamefont {Gamble}, \citenamefont {Rudinger}, \citenamefont
  {Scholten}, \citenamefont {Young},\ and\ \citenamefont
  {Blume-Kohout}}]{nielsen2020gate}%
  \BibitemOpen
  \bibfield  {author} {\bibinfo {author} {\bibfnamefont {E.}~\bibnamefont
  {Nielsen}}, \bibinfo {author} {\bibfnamefont {J.~K.}\ \bibnamefont {Gamble}},
  \bibinfo {author} {\bibfnamefont {K.}~\bibnamefont {Rudinger}}, \bibinfo
  {author} {\bibfnamefont {T.}~\bibnamefont {Scholten}}, \bibinfo {author}
  {\bibfnamefont {K.}~\bibnamefont {Young}},\ and\ \bibinfo {author}
  {\bibfnamefont {R.}~\bibnamefont {Blume-Kohout}},\ }\bibfield  {title}
  {\bibinfo {title} {Gate set tomography},\ }\href@noop {} {\bibfield
  {journal} {\bibinfo  {journal} {arXiv preprint arXiv:2009.07301}\ } (\bibinfo
  {year} {2020})}\BibitemShut {NoStop}%
\bibitem [{\citenamefont {Endo}\ \emph {et~al.}(2020)\citenamefont {Endo},
  \citenamefont {Cai}, \citenamefont {Benjamin},\ and\ \citenamefont
  {Yuan}}]{endo2020hybrid}%
  \BibitemOpen
  \bibfield  {author} {\bibinfo {author} {\bibfnamefont {S.}~\bibnamefont
  {Endo}}, \bibinfo {author} {\bibfnamefont {Z.}~\bibnamefont {Cai}}, \bibinfo
  {author} {\bibfnamefont {S.~C.}\ \bibnamefont {Benjamin}},\ and\ \bibinfo
  {author} {\bibfnamefont {X.}~\bibnamefont {Yuan}},\ }\bibfield  {title}
  {\bibinfo {title} {Hybrid quantum-classical algorithms and quantum error
  mitigation},\ }\href@noop {} {\bibfield  {journal} {\bibinfo  {journal}
  {arXiv preprint arXiv:2011.01382}\ } (\bibinfo {year} {2020})}\BibitemShut
  {NoStop}%
\bibitem [{\citenamefont {Bravyi}\ and\ \citenamefont
  {Gosset}(2016)}]{bravyi2016improved}%
  \BibitemOpen
  \bibfield  {author} {\bibinfo {author} {\bibfnamefont {S.}~\bibnamefont
  {Bravyi}}\ and\ \bibinfo {author} {\bibfnamefont {D.}~\bibnamefont
  {Gosset}},\ }\bibfield  {title} {\bibinfo {title} {Improved classical
  simulation of quantum circuits dominated by clifford gates},\ }\href@noop {}
  {\bibfield  {journal} {\bibinfo  {journal} {Physical review letters}\
  }\textbf {\bibinfo {volume} {116}},\ \bibinfo {pages} {250501} (\bibinfo
  {year} {2016})}\BibitemShut {NoStop}%
\bibitem [{\citenamefont {Piveteau}\ \emph {et~al.}(2021)\citenamefont
  {Piveteau}, \citenamefont {Sutter}, \citenamefont {Bravyi}, \citenamefont
  {Gambetta},\ and\ \citenamefont {Temme}}]{piveteau2021error}%
  \BibitemOpen
  \bibfield  {author} {\bibinfo {author} {\bibfnamefont {C.}~\bibnamefont
  {Piveteau}}, \bibinfo {author} {\bibfnamefont {D.}~\bibnamefont {Sutter}},
  \bibinfo {author} {\bibfnamefont {S.}~\bibnamefont {Bravyi}}, \bibinfo
  {author} {\bibfnamefont {J.~M.}\ \bibnamefont {Gambetta}},\ and\ \bibinfo
  {author} {\bibfnamefont {K.}~\bibnamefont {Temme}},\ }\bibfield  {title}
  {\bibinfo {title} {Error mitigation for universal gates on encoded qubits},\
  }\href@noop {} {\bibfield  {journal} {\bibinfo  {journal} {arXiv preprint
  arXiv:2103.04915}\ } (\bibinfo {year} {2021})}\BibitemShut {NoStop}%
\bibitem [{\citenamefont {Wang}\ \emph {et~al.}(2003)\citenamefont {Wang},
  \citenamefont {Harrington},\ and\ \citenamefont
  {Preskill}}]{wang2003confinement}%
  \BibitemOpen
  \bibfield  {author} {\bibinfo {author} {\bibfnamefont {C.}~\bibnamefont
  {Wang}}, \bibinfo {author} {\bibfnamefont {J.}~\bibnamefont {Harrington}},\
  and\ \bibinfo {author} {\bibfnamefont {J.}~\bibnamefont {Preskill}},\
  }\bibfield  {title} {\bibinfo {title} {Confinement-higgs transition in a
  disordered gauge theory and the accuracy threshold for quantum memory},\
  }\href@noop {} {\bibfield  {journal} {\bibinfo  {journal} {Annals of
  Physics}\ }\textbf {\bibinfo {volume} {303}},\ \bibinfo {pages} {31}
  (\bibinfo {year} {2003})}\BibitemShut {NoStop}%
\bibitem [{\citenamefont {Delfosse}\ and\ \citenamefont
  {Nickerson}(2017)}]{delfosse2017almost}%
  \BibitemOpen
  \bibfield  {author} {\bibinfo {author} {\bibfnamefont {N.}~\bibnamefont
  {Delfosse}}\ and\ \bibinfo {author} {\bibfnamefont {N.~H.}\ \bibnamefont
  {Nickerson}},\ }\bibfield  {title} {\bibinfo {title} {Almost-linear time
  decoding algorithm for topological codes},\ }\href@noop {} {\bibfield
  {journal} {\bibinfo  {journal} {arXiv preprint arXiv:1709.06218}\ } (\bibinfo
  {year} {2017})}\BibitemShut {NoStop}%
\bibitem [{\citenamefont {Dennis}\ \emph {et~al.}(2002)\citenamefont {Dennis},
  \citenamefont {Kitaev}, \citenamefont {Landahl},\ and\ \citenamefont
  {Preskill}}]{dennis2002topological}%
  \BibitemOpen
  \bibfield  {author} {\bibinfo {author} {\bibfnamefont {E.}~\bibnamefont
  {Dennis}}, \bibinfo {author} {\bibfnamefont {A.}~\bibnamefont {Kitaev}},
  \bibinfo {author} {\bibfnamefont {A.}~\bibnamefont {Landahl}},\ and\ \bibinfo
  {author} {\bibfnamefont {J.}~\bibnamefont {Preskill}},\ }\bibfield  {title}
  {\bibinfo {title} {Topological quantum memory},\ }\href@noop {} {\bibfield
  {journal} {\bibinfo  {journal} {Journal of Mathematical Physics}\ }\textbf
  {\bibinfo {volume} {43}},\ \bibinfo {pages} {4452} (\bibinfo {year}
  {2002})}\BibitemShut {NoStop}%
\bibitem [{\citenamefont {Edmonds}(1965)}]{edmonds1965paths}%
  \BibitemOpen
  \bibfield  {author} {\bibinfo {author} {\bibfnamefont {J.}~\bibnamefont
  {Edmonds}},\ }\bibfield  {title} {\bibinfo {title} {Paths, trees, and
  flowers},\ }\href@noop {} {\bibfield  {journal} {\bibinfo  {journal}
  {Canadian Journal of mathematics}\ }\textbf {\bibinfo {volume} {17}},\
  \bibinfo {pages} {449} (\bibinfo {year} {1965})}\BibitemShut {NoStop}%
\bibitem [{\citenamefont {Strikis}\ \emph {et~al.}(2020)\citenamefont
  {Strikis}, \citenamefont {Qin}, \citenamefont {Chen}, \citenamefont
  {Benjamin},\ and\ \citenamefont {Li}}]{strikis2020learning}%
  \BibitemOpen
  \bibfield  {author} {\bibinfo {author} {\bibfnamefont {A.}~\bibnamefont
  {Strikis}}, \bibinfo {author} {\bibfnamefont {D.}~\bibnamefont {Qin}},
  \bibinfo {author} {\bibfnamefont {Y.}~\bibnamefont {Chen}}, \bibinfo {author}
  {\bibfnamefont {S.~C.}\ \bibnamefont {Benjamin}},\ and\ \bibinfo {author}
  {\bibfnamefont {Y.}~\bibnamefont {Li}},\ }\bibfield  {title} {\bibinfo
  {title} {Learning-based quantum error mitigation},\ }\href@noop {} {\bibfield
   {journal} {\bibinfo  {journal} {arXiv preprint arXiv:2005.07601}\ }
  (\bibinfo {year} {2020})}\BibitemShut {NoStop}%
\bibitem [{\citenamefont {Aaronson}\ and\ \citenamefont
  {Gottesman}(2004)}]{aaronson2004improved}%
  \BibitemOpen
  \bibfield  {author} {\bibinfo {author} {\bibfnamefont {S.}~\bibnamefont
  {Aaronson}}\ and\ \bibinfo {author} {\bibfnamefont {D.}~\bibnamefont
  {Gottesman}},\ }\bibfield  {title} {\bibinfo {title} {Improved simulation of
  stabilizer circuits},\ }\href@noop {} {\bibfield  {journal} {\bibinfo
  {journal} {Physical Review A}\ }\textbf {\bibinfo {volume} {70}},\ \bibinfo
  {pages} {052328} (\bibinfo {year} {2004})}\BibitemShut {NoStop}%
\bibitem [{\citenamefont {Ekert}\ \emph {et~al.}(2002)\citenamefont {Ekert},
  \citenamefont {Alves}, \citenamefont {Oi}, \citenamefont {Horodecki},
  \citenamefont {Horodecki},\ and\ \citenamefont {Kwek}}]{ekert2002direct}%
  \BibitemOpen
  \bibfield  {author} {\bibinfo {author} {\bibfnamefont {A.~K.}\ \bibnamefont
  {Ekert}}, \bibinfo {author} {\bibfnamefont {C.~M.}\ \bibnamefont {Alves}},
  \bibinfo {author} {\bibfnamefont {D.~K.}\ \bibnamefont {Oi}}, \bibinfo
  {author} {\bibfnamefont {M.}~\bibnamefont {Horodecki}}, \bibinfo {author}
  {\bibfnamefont {P.}~\bibnamefont {Horodecki}},\ and\ \bibinfo {author}
  {\bibfnamefont {L.~C.}\ \bibnamefont {Kwek}},\ }\bibfield  {title} {\bibinfo
  {title} {Direct estimations of linear and nonlinear functionals of a quantum
  state},\ }\href@noop {} {\bibfield  {journal} {\bibinfo  {journal} {Physical
  review letters}\ }\textbf {\bibinfo {volume} {88}},\ \bibinfo {pages}
  {217901} (\bibinfo {year} {2002})}\BibitemShut {NoStop}%
\bibitem [{\citenamefont {Pednault}\ \emph {et~al.}(2019)\citenamefont
  {Pednault}, \citenamefont {Gunnels}, \citenamefont {Nannicini}, \citenamefont
  {Horesh},\ and\ \citenamefont {Wisnieff}}]{pednault2019leveraging}%
  \BibitemOpen
  \bibfield  {author} {\bibinfo {author} {\bibfnamefont {E.}~\bibnamefont
  {Pednault}}, \bibinfo {author} {\bibfnamefont {J.~A.}\ \bibnamefont
  {Gunnels}}, \bibinfo {author} {\bibfnamefont {G.}~\bibnamefont {Nannicini}},
  \bibinfo {author} {\bibfnamefont {L.}~\bibnamefont {Horesh}},\ and\ \bibinfo
  {author} {\bibfnamefont {R.}~\bibnamefont {Wisnieff}},\ }\bibfield  {title}
  {\bibinfo {title} {Leveraging secondary storage to simulate deep 54-qubit
  sycamore circuits},\ }\href@noop {} {\bibfield  {journal} {\bibinfo
  {journal} {arXiv preprint arXiv:1910.09534}\ } (\bibinfo {year}
  {2019})}\BibitemShut {NoStop}%
\bibitem [{\citenamefont {Suzuki}(1991)}]{suzuki1991general}%
  \BibitemOpen
  \bibfield  {author} {\bibinfo {author} {\bibfnamefont {M.}~\bibnamefont
  {Suzuki}},\ }\bibfield  {title} {\bibinfo {title} {General theory of fractal
  path integrals with applications to many-body theories and statistical
  physics},\ }\href@noop {} {\bibfield  {journal} {\bibinfo  {journal} {Journal
  of Mathematical Physics}\ }\textbf {\bibinfo {volume} {32}},\ \bibinfo
  {pages} {400} (\bibinfo {year} {1991})}\BibitemShut {NoStop}%
\bibitem [{\citenamefont {Low}\ and\ \citenamefont
  {Chuang}(2019)}]{low2019hamiltonian}%
  \BibitemOpen
  \bibfield  {author} {\bibinfo {author} {\bibfnamefont {G.~H.}\ \bibnamefont
  {Low}}\ and\ \bibinfo {author} {\bibfnamefont {I.~L.}\ \bibnamefont
  {Chuang}},\ }\bibfield  {title} {\bibinfo {title} {Hamiltonian simulation by
  qubitization},\ }\href@noop {} {\bibfield  {journal} {\bibinfo  {journal}
  {Quantum}\ }\textbf {\bibinfo {volume} {3}},\ \bibinfo {pages} {163}
  (\bibinfo {year} {2019})}\BibitemShut {NoStop}%
\bibitem [{\citenamefont {Berry}\ \emph {et~al.}(2015)\citenamefont {Berry},
  \citenamefont {Childs}, \citenamefont {Cleve}, \citenamefont {Kothari},\ and\
  \citenamefont {Somma}}]{berry2015simulating}%
  \BibitemOpen
  \bibfield  {author} {\bibinfo {author} {\bibfnamefont {D.~W.}\ \bibnamefont
  {Berry}}, \bibinfo {author} {\bibfnamefont {A.~M.}\ \bibnamefont {Childs}},
  \bibinfo {author} {\bibfnamefont {R.}~\bibnamefont {Cleve}}, \bibinfo
  {author} {\bibfnamefont {R.}~\bibnamefont {Kothari}},\ and\ \bibinfo {author}
  {\bibfnamefont {R.~D.}\ \bibnamefont {Somma}},\ }\bibfield  {title} {\bibinfo
  {title} {Simulating hamiltonian dynamics with a truncated taylor series},\
  }\href@noop {} {\bibfield  {journal} {\bibinfo  {journal} {Physical review
  letters}\ }\textbf {\bibinfo {volume} {114}},\ \bibinfo {pages} {090502}
  (\bibinfo {year} {2015})}\BibitemShut {NoStop}%
\bibitem [{\citenamefont {Low}\ and\ \citenamefont
  {Chuang}(2017)}]{low2017optimal}%
  \BibitemOpen
  \bibfield  {author} {\bibinfo {author} {\bibfnamefont {G.~H.}\ \bibnamefont
  {Low}}\ and\ \bibinfo {author} {\bibfnamefont {I.~L.}\ \bibnamefont
  {Chuang}},\ }\bibfield  {title} {\bibinfo {title} {Optimal hamiltonian
  simulation by quantum signal processing},\ }\href@noop {} {\bibfield
  {journal} {\bibinfo  {journal} {Physical review letters}\ }\textbf {\bibinfo
  {volume} {118}},\ \bibinfo {pages} {010501} (\bibinfo {year}
  {2017})}\BibitemShut {NoStop}%
\bibitem [{\citenamefont {Endo}\ \emph {et~al.}(2019)\citenamefont {Endo},
  \citenamefont {Zhao}, \citenamefont {Li}, \citenamefont {Benjamin},\ and\
  \citenamefont {Yuan}}]{endo2019mitigating}%
  \BibitemOpen
  \bibfield  {author} {\bibinfo {author} {\bibfnamefont {S.}~\bibnamefont
  {Endo}}, \bibinfo {author} {\bibfnamefont {Q.}~\bibnamefont {Zhao}}, \bibinfo
  {author} {\bibfnamefont {Y.}~\bibnamefont {Li}}, \bibinfo {author}
  {\bibfnamefont {S.}~\bibnamefont {Benjamin}},\ and\ \bibinfo {author}
  {\bibfnamefont {X.}~\bibnamefont {Yuan}},\ }\bibfield  {title} {\bibinfo
  {title} {Mitigating algorithmic errors in a hamiltonian simulation},\
  }\href@noop {} {\bibfield  {journal} {\bibinfo  {journal} {Physical Review
  A}\ }\textbf {\bibinfo {volume} {99}},\ \bibinfo {pages} {012334} (\bibinfo
  {year} {2019})}\BibitemShut {NoStop}%
\bibitem [{\citenamefont {Vazquez}\ \emph {et~al.}(2020)\citenamefont
  {Vazquez}, \citenamefont {Hiptmair},\ and\ \citenamefont
  {Woerner}}]{vazquez2020enhancing}%
  \BibitemOpen
  \bibfield  {author} {\bibinfo {author} {\bibfnamefont {A.~C.}\ \bibnamefont
  {Vazquez}}, \bibinfo {author} {\bibfnamefont {R.}~\bibnamefont {Hiptmair}},\
  and\ \bibinfo {author} {\bibfnamefont {S.}~\bibnamefont {Woerner}},\
  }\bibfield  {title} {\bibinfo {title} {Enhancing the quantum linear systems
  algorithm using richardson extrapolation},\ }\href@noop {} {\bibfield
  {journal} {\bibinfo  {journal} {arXiv preprint arXiv:2009.04484}\ } (\bibinfo
  {year} {2020})}\BibitemShut {NoStop}%
\bibitem [{\citenamefont {Mitarai}\ and\ \citenamefont
  {Fujii}(2019)}]{mitarai2019constructing}%
  \BibitemOpen
  \bibfield  {author} {\bibinfo {author} {\bibfnamefont {K.}~\bibnamefont
  {Mitarai}}\ and\ \bibinfo {author} {\bibfnamefont {K.}~\bibnamefont
  {Fujii}},\ }\bibfield  {title} {\bibinfo {title} {Constructing a virtual
  two-qubit gate from single-qubit operations},\ }\href@noop {} {\bibfield
  {journal} {\bibinfo  {journal} {arXiv preprint arXiv:1909.07534}\ } (\bibinfo
  {year} {2019})}\BibitemShut {NoStop}%
\bibitem [{\citenamefont {Mitarai}\ and\ \citenamefont
  {Fujii}(2020)}]{mitarai2020overhead}%
  \BibitemOpen
  \bibfield  {author} {\bibinfo {author} {\bibfnamefont {K.}~\bibnamefont
  {Mitarai}}\ and\ \bibinfo {author} {\bibfnamefont {K.}~\bibnamefont
  {Fujii}},\ }\bibfield  {title} {\bibinfo {title} {Overhead of the
  non-local-to-local channel decomposition by quasiprobability sampling},\
  }\href@noop {} {\bibfield  {journal} {\bibinfo  {journal} {arXiv preprint
  arXiv:2006.11174}\ } (\bibinfo {year} {2020})}\BibitemShut {NoStop}%
\bibitem [{\citenamefont {McClean}\ \emph {et~al.}(2020)\citenamefont
  {McClean}, \citenamefont {Jiang}, \citenamefont {Rubin}, \citenamefont
  {Babbush},\ and\ \citenamefont {Neven}}]{mcclean2020decoding}%
  \BibitemOpen
  \bibfield  {author} {\bibinfo {author} {\bibfnamefont {J.~R.}\ \bibnamefont
  {McClean}}, \bibinfo {author} {\bibfnamefont {Z.}~\bibnamefont {Jiang}},
  \bibinfo {author} {\bibfnamefont {N.~C.}\ \bibnamefont {Rubin}}, \bibinfo
  {author} {\bibfnamefont {R.}~\bibnamefont {Babbush}},\ and\ \bibinfo {author}
  {\bibfnamefont {H.}~\bibnamefont {Neven}},\ }\bibfield  {title} {\bibinfo
  {title} {Decoding quantum errors with subspace expansions},\ }\href@noop {}
  {\bibfield  {journal} {\bibinfo  {journal} {Nature Communications}\ }\textbf
  {\bibinfo {volume} {11}},\ \bibinfo {pages} {1} (\bibinfo {year}
  {2020})}\BibitemShut {NoStop}%
\bibitem [{\citenamefont {McClean}\ \emph {et~al.}(2017)\citenamefont
  {McClean}, \citenamefont {Kimchi-Schwartz}, \citenamefont {Carter},\ and\
  \citenamefont {De~Jong}}]{mcclean2017hybrid}%
  \BibitemOpen
  \bibfield  {author} {\bibinfo {author} {\bibfnamefont {J.~R.}\ \bibnamefont
  {McClean}}, \bibinfo {author} {\bibfnamefont {M.~E.}\ \bibnamefont
  {Kimchi-Schwartz}}, \bibinfo {author} {\bibfnamefont {J.}~\bibnamefont
  {Carter}},\ and\ \bibinfo {author} {\bibfnamefont {W.~A.}\ \bibnamefont
  {De~Jong}},\ }\bibfield  {title} {\bibinfo {title} {Hybrid quantum-classical
  hierarchy for mitigation of decoherence and determination of excited
  states},\ }\href@noop {} {\bibfield  {journal} {\bibinfo  {journal} {Physical
  Review A}\ }\textbf {\bibinfo {volume} {95}},\ \bibinfo {pages} {042308}
  (\bibinfo {year} {2017})}\BibitemShut {NoStop}%
\bibitem [{\citenamefont {Xiong}\ \emph {et~al.}(2020)\citenamefont {Xiong},
  \citenamefont {Chandra}, \citenamefont {Ng},\ and\ \citenamefont
  {Hanzo}}]{xiong2020sampling}%
  \BibitemOpen
  \bibfield  {author} {\bibinfo {author} {\bibfnamefont {Y.}~\bibnamefont
  {Xiong}}, \bibinfo {author} {\bibfnamefont {D.}~\bibnamefont {Chandra}},
  \bibinfo {author} {\bibfnamefont {S.~X.}\ \bibnamefont {Ng}},\ and\ \bibinfo
  {author} {\bibfnamefont {L.}~\bibnamefont {Hanzo}},\ }\bibfield  {title}
  {\bibinfo {title} {Sampling overhead analysis of quantum error mitigation:
  Uncoded vs. coded systems},\ }\href@noop {} {\bibfield  {journal} {\bibinfo
  {journal} {IEEE Access}\ } (\bibinfo {year} {2020})}\BibitemShut {NoStop}%
\bibitem [{\citenamefont {Lostaglio}\ and\ \citenamefont
  {Ciani}(2021)}]{lostaglio2021error}%
  \BibitemOpen
  \bibfield  {author} {\bibinfo {author} {\bibfnamefont {M.}~\bibnamefont
  {Lostaglio}}\ and\ \bibinfo {author} {\bibfnamefont {A.}~\bibnamefont
  {Ciani}},\ }\bibfield  {title} {\bibinfo {title} {Error mitigation and
  quantum-assisted simulation in the error corrected regime},\ }\href@noop {}
  {\bibfield  {journal} {\bibinfo  {journal} {arXiv preprint arXiv:2103.07526}\
  } (\bibinfo {year} {2021})}\BibitemShut {NoStop}%
\bibitem [{\citenamefont {Bravyi}\ and\ \citenamefont
  {Kitaev}(1998)}]{bravyi1998quantum}%
  \BibitemOpen
  \bibfield  {author} {\bibinfo {author} {\bibfnamefont {S.~B.}\ \bibnamefont
  {Bravyi}}\ and\ \bibinfo {author} {\bibfnamefont {A.~Y.}\ \bibnamefont
  {Kitaev}},\ }\bibfield  {title} {\bibinfo {title} {Quantum codes on a lattice
  with boundary},\ }\href@noop {} {\bibfield  {journal} {\bibinfo  {journal}
  {arXiv preprint quant-ph/9811052}\ } (\bibinfo {year} {1998})}\BibitemShut
  {NoStop}%
\bibitem [{\citenamefont {Brown}\ \emph {et~al.}(2017)\citenamefont {Brown},
  \citenamefont {Laubscher}, \citenamefont {Kesselring},\ and\ \citenamefont
  {Wootton}}]{brown2017poking}%
  \BibitemOpen
  \bibfield  {author} {\bibinfo {author} {\bibfnamefont {B.~J.}\ \bibnamefont
  {Brown}}, \bibinfo {author} {\bibfnamefont {K.}~\bibnamefont {Laubscher}},
  \bibinfo {author} {\bibfnamefont {M.~S.}\ \bibnamefont {Kesselring}},\ and\
  \bibinfo {author} {\bibfnamefont {J.~R.}\ \bibnamefont {Wootton}},\
  }\bibfield  {title} {\bibinfo {title} {Poking holes and cutting corners to
  achieve clifford gates with the surface code},\ }\href@noop {} {\bibfield
  {journal} {\bibinfo  {journal} {Physical Review X}\ }\textbf {\bibinfo
  {volume} {7}},\ \bibinfo {pages} {021029} (\bibinfo {year}
  {2017})}\BibitemShut {NoStop}%
\bibitem [{\citenamefont {Gidney}\ and\ \citenamefont
  {Fowler}(2019)}]{gidney2019efficient}%
  \BibitemOpen
  \bibfield  {author} {\bibinfo {author} {\bibfnamefont {C.}~\bibnamefont
  {Gidney}}\ and\ \bibinfo {author} {\bibfnamefont {A.~G.}\ \bibnamefont
  {Fowler}},\ }\bibfield  {title} {\bibinfo {title} {Efficient magic state
  factories with a catalyzed {$|CCZ\rangle$ to $2| T\rangle$} transformation},\
  }\href@noop {} {\bibfield  {journal} {\bibinfo  {journal} {Quantum}\ }\textbf
  {\bibinfo {volume} {3}},\ \bibinfo {pages} {135} (\bibinfo {year}
  {2019})}\BibitemShut {NoStop}%
\bibitem [{\citenamefont {Li}(2015)}]{li2015magic}%
  \BibitemOpen
  \bibfield  {author} {\bibinfo {author} {\bibfnamefont {Y.}~\bibnamefont
  {Li}},\ }\bibfield  {title} {\bibinfo {title} {A magic state’s fidelity can
  be superior to the operations that created it},\ }\href@noop {} {\bibfield
  {journal} {\bibinfo  {journal} {New Journal of Physics}\ }\textbf {\bibinfo
  {volume} {17}},\ \bibinfo {pages} {023037} (\bibinfo {year}
  {2015})}\BibitemShut {NoStop}%
\bibitem [{\citenamefont {Trout}\ and\ \citenamefont
  {Brown}(2015)}]{trout2015magic}%
  \BibitemOpen
  \bibfield  {author} {\bibinfo {author} {\bibfnamefont {C.~J.}\ \bibnamefont
  {Trout}}\ and\ \bibinfo {author} {\bibfnamefont {K.~R.}\ \bibnamefont
  {Brown}},\ }\bibfield  {title} {\bibinfo {title} {Magic state distillation
  and gate compilation in quantum algorithms for quantum chemistry},\
  }\href@noop {} {\bibfield  {journal} {\bibinfo  {journal} {International
  Journal of Quantum Chemistry}\ }\textbf {\bibinfo {volume} {115}},\ \bibinfo
  {pages} {1296} (\bibinfo {year} {2015})}\BibitemShut {NoStop}%
\bibitem [{\citenamefont {Kolmogorov}(2009)}]{kolmogorov2009blossom}%
  \BibitemOpen
  \bibfield  {author} {\bibinfo {author} {\bibfnamefont {V.}~\bibnamefont
  {Kolmogorov}},\ }\bibfield  {title} {\bibinfo {title} {Blossom v: a new
  implementation of a minimum cost perfect matching algorithm},\ }\href@noop {}
  {\bibfield  {journal} {\bibinfo  {journal} {Mathematical Programming
  Computation}\ }\textbf {\bibinfo {volume} {1}},\ \bibinfo {pages} {43}
  (\bibinfo {year} {2009})}\BibitemShut {NoStop}%
\bibitem [{\citenamefont {Suzuki}\ \emph {et~al.}(2021)\citenamefont {Suzuki},
  \citenamefont {Kawase}, \citenamefont {Masumura}, \citenamefont {Hiraga},
  \citenamefont {Nakadai}, \citenamefont {Chen}, \citenamefont {Nakanishi},
  \citenamefont {Mitarai}, \citenamefont {Imai}, \citenamefont {Tamiya} \emph
  {et~al.}}]{suzuki2021qulacs}%
  \BibitemOpen
  \bibfield  {author} {\bibinfo {author} {\bibfnamefont {Y.}~\bibnamefont
  {Suzuki}}, \bibinfo {author} {\bibfnamefont {Y.}~\bibnamefont {Kawase}},
  \bibinfo {author} {\bibfnamefont {Y.}~\bibnamefont {Masumura}}, \bibinfo
  {author} {\bibfnamefont {Y.}~\bibnamefont {Hiraga}}, \bibinfo {author}
  {\bibfnamefont {M.}~\bibnamefont {Nakadai}}, \bibinfo {author} {\bibfnamefont
  {J.}~\bibnamefont {Chen}}, \bibinfo {author} {\bibfnamefont {K.~M.}\
  \bibnamefont {Nakanishi}}, \bibinfo {author} {\bibfnamefont {K.}~\bibnamefont
  {Mitarai}}, \bibinfo {author} {\bibfnamefont {R.}~\bibnamefont {Imai}},
  \bibinfo {author} {\bibfnamefont {S.}~\bibnamefont {Tamiya}}, \emph
  {et~al.},\ }\bibfield  {title} {\bibinfo {title} {Qulacs: a fast and
  versatile quantum circuit simulator for research purpose},\ }\href@noop {}
  {\bibfield  {journal} {\bibinfo  {journal} {Quantum}\ }\textbf {\bibinfo
  {volume} {5}},\ \bibinfo {pages} {559} (\bibinfo {year} {2021})}\BibitemShut
  {NoStop}%
\bibitem [{\citenamefont {Bennink}\ \emph {et~al.}(2017)\citenamefont
  {Bennink}, \citenamefont {Ferragut}, \citenamefont {Humble}, \citenamefont
  {Laska}, \citenamefont {Nutaro}, \citenamefont {Pleszkoch},\ and\
  \citenamefont {Pooser}}]{bennink2017unbiased}%
  \BibitemOpen
  \bibfield  {author} {\bibinfo {author} {\bibfnamefont {R.~S.}\ \bibnamefont
  {Bennink}}, \bibinfo {author} {\bibfnamefont {E.~M.}\ \bibnamefont
  {Ferragut}}, \bibinfo {author} {\bibfnamefont {T.~S.}\ \bibnamefont
  {Humble}}, \bibinfo {author} {\bibfnamefont {J.~A.}\ \bibnamefont {Laska}},
  \bibinfo {author} {\bibfnamefont {J.~J.}\ \bibnamefont {Nutaro}}, \bibinfo
  {author} {\bibfnamefont {M.~G.}\ \bibnamefont {Pleszkoch}},\ and\ \bibinfo
  {author} {\bibfnamefont {R.~C.}\ \bibnamefont {Pooser}},\ }\bibfield  {title}
  {\bibinfo {title} {Unbiased simulation of near-clifford quantum circuits},\
  }\href@noop {} {\bibfield  {journal} {\bibinfo  {journal} {Physical Review
  A}\ }\textbf {\bibinfo {volume} {95}},\ \bibinfo {pages} {062337} (\bibinfo
  {year} {2017})}\BibitemShut {NoStop}%
\bibitem [{\citenamefont {Hakkaku}\ \emph {et~al.}(2021)\citenamefont
  {Hakkaku}, \citenamefont {Mitarai},\ and\ \citenamefont
  {Fujii}}]{hakkaku2021sampling}%
  \BibitemOpen
  \bibfield  {author} {\bibinfo {author} {\bibfnamefont {S.}~\bibnamefont
  {Hakkaku}}, \bibinfo {author} {\bibfnamefont {K.}~\bibnamefont {Mitarai}},\
  and\ \bibinfo {author} {\bibfnamefont {K.}~\bibnamefont {Fujii}},\ }\bibfield
   {title} {\bibinfo {title} {A sampling-based quasi-probability simulation for
  fault-tolerant quantum error correction on the surface codes under coherent
  noise},\ }\href@noop {} {\bibfield  {journal} {\bibinfo  {journal} {arXiv
  preprint arXiv:2105.04478}\ } (\bibinfo {year} {2021})}\BibitemShut {NoStop}%
\end{thebibliography}%

\end{document}